\documentclass[12pt]{iopart}

\usepackage{iopams}

\usepackage[utf8]{inputenc}
\usepackage{CJKutf8}
\usepackage{kotex}
\usepackage{graphicx}

\usepackage{algpseudocode} 
\usepackage{mathrsfs}

\usepackage{amsfonts}
\usepackage{amssymb}
\usepackage{accents}
\usepackage{bm}
\usepackage{color}
\usepackage{graphicx}
\usepackage{amsthm}
\usepackage{enumerate}
\usepackage{mathrsfs}
\usepackage{epsfig}
\usepackage{caption}
\usepackage{subcaption}
\usepackage{hyperref}
\usepackage{multirow}
\usepackage[dvipsnames]{xcolor}
\usepackage{algorithm}
\usepackage{algpseudocode}
\usepackage{orcidlink}
\usepackage{ulem}

\usepackage{pgf,tikz,colortbl}
\usetikzlibrary{arrows}
\usetikzlibrary{babel}
\usetikzlibrary{calc}
\usepackage{zref-savepos}

\newcounter{NoTableEntry}
\renewcommand*{\theNoTableEntry}{NTE-\the\value{NoTableEntry}}

\newcommand*{\notableentry}{%
  \multicolumn{1}{@{}c@{}}{%
    \stepcounter{NoTableEntry}%
    \vadjust pre{\zsavepos{\theNoTableEntry t}}
    \vadjust{\zsavepos{\theNoTableEntry b}}
    \zsavepos{\theNoTableEntry l}
    \hspace{0pt plus 1filll}%
    \zsavepos{\theNoTableEntry r}
    \tikz[overlay]{%
      \draw[black]
        let
          \n{llx}={\zposx{\theNoTableEntry l}sp-\zposx{\theNoTableEntry r}sp},
          \n{urx}={0},
          \n{lly}={\zposy{\theNoTableEntry b}sp-\zposy{\theNoTableEntry r}sp},
          \n{ury}={\zposy{\theNoTableEntry t}sp-\zposy{\theNoTableEntry r}sp}
        in
        (\n{llx}, \n{lly}) -- (\n{urx}, \n{ury})
        (\n{llx}, \n{ury}) -- (\n{urx}, \n{lly})
      ;
    }%
  }%
}

\def\sizeV{N}
\def\sizeW{M}
\def\V{\mathcal{V}}
\def\W{\mathcal{W}}
\def\E{\mathcal{E}}
\def\Hy{\mathcal{H}}
\def\IRight{\mathbf{I}^{\rightarrow}}
\def\ILeft{\mathbf{I}^{\leftarrow}}
\def\I{\mathbf{I}}
\def\degreeOut{k^{\rm out}}
\def\degreeIn{k^{\rm in}}
\def\cardOut{\chi^{\rm out}}
\def\cardIn{\chi^{\rm in}}
\def\define{:=}

\graphicspath{{Figures/}}

\begin{document}

\title[]{Connected components in networks with higher-order interactions}

\author{Gyeong-Gyun Ha(하경균)$^*$\,\orcidlink{0009-0009-9298-1806}, Izaak Neri\,\orcidlink{0000-0001-9529-5742} and Alessia Annibale\,\orcidlink{0000-0003-4010-6742}}

\address{Department of Mathematics, King's College London, Strand, London, WC2R 2LS, UK}
\address{$^*$Author to whom any correspondence should be addressed.}
\ead{gyeong-gyun.ha@kcl.ac.uk}
\vspace{10pt}

\begin{abstract}
We address the problem of defining connected components in hypergraphs, which are models for systems with higher-order interactions.    For graphs with dyadic interactions, connected components are defined in terms of paths connecting nodes along the graph.   However, defining connected components in hypergraphs  is a more involved problem,  as  one needs to consider the higher-order nature of the  interactions associated with the hyperedge.  Higher-order interactions can be taken into consideration through a logic associated with the hyperedges, two examples being OR-logic and AND-logic; these logical operations can be considered  two limiting cases corresponding to non-cooperative and fully cooperative interactions, respectively.  In this paper we show how connected components can be defined in hypergraphs with OR- or AND-logic.    While  OR-logic and  AND-logic   provide the same connected components for nondirected hypergraphs, for directed hypergraphs the strongly connected component of AND-logic is a subset of the OR-logic strongly connected component.  Interestingly,  higher-order interactions change the general  topological properties of connected components in directed hypergraphs.     Notably, while for directed graphs the strongly connected component is the intersection of its  in- and out-component, in hypergraphs with AND-logic the intersection of in- and out-component does not equal the   strongly connected component.  We develop a theory for the fraction of nodes that are part of the largest connected component and through comparison with real-world data we show that degree-cardinality correlations play a significant role. 
\end{abstract}

%
%
%
%
%

\section{Introduction}

Network science has traditionally focused on dyadic interactions, where links connect pairs of nodes~\cite{newman2006structure, barabasi,dorogovtsev2022nature}. However, real-world systems often exhibit multi-party interactions  that can  be represented as hyperedges in a hypergraph.   Multi-party interactions can be cooperative, and we refer to them as higher-order interactions~\cite{battiston2020networks}.    Examples of higher-order interactions are  social interactions, as individuals can behave differently  tête-à-tête  than in large groups~\cite{iacopini2024temporal, iacopini2024not}, and gene-regulatory interactions as a gene  may require the presence of multiple transcription factors for activation~\cite{moon2012genetic, hannam2019percolation}.  
At present it remains  challenging to study dynamical   systems  with higher-order interactions, as these  involve nonlinear effects.

For networks with dyadic interactions, connected components play an important role in the dynamics of processes defined on them.    For nondirected graphs, a   connected component is a sub-graph for which there exist a path between   any pair of its nodes~\cite{erdos1960evolution,bollobas1984evolution}.   At high connectivity, the largest connected component of a random graph grows  linearly with the total number of nodes, and we speak of a giant component~\cite{newman1}.    
The existence of a giant component is a requirement for the observation of various emergent or collective phenomena on networks, such as a ferromagnetic or spin-glass phase transition in spin models on random graphs, see e.g. Chapter 5 in Ref.~\cite{hartmann2006phase} and ~\cite{Annibale_2010}, or large scale epidemic outbreaks on networks of contacts~\cite{callaway2000network,newman2002spread, newman2007component}.     
For directed networks,  the relevant concept is the giant strongly connected component.  A  sub-graph is strongly connected if  every node can be reached from any other node within the sub-graph, and vice versa, meaning that every node in the sub-graph can reach every other node~\cite{broder2000graph, newman1, dorogovtsev2001giant, kryven2016emergence,timar2017mapping}.     The existence of a  giant strongly connected component is a requirement for  observing emergent phenomena on large directed graphs, for example,  phase transitions in spin models on  large directed graphs, including transitions from  a paramagnetic to   a ferromagnetic phase~\cite{derrida1987exactly,hatchett2004parallel, neri2009cavity,aurell2017cavity} or  transitions from an ordered to  a chaotic phase~\cite{derrida1987dynamical, derrida1987exactly,hatchett2004parallel, neri2009cavity};    spectra of large  random directed graphs that have continuous components  and delocalised eigenvectors~\cite{neri2020linear};  and  dynamical systems with a large number of attractors, including fixed points, periodic cycles, or  chaotic attractors~\cite{correale2006computational}.

To extend the  theory of connected components to higher-order networks we need to model the higher-order interactions.  
The most straightforward approach is to represent higher-order interactions as a second set of nodes, and in this way one recovers a bipartite graph to which the  definitions  of connected components of graphs apply, by treating   both set of nodes as vertices of the bipartite graph. In this case, a node belongs to the connected component if at least one of its neighbours belongs to the connected component.  Therefore,  we refer to this approach as the OR-logic approach.     However, such an approach does not consider the possibility of cooperativity.   Therefore we consider a second approach for which a hyperedge belongs to a connected component only if all of its in-neighbours belong to the connected component.    Such connected components are  motivated by gene regulatory networks~\cite{hannam2019percolation,torrisi2020percolation}, as  genes require sometimes the presence of multiple transcription factors for activation.   Note that Ref.~\cite{bianconi2024theory} defines a similar concept for percolation theory on hypergraphs.   

In this Paper, we  formalise connected components within OR-logic and AND-logic for both nondirected and directed hypergraphs.   While for nondirected hypergraphs these are the same, we show that for directed hypergraphs AND-logic yields different components from OR-logic.   Furthermore, we derive  generic topological properties of AND-logic components and discuss how they are distinct from those within OR-logic.  We also develop an algorithm to determine the AND-logic connected components of directed hypergraphs.  Subsequently,  we investigate the size and properties of the largest connected component within OR-logic and AND-logic, in both nondirected and directed hypergraphs.  We develop a theory based on the cavity method that applies to infinitely large random hypergraphs, and we compare obtained theoretical results with data from empirical and synthetic hypergraphs.
We find that degree-cardinality correlations play an important role for characterising largest connected components in real-world hypergraphs.

The paper is structured as follows. 
In Sec.~\ref{ch:definitions}, we define  hypergraphs and introduce the notation  used in this paper. In Sec.~\ref{sec:conn}, we define  connected components in nondirected and directed hypergraphs with OR-logic and AND-logic,  we derive generic properties of those connected components, and we develop an algorithm to find the AND-logic connected components of hypergraphs. In Secs.~\ref{ch:nondir_GC} and  \ref{ch:dir_GC}, we analyse the connected components of nondirected and directed hypergraphs, respectively. Concretely, we use the cavity method to estimate the fraction of nodes that belong to the largest connected component of an ensemble of random hypergraphs that have prescribed correlations between the degrees and  cardinalities of neighbouring vertices. We compare these estimates with empirical numerics found in real-world hypergraphs. Conclusions are given in Sec.~\ref{ch:discussion}, and the paper ends with several appendices containing technical details.

\section{Hypergraphs: basic definitions}\label{ch:definitions}

A hypergraph is a triplet $\Hy=(\V, \W, \E)$   consisting of a set $\V$ of $\sizeV=|\V|$ nodes, a set of $\W$ of $\sizeW=|\W|$ hyperedges, and a set $\E$ of directed links for which $\E\subseteq (\V\times \W) \cup (\W\times \V)$~\cite{bretto2013hypergraph}.   We call $\V\cup\W$ the set of vertices, and hence a vertex can be both a node or a hyperedge.    We denote nodes by roman indices, $a,b\in \V$, and hyperedges by Greek indices $\alpha,\beta\in \W$.     The set of directed links $\E$ consists of pairs $(a,\alpha)$ with $a\in \V$ and $\alpha\in \W$ and pairs $(\alpha,a)$ with  $\alpha\in \W$ and $a\in \V$.   For a hypergraph, each pair $(a,\alpha)$ occurs at most once in the set $\E$, and the hypergraph is  nondirected when $(a,\alpha)\in \E$ implies that also  $(\alpha,a)\in \E$ .      A sub-hypergraph $\mathcal{H}'=(\V',\W',\E')$ of $\mathcal{H}=(\V,\W,\E)$ is a hypergraph that satisfies $\V'\subseteq \V$, $\W'\subseteq\W$
and $\E'\subseteq \E$ 
 and we denote this property by $\Hy'\subseteq \Hy$.  

We represent directed hypergraphs with a pair of incidence matrices $\mathbf{I}^{\leftrightarrow} \define (\IRight, \ILeft)$, whose entries are defined by 
\begin{equation}
I^{\rightarrow}_{i\alpha}  \define \left\{\begin{array}{ccc} 1 &{\rm if}& (i,\alpha)\in \E ,\\ 0 &{\rm if} & (i,\alpha)\notin \E,\end{array}\right.
\end{equation}
and 
\begin{equation}
I^{\leftarrow}_{i\alpha}  \define \left\{\begin{array}{ccc} 1 &{\rm if}& (\alpha, i)\in \E ,\\ 0 &{\rm if} & (\alpha,i)\notin \E.\end{array}\right.
\end{equation}

\begin{figure}
 \centering
 \setlength{\unitlength}{0.1\textwidth}
 \includegraphics[width=\textwidth]{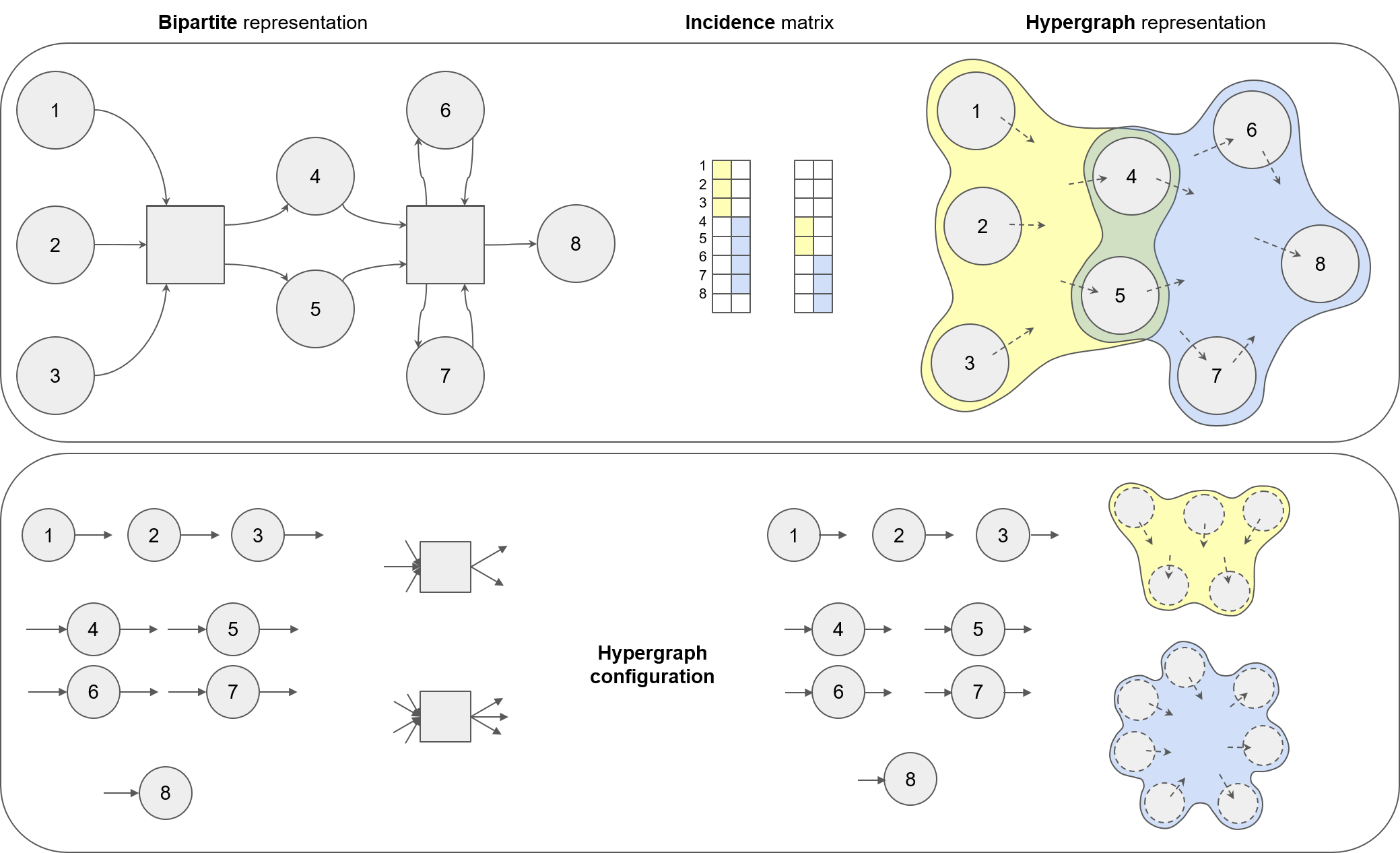}
 \put(-2.4,5.25){\footnotesize$\alpha$}
 \put(-1.7,5.25){\footnotesize$\beta$}
 \put(-8.75,4.3){\scriptsize$\alpha$}
 \put(-6.87,4.3){\scriptsize$\beta$}
 \put(-4.9,3.65){\scriptsize$\IRight$}
 \put(-4.3,3.65){\scriptsize$\ILeft$}
 \put(-4.9,5.){\tiny$\alpha$}
 \put(-4.75,5.){\tiny$\beta$}
 \put(-4.3,5.){\tiny$\alpha$}
 \put(-4.15,5.){\tiny$\beta$}
 \put(-9.3,2.6){\scriptsize Nodes set $ \V$}
 \put(-9.5,1.9){\tiny $1$ out-degree nodes}
 \put(-9.7,0.77){\tiny $1$ in- and out-degree nodes}
 \put(-9.3,0.1){\tiny $1$ in-degree node}
 \put(-7.5,2.6){\scriptsize Hyperedge set $ \W$}
 \put(-7.7,1.6){\tiny $3$ in- and $2$ out-cardinality}
 \put(-7.2,1.4){\tiny hyperedge}
 \put(-7.7,0.57){\tiny $4$ in- and $3$ out-cardinality}
 \put(-7.2,0.37){\tiny hyperedge}
 \put(-4.1,2.6){\scriptsize Nodes set $ \V$}
 \put(-4.3,1.9){\tiny $1$ out-degree nodes}
 \put(-4.5,0.81){\tiny $1$ in- and out-degree nodes}
 \put(-4.1,0.25){\tiny $1$ in-degree node}
 \put(-2.2,2.7){\scriptsize Hyperedge set $ \W$}
 \put(-2.4,1.6){\tiny $3$ in- and $2$ out-cardinality}
 \put(-1.0,1.45){\tiny hyperedge}
 \put(-3.0,0.07){\tiny $4$ in- and $3$ out-cardinality hyperedge}
 \put(-1.0,2.){\scriptsize$\alpha$}
 \put(-0.7,0.8){\scriptsize$\beta$}
 \put(-6.87,2){\scriptsize$\alpha$}
 \put(-6.87,0.94){\scriptsize$\beta$}
 \caption{{\it Illustration of  the different representations of a hypergraph.} The upper panel shows  three ways of representing a hypergraph, namely, as a bipartite graph, as a pair of incidence matrices, and as a graph with higher-order interactions. The lower panel shows the configuration of nodes and hyperedges corresponding with the graph shown in the upper panel.}
 \label{fig:description of the notation}
\end{figure}
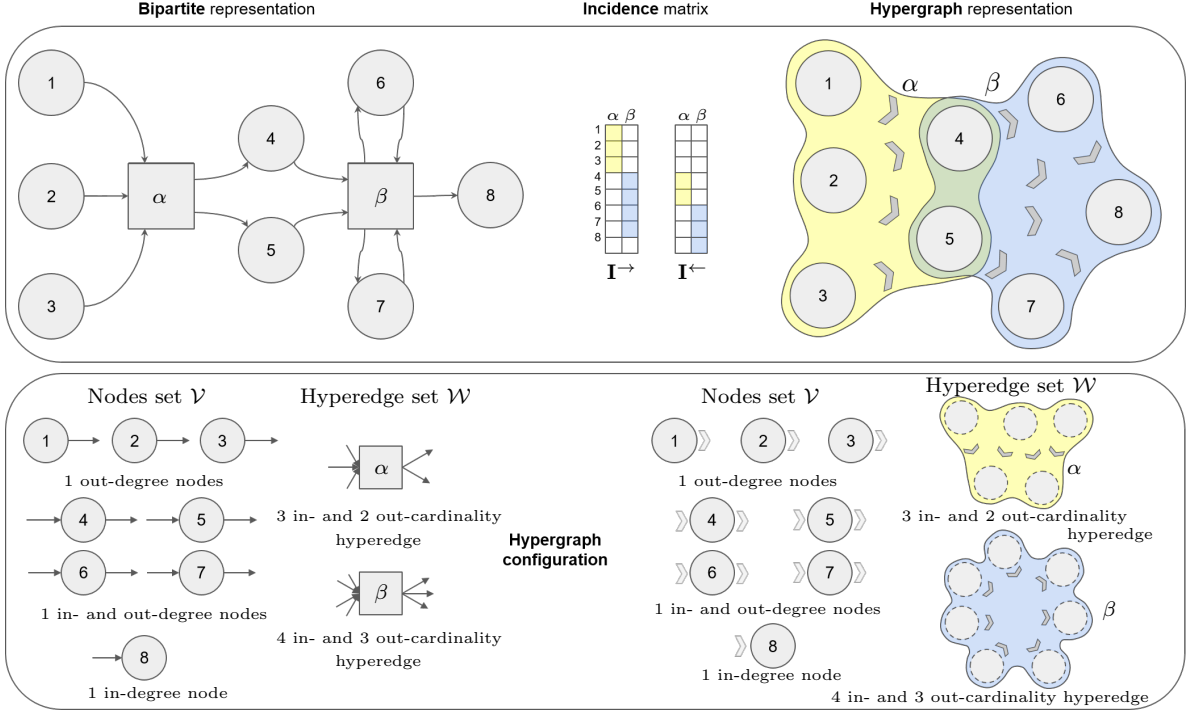

Consequently, a hypergraph can also be represented as a bipartite graph whose vertices are the nodes and the hyperedges of the hypergraph and whose edges are the links of the hypergraph. Figure \ref{fig:description of the notation} shows an example of a hypergraph represented as a bipartite graph and a pair of incidence matrices.

We define some basic network observables that we use   in this paper. 
The {\it out-degree} and the {\it in-degree} of node $i\in\V$  is defined by  
\begin{equation}
\degreeOut_i(\IRight) \define \sum^\sizeW_{\alpha=1}I^{\rightarrow}_{i\alpha}   \  \ {\rm and} \ \ \degreeIn_i(\ILeft) \define \sum^\sizeW_{\alpha=1} I^{\leftarrow}_{ i \alpha} .
\end{equation}
Analogously, we define the {\it out-cardinality}  and the {\it in-cardinality} by  
\begin{equation}
\cardOut_\alpha(\ILeft) \define \sum^\sizeV_{i=1} I^{\leftarrow}_{i \alpha }  \  \ {\rm and} \  \ \cardIn_\alpha(\IRight) \define \sum^\sizeV_{i=1} I^{\rightarrow}_{i \alpha}, 
\end{equation}
respectively.      In what follows, summations over the Roman  indices run from $1$ till $N$ and those over the Greek indices run from $1$ till $M$, unless otherwise specified.  

We use vector notation for degree and cardinality sequences, i.e., 
\begin{equation}
\vec{k}^{\rm in}(\mathbf{I}^{\leftarrow }) :=  (k^{\rm in}_1(\mathbf{I}^{\leftarrow }),k^{\rm in}_2(\mathbf{I}^{\leftarrow }),\ldots,k^{\rm in}_N(\mathbf{I}^{\leftarrow }))
\end{equation}
and 
\begin{equation}
\vec{\chi}^{\rm in}(\mathbf{I}^{\rightarrow }) :=  (\chi^{\rm in}_1(\mathbf{I}^{\rightarrow }),\chi^{\rm in}_2(\mathbf{I}^{\rightarrow }),\ldots,\chi^{\rm in}_M(\mathbf{I}^{\rightarrow })),
\end{equation}
and similar for $\vec{k}^{\rm out}(\mathbf{I}^{\rightarrow })$ and $\vec{\chi}^{\rm out}(\mathbf{I}^{\leftarrow })$.

Next, we define the set of hyperedges incident to the node $i$ as the union
\begin{equation}
\partial_{i}(\mathbf{I}^{\leftrightarrow})
\define \partial^{\rm out}_{i}(\mathbf{I}^{\rightarrow}) \cup\partial^{\rm in}_{i}(\mathbf{I}^{\leftarrow})
\end{equation}
of the two hyperedge neighbourhood sets 
\begin{equation}
\partial^{\rm out}_{i}(\mathbf{I}^{\rightarrow})  \define \{\alpha\in \W| I^{\rightarrow}_{i\alpha}\neq0 \},\ {\rm and} \ 
\partial^{\rm in}_{i}(\mathbf{I}^{\leftarrow})  \define \{\alpha\in \W| I^{\leftarrow}_{i\alpha}\neq0 \}.
\end{equation}
Analogously, we  define the set of nodes incident to the hyperedge $\alpha$ as
\begin{equation}
\partial_{\alpha}(\mathbf{I}^{\leftrightarrow})
\define \partial^{\rm out}_{\alpha}(\mathbf{I}^{\leftarrow}) \cup\partial^{\rm in}_{\alpha}(\mathbf{I}^{\rightarrow})
\end{equation}
where
\begin{equation}
\partial^{\rm out}_{\alpha}(\mathbf{I}^{\leftarrow})  \define \{i\in \V| I^{\leftarrow}_{i\alpha}\neq0 \}, \ {\rm and} \  \partial^{\rm in}_{\alpha}(\mathbf{I}^{\rightarrow})  \define \{i\in \V| I^{\rightarrow}_{i\alpha}\neq0 \}.
\end{equation}

For a nondirected hypergraph $\Hy$, the incidence matrices are identical, i.e.,
$\IRight = \ILeft$. In this case, we use an incidence matrix without arrows i.e., $\IRight=\ILeft = \I$.  For nondirected hypergraphs,  there is no distinction between in-degrees and out-degrees (as well as in-cardinalities and out-cardinalities) and we   denote them by $ k_i(\I)$ and $\chi_{\alpha}(\I)$, respectively.   Analougsly, we have a single degree sequence $\vec{k}(\mathbf{I})$ and cardinality sequence $\vec{\chi}(\mathbf{I})$.

\section{Connected components in hypergraphs}
\label{sec:conn}

Connected components of hypergraphs are sub-hypergraphs that consist of nodes that are connected by paths.    While for graphs it is straightforward to define a path as  a sequence of connecting edges  starting at one node and ending in the other node, this is not the case for  hypergraphs, as  hyperedges represent  higher-order interactions.     Hence, depending on the relevant real-world application there may exist different rules that activate hyperedges.   For example, in the case of gene regulatory networks, it can be  that the
 transcription factor  encoded by one gene activates the expression of another gene, while in other cases it is required that the transcription factors of several genes need to be present for the activation of a target gene  
~\cite{torrisi2020percolation}. 
We refer to the implemented rule for the higher-order interaction as the  hyperedge logic.   Here, we investigate two kind of logical operations associated to the hyperedges, namely, OR-logic in Sec.~\ref{ch:con_components} and  AND-logic in Sec.~\ref{sec:component_AND}.  An OR-logic hyperedge is part of a connected component as soon as one of its in-neighbours belongs to the connected component, whereas an AND-logic hyperedge requires that all in-neighbours belong to the connected component. 

We define connected components through equivalence relations between the vertices of a hypergraph. To this end, we use a binary relation $x \sim y$ from a vertex $x$ to a vertex $y$. We say that $i\sim j$, if there exists a {\it path} in $\mathcal{H}$ that starts in node $i$ and ends in node $j$.   In other words, $i\sim j$ if  there exists a sequence
  \begin{equation}
  i\rightarrow \alpha_1\rightarrow a_1\rightarrow \alpha_2\rightarrow \ldots \alpha_\ell \rightarrow j
  \end{equation} 
  such that 
  \begin{equation}
  I^{\rightarrow}_{i\alpha_1} I^{\leftarrow}_{a_1 \alpha_1} I^{\rightarrow}_{a_1\alpha_2}\ldots  I^{\leftarrow}_{j \alpha_\ell}  =1. 
  \end{equation}  
  Analogously, we  define $\alpha\sim\beta$ from a hyperedge $\alpha$  to a  hyperedge $\beta$,  $\alpha\sim i$ from a hyperedge $\alpha$ to a node $i$, and  $i\sim \alpha$ from a node  $i$ to a hyperedge $\alpha$. 

\subsection{Connected components of hypergraphs  without cooperativity (OR-logic)}\label{ch:con_components}
In this Section, we define connected components with OR-logic for nondirected hypergraphs (Sec.~\ref{sec:component_nondir}) and directed hypergraphs (Sec.~\ref{sec:component_dir_OR}).

\subsubsection{Nondirected hypergraphs}\label{sec:component_nondir}$\\$
For nondirected hypergraphs $\Hy= (\V,\W,\E)$, if $x
\sim y$ then also $y
\sim x$, with $x,y\in \V\cup\W$ two vertices in the hypergraph.  Therefore, the binary relation $
\sim$ is an equivalence relation, and we say that two vertices $x$ and $y$ are {\it connected} if $x
\sim y$.  
We  say that a nondirected hypergraph $\Hy$ is {\it connected} if  all pairs $(x,y)$  with  $x,y\in \V\cup\W$  are   connected.  

A  {\it connected component}  $\Hy_{\rm c} = (\V_c,\W_c,\E_c)$ of $\Hy$ is a connected sub-hypergraph  of $\Hy$  for which there exist no other  connected sub-hypergraph  of $\Hy$ that contains $\Hy_{\rm c}$.  The sets  $\V_c\cup \W_c$ associated with the connected components  of $\Hy$ are  the  equivalence classes of $\sim$ in $\V\cup \W$.

The {\it largest  connected component} $\Hy^\ast = (\V^\ast,\W^\ast,\E^\ast)$ of a hypergraph $\Hy$ is the connected component with  the largest number $n^\ast = |\V^\ast|$ of nodes; note that we could also define the largest connected component as the connected component that has the largest number of hyperedges.  In the limit of large $N$, we quantify the size of the largest connected component with  $f(\mathbf{I})$,  the relative number of nodes 
\begin{equation}
f(\mathbf{I}) := \frac{n^\ast(\mathbf{I})}{N},
\label{eq:largest_component}
\end{equation}
 that belong to the largest connected component. 

The connected component of a hypergraph, including the largest one, can be obtained with  breadth-first search  or depth-first search algorithms~\cite{hartmann2006phase}.   These algorithms readily apply to  hypergraphs by representing the hypergraph as a bipartite graph of nodes and hyperedges~\cite{bretto2013hypergraph}.    

\subsubsection{Directed hypergraphs with OR-logic} \label{sec:component_dir_OR}$\\$
For directed hypergraphs $\Hy= (\V,\W,\E)$, $x\sim y$ does not imply that $y\sim x$.  Thus, $\sim$ is not an equivalence relation and cannot be used to define connected components.  Nevertheless, following Tarjan~\cite{tarjan1972depth},  we can define another equivalence relation between nodes that we call  OR-logic strongly connectedness.  We say that two vertices $x,y\in \V \cup \W$ are  {\it OR-logic strongly connected}, denoted by   $x \sim^{\rm OR}_{\rm S}y$,  if both  $x\sim y$ and $y\sim x$.  A hypergraph $\mathcal{H}$  is OR-logic strongly connected if  any pair of vertices in $\mathcal{H}$ are   OR-logic strongly connected.

The binary relation $\sim^{\rm OR}_{\rm S}$ is an equivalence relation on $\V\cup \W$.   Therefore it partitions the set  of vertices $\V\cup \W$ into equivalence classes, which determine  the strongly connected components of directed hypergraphs.  We define the  OR-logic strongly connected components of $\mathcal{H}$ as the sub-hypergraphs $\mathcal{H}^{\rm OR}_{\rm s} = (\V^{\rm OR}_{\rm s}, \W^{\rm OR}_{\rm s}, \E^{\rm OR}_{\rm s})$ that are  OR-logic strongly connected and for which   there exist no other    OR-logic strongly  connected sub-hypergraph of $\mathcal{H}$ that contains $\mathcal{H}^{\rm OR}_{\rm s}$.

Each  $\Hy^{\rm OR}_{\rm s}$ has an in-component, an out-component, and a weakly connected component. The {\it in-component} consists of all vertices $x\in \V\cup\W$  for which there exist a  vertex $y\in \V^{\rm OR}_{\rm s}\cup\W^{\rm OR}_{\rm s}$ with $x\sim y$; analogously, the {\it out-component}  consists of all vertices $x\in \V\cup\W$  for which there exist a vertex $y\in \V^{\rm OR}_{\rm s}\cup\W^{\rm OR}_{\rm s}$ with $y\sim x$.  Lastly, the {\it weakly connected component}  is a connected component of the nondirected hypergraph  $\mathcal{H}^{\rm OR}_{\rm w}$ obtained from $\mathcal{H}$ by making all hyperedges nondirected.   Specifically, the weakly connected component of $\Hy^{\rm OR}_{\rm s}$ is the connected component of 
$\mathcal{H}^{\rm OR}_{\rm w}$ that contains $\V^{\rm OR}_{\rm s}$.   

To determine the size of the largest strongly connected component and its related sub-graphs, we define the quantities
\begin{equation}
f^{\mathfrak{a}}_{\rm OR}(\mathbf{I}^{\leftrightarrow}) := \frac{n^{\mathfrak{a}}_{\rm OR}(\mathbf{I}^{\leftrightarrow})}{N}, \label{eq:fSc}
\end{equation}
with $\mathfrak{a}\in \left\{{\rm sc},{\rm oc}, {\rm ic}, {\rm t}, {\rm wc}\right\}$, corresponding with the relative number of nodes in the largest strongly connected component (sc), largest out-component (oc), largest in-component (ic), the  tendrils (t), and the largest weakly connected component (wc);  the tendrils denote all nodes that are part of the largest weakly connected component, but not part of the largest in-component or out-component.

The OR-logic strongly connected components of a given hypergraph can  be computed  with either Tarjan's algorithm ~\cite{tarjan1972depth} or Kosaraju's algorithm~\cite{sharir1981strong}.   These algorithms  readily apply to  OR-logic strongly connected components of directed hypergraphs by representing the  hypergraph as a bipartite graph of nodes and hyperedges~\cite{allamigeon2011strongly}. 

\subsection{Connected components with cooperativity (AND-logic)} \label{sec:component_AND}
In systems with higher-order interactions it is sometimes the case that  interactions, modelled by hyperedges in a hypergraph, are active if and only if   all   nodes involved are active.   To model connected components in hypergraphs with such  cooperative interactions, we define   in Sec.~\ref{sec:component_dir_AND} connected components with   AND-logic~\cite{torrisi2020percolation}, and  in Sec.~\ref{sec:component_algo} we introduce numerical algorithms for determining AND-logic connected components in directed hypergraphs.  In   Sec.~\ref{sec:ICOC_SCC_AND_logic}, we discuss the distinction between   AND-logic strongly connected component and the  intersection between the in- and out-components of directed hypergraphs.

\subsubsection{Definition of AND-logic  connected components} \label{sec:component_dir_AND}$\\$
 Consider a hypergraph  $\mathcal{H}=(\mathcal{V},\mathcal{W},\mathcal{E})$  and let below $x\sim^{\rm OR}_{\rm s}y$ denote OR-logic strongly connectedness of two vertices $x,y \in \V\cup \W$.   We say that a sub-hypergraph $\mathcal{H}'=(\V',\W',\E')$ is AND-logic strongly connected in $\Hy$ if
\begin{enumerate}
\item for all pairs of vertices $x,y\in\V'\cup \W'$, it holds that $x\sim^{\rm OR}_{\rm S}y$;   
\item for all  hyperedges $\alpha\in \W'$ and for all nodes $i,j\in \partial^{\rm in}_\alpha(\mathcal{H})$ it holds that 
\begin{equation}
i\sim^{\rm OR}_{\rm S}j.
\end{equation}
\end{enumerate}
Note that for point (ii) it is not sufficient to consider all nodes $i,j\in \partial^{\rm in}_\alpha(\mathcal{H}')$, as the latter condition is already satisfied by the condition (i).
We call this an AND-logic strongly connected graph, as a path between two vertices $x$ and $y$ only matters if all the in-neighbours along that path are also strongly connected to $x$ and $y$.

  If there exists a sub-graph $\mathcal{H}'$ that is AND-logic strongly connected, and  if $x,y\in \V'\cup \W'$, then we say that the  vertices $x\in \V\cup \W$ and $y\in \V\cup \W$  are {\it AND-logic strongly connected}.    We denote AND-logic strongly connectedness of two vertices $x$ and $y$ by  
\begin{equation}
    x \sim^{\rm AND}_{\rm S} y .
\end{equation}

\begin{figure}
 \centering
 \setlength{\unitlength}{0.1\textwidth}
 \includegraphics[width=0.9\textwidth]{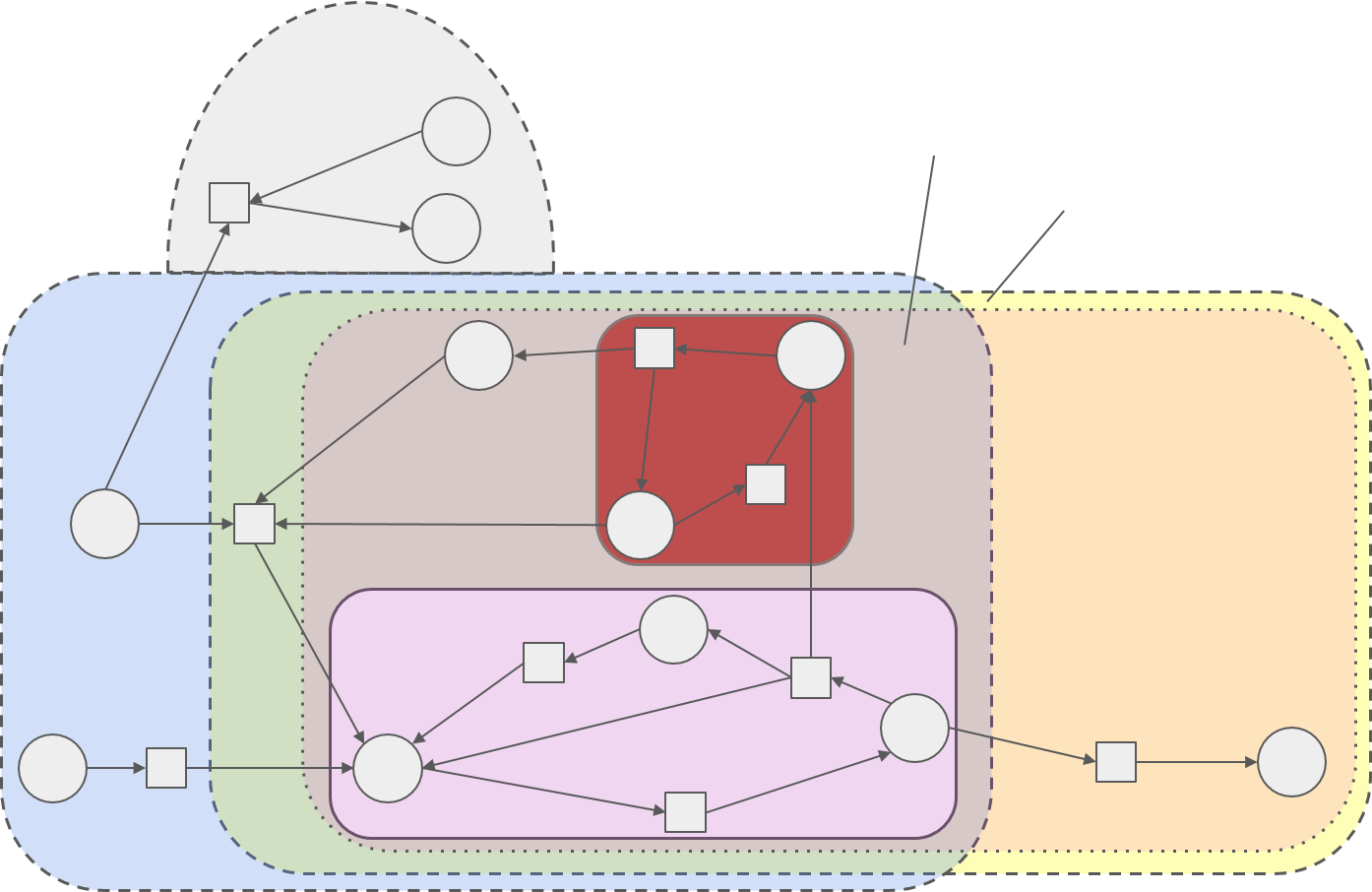}
 \put(-7.55,3.3){\normalsize $\Hy^{\rm OR}_{\rm s}$}
 \put(-4.76,2.95){\normalsize $\Hy^{(1),\rm  AND}_{\rm s}$}
 \put(-6.5,1.6){\normalsize $\Hy^{(2), \rm AND}_{\rm s}$}
 \put(-8.85,3.47){\normalsize $\Hy^{\rm OR, AND}_{\rm in}$}
 \put(-1.2,3.4){\normalsize $\Hy^{\rm AND}_{\rm out}$}
 \put(-2.0,4.6){\normalsize $\Hy^{\rm OR}_{\rm out}$}
 \put(-3.7,5.0){\normalsize $\Hy^{\rm AND}_{\rm in}\cap\Hy^{\rm AND}_{\rm out}$}
 \put(-7.0,5.4){\footnotesize tendrills}
 \caption{{\it Connected components in an example of a directed  hypergraph.} The circles represent nodes, and the squares represent hyperedges. The hypergraph has two AND-logic strongly connected components $\Hy^{(1),\rm AND}_{\rm s}$ and $\Hy^{(2),\rm AND}_{\rm s}$ that are larger than a single vertex.   These two AND-logic strongly connected components  have the same in-components $\Hy^{\rm AND}_{\rm in}$, out-components $\Hy^{\rm AND}_{\rm out}$, and weakly connected components $\Hy^{\rm AND}_{\rm w}$, which are as shown in the figure.   The hypergraph has one OR-logic strongly connected component  
 $\Hy^{\rm OR}_{\rm s} = \Hy^{\rm OR}_{\rm in} \cap  \Hy^{\rm OR}_{\rm out}  $ that is larger than a single vertex.   The  $\Hy^{\rm OR}_{\rm s}$  consists of the two indicated $\Hy^{(1),\rm AND}_{\rm s}$,  $\Hy^{(2),\rm AND}_{\rm s}$, one additional hyperedge, and one additional node.   As shown,  $\Hy^{\rm AND}_{\rm in}=\Hy^{\rm OR}_{\rm in}$.    On the other hand, $\Hy^{\rm AND}_{\rm out}$ is a sub-hypergraph of $\Hy^{\rm OR}_{\rm out}$.  In this example,   $\Hy=\Hy^{\rm AND}_{\rm w}=\Hy^{\rm OR}_{\rm w}$. }
 \label{fig:example_conn_dir}
\end{figure}
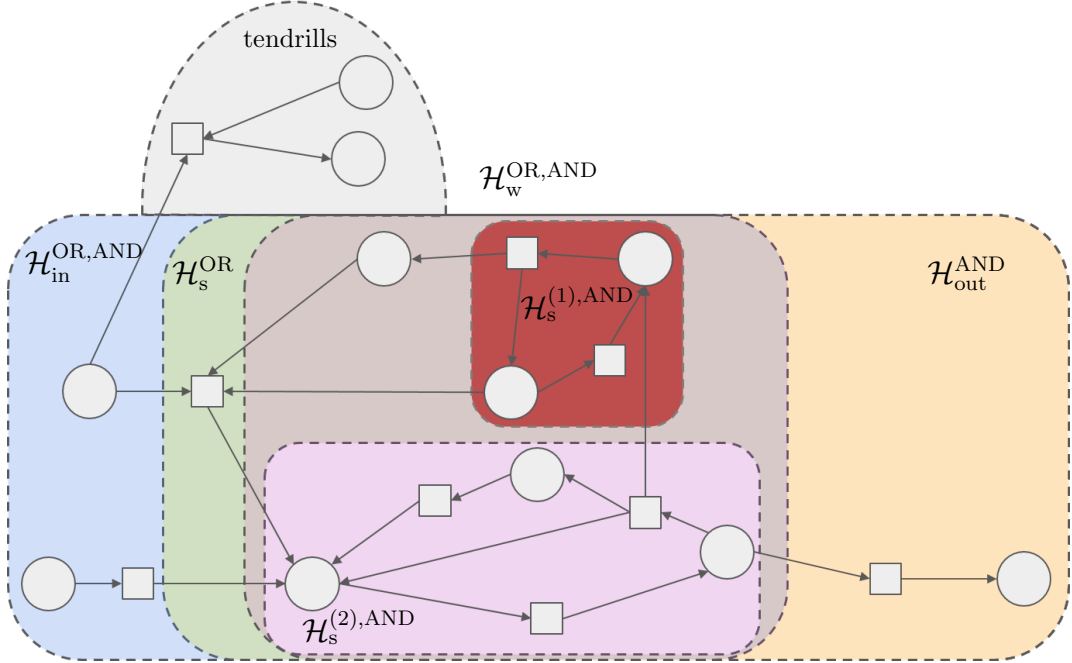

If we assume that $x\sim^{\rm AND}_{\rm S}x$ for any vertex $x\in \V\cup\W$, then   
the relation $\sim^{\rm AND}_{\rm S}$ is  an  equivalence relation on the set $\V\cup\W$.  Therefore it partitions the set $\V\cup\W$ into equivalence classes $(\mathcal{V}^{\rm AND}_{\rm s},\mathcal{W}^{\rm AND}_{\rm s})$.   We call the sub-hypergraphs corresponding with those equivalence classes {\it AND-logic strongly connected components} and we denote them by  $\mathcal{H}^{\rm AND}_{\rm s} = (\mathcal{V}^{\rm AND}_{\rm s},\mathcal{W}^{\rm AND}_{\rm s},\mathcal{E}^{\rm AND}_{\rm s})$. For example, Fig.~\ref{fig:example_conn_dir} shows a hypergraph  that has  two strongly connected components with the AND-logic that are non-trivial (i.e. they have more than one vertex).

Due to  condition (ii), the definition of the AND-logic strongly connected component is more restrictive than that for the OR-logic strongly connected component, which is simply defined by condition (i). Hence, $\mathcal{H}_{\rm s}^{\rm AND}$ is  a sub-hypergraph of $\mathcal{H}_{\rm s}^{\rm OR}$.  In particular, in the example of  Fig.~\ref{fig:example_conn_dir} there is one OR-logic strongly connected component $\Hy^{\rm OR}_{\rm s}$ that is larger than a single vertex, and hence $\Hy^{(1),\rm AND}_{\rm s}\subset \Hy^{\rm OR}_{\rm s}$ and  $\Hy^{(2),\rm AND}_{\rm s}\subset \Hy^{\rm OR}_{\rm s}$.

Next, we define the out-components and in-components associated with a sub-hypergraph $\mathcal{H}_{\rm s}^{\rm AND}$ that is AND-logic strongly connected.  The AND-logic in-component of $\mathcal{H}_{\rm s}^{\rm AND}$ is  the {\it largest} hypergraph $\mathcal{H}^{\rm AND}_{\rm in}=(\mathcal{V}^{\rm AND}_{\rm in},\mathcal{W}^{\rm AND}_{\rm in},\mathcal{E}^{\rm AND}_{\rm in})$ for which it holds that
\begin{enumerate}
\item[(i)]
for all vertices $x\in \mathcal{V}^{\rm AND}_{\rm in}\cup \W^{\rm AND}_{\rm in}$  there exists a $y\in \mathcal{V}^{\rm AND}_{\rm s}\cup\mathcal{W}^{\rm AND}_{\rm s}$ so that $x \sim y$;
\item[(ii)] for all $\alpha\in \W^{\rm AND}_{\rm in}$ and for all $j\in \partial^{\rm in}_\alpha(\Hy)$ it holds that $j\in \mathcal{V}^{\rm AND}_{\rm in}$.   
\end{enumerate}
It follows from the definition of $\mathcal{H}^{\rm AND}_{\rm in}$ as a {\it maximal set} of nodes with 
an incident path to nodes in $\mathcal{H}'$
that    
condition (ii) is automatically satisfied. As a consequence, the AND-logic in-component coincides with the OR-logic in-component, which is defined merely by condition (i).  We show this in Fig.~\ref{fig:example_conn_dir} for the example.

The AND-logic out-component consists of the {\it largest} hypergraph $\mathcal{H}^{\rm AND}_{\rm out}=(\mathcal{V}^{\rm AND}_{\rm out},\mathcal{W}^{\rm AND}_{\rm out},\mathcal{E}^{\rm AND}_{\rm out})$ for which it holds that 
\begin{enumerate}
\item[(i)]
for all vertices $x\in \mathcal{V}^{\rm AND}_{\rm out}\cup \mathcal{W}^{\rm AND}_{\rm out}$ there exists a $y\in \mathcal{V}^{\rm AND}_{\rm s}\cup\mathcal{W}^{\rm AND}_{\rm s}$  so that $y\sim x$;
\item[(ii)] for all $\alpha\in \W^{\rm AND}_{\rm out}$ and for all $j\in \partial^{\rm in}_{\alpha}(\Hy)$ it  holds that $j\in \mathcal{V}^{\rm AND}_{\rm out}$.   
\end{enumerate}
Thus, the out-component $\Hy^{\rm AND}_{\rm out}$ is  a sub-hypergraph of $\Hy_{\rm out}^{\rm OR}$, as also shown in the example of Fig.~\ref{fig:example_conn_dir}.

Note that for nondirected hypergraphs   OR- and AND-logic connected components are identical.  In the OR-logic a hyperedge is part of the connected component if at least one of its neighbours belongs to it, while in the AND-logic, a hyperedge is included only if all its neighbours are also part of the component. For nondirected hypergraphs, however, the bidirectional relationships between nodes ensure that if one node can influence another under OR-logic, the reverse is also true, and therefore  the conditions for AND-logic are always satisfied.

\begin{figure}
 \centering
 \setlength{\unitlength}{0.1\textwidth}
 \includegraphics[width=0.5\textwidth]{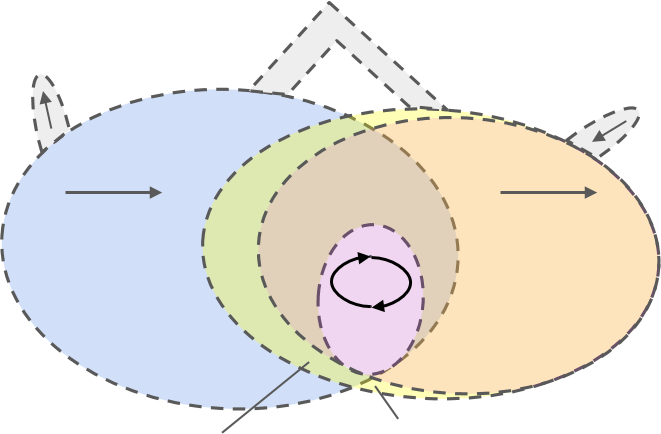}
 \put(-3.6,-0.23){\normalsize $\Hy^{\rm OR}_{\rm s}$}
 \put(-2.5,0.66){\normalsize $\Hy^{\rm AND}_{\rm s}$}
 \put(-4.7,0.9){\large $\Hy^{\rm OR, AND}_{\rm in}$}
 \put(-1.2,0.9){\large $\Hy^{\rm AND}_{\rm out}$}
 \put(-1.8,2.8){\normalsize $\Hy^{\rm OR, AND}_{\rm w}$}
 \put(-2.05,-0.1){\normalsize $\Hy^{\rm OR}_{\rm out}$}
 \caption{{\it Schematic illustration of  the relations between connected components in  directed hypergraphs.}  The coloured areas represent: in-components  $\Hy^{\rm OR}_{\rm in}=\Hy^{\rm AND}_{\rm in}$ (blue, green, brown, and magenta),   the out-component $\Hy^{\rm AND}_{\rm out}$ (orange, brown, and magenta),  the out-component $\Hy^{\rm OR}_{\rm out}$ (yellow, orange, green, brown, and magenta), the OR-logic strongly connected component $\Hy^{\rm OR}_{\rm s} = \Hy^{\rm OR}_{\rm in}\cap \Hy^{\rm OR}_{\rm out}$  (green, brown, and magenta), the intersection $\Hy^{\rm AND}_{\rm in}\cap \Hy^{\rm AND}_{\rm out}$ (brown and magenta), the AND-logic strongly connected component $\Hy^{\rm AND}_{\rm s}\subseteq \Hy^{\rm AND}_{\rm in}\cap \Hy^{\rm AND}_{\rm out}$ (magenta), and the weakly connected components $\Hy^{\rm OR}_{\rm w} = \Hy^{\rm AND}_{\rm w}$ (all areas including the grey parts).  } 
 \label{fig:AND_component}
\end{figure}

Figure~\ref{fig:AND_component} sketches the general topology of an  AND-logic strongly connected component $\Hy^{\rm AND}_{\rm s}$ and its corresponding OR-logic strongly connected component $\Hy^{\rm OR}_{\rm s}$   for which $\Hy^{\rm AND}_{\rm s}\subseteq \Hy^{\rm OR}_{\rm s}$.  For such a pair of strongly connected components the following relations hold: (i) $\Hy^{\rm AND}_{\rm in}=\Hy^{\rm OR}_{\rm in}$; (ii)  $\Hy^{\rm AND}_{\rm out}\subseteq \Hy^{\rm OR}_{\rm out}$; (iii)  $\Hy^{\rm OR}_{\rm w}=\Hy^{\rm AND}_{\rm w}$; (iv) $\Hy^{\rm OR}_{\rm s}=\Hy^{\rm OR}_{\rm in}\cap \Hy^{\rm OR}_{\rm out}$, where $\Hy^{\rm OR}_{\rm in}\cap \Hy^{\rm OR}_{\rm out} = (\V^{\rm OR}_{\rm in}\cap \V^{\rm OR}_{\rm out},\W^{\rm OR}_{\rm in}\cap \W^{\rm OR}_{\rm out},\E^{\rm OR}_{\rm in}\cap \E^{\rm OR}_{\rm out})$ is the intersection between the in- and out-components; (v) $\Hy^{\rm AND}_{\rm s}\subseteq(\Hy^{\rm AND}_{\rm in}\cap \Hy^{\rm AND}_{\rm out})$.     Note that differently from OR-logic strongly connected components, within AND-logic the strongly connected component is not the intersection of the in- and out-component. For example,  in Fig.~\ref{fig:example_conn_dir} $\Hy^{\rm OR}_{\rm s} = (\Hy^{\rm OR}_{\rm in} \cap \Hy^{\rm OR}_{\rm out})$, whereas  $\Hy^{(1),\rm AND}_{\rm s}\subset  (\Hy^{\rm AND}_{\rm in} \cap \Hy^{AND}_{\rm out})$ and  $\Hy^{(2),\rm AND}_{\rm s}\subset  (\Hy^{\rm AND}_{\rm in} \cap \Hy^{AND}_{\rm out})$.   Hence, in this example the intersection of the AND-logic in- and out-components (brown area) contains two  AND-logic strongly connected components (and some additional vertices).

Analogously to the OR-logic connected components,  we quantify the relative sizes of the AND-logic components   with the quantity $f^{\mathfrak{a}}_{\rm AND}(\mathbf{I}^{\leftrightarrow})$, see Eq.~(\ref{eq:fSc}).

\subsubsection{Algorithms for AND-logic connected components} \label{sec:component_algo}$\\$
For AND-logic strongly connected components, Torrisi et al.~developed an algorithm that yields an  AND-logic strongly connected component~\cite{torrisi2020percolation}.   However, the AND-logic strongly connected component returned by this algorithm is not guaranteed to be the largest one.     Here, we adapt Torrisi's algorithm  so that it is guaranteed to yield the largest AND-logic strongly connected component, as well as  its  in- and out-components.    The algorithm has three phases that are described below:

\begin{algorithm} 
	\caption{\textsc{FindLargestAND-SCC}(Hypergraph $\Hy$, Largest AND-SCC $\Hy^{\rm AND}_{\rm s}$)}
	\begin{algorithmic}[1]
		\State \{$\Hy^{\rm OR}_{\rm s}$\} $\leftarrow$ \textsc{Tarjan}($\Hy$) \Comment{determine OR-SCCs}
		\State $Q$ $\leftarrow$ \{$\Hy^{\rm OR}_{\rm s}$\} \Comment{All OR-SCCs will be examined (in descending order)}
		\While {not done} 
  		\State $\mathcal{H}'$ $\leftarrow$ $Q$.remove(largest $\Hy^{\rm OR}_{\rm s}$) \Comment{Determine largest sub-hypergraph to examine}
        \State $\Hy_{\rm pruned}$ $\leftarrow$\textsc{RemoveHyperedges}($\Hy$,$\mathcal{H}'$) \Comment{Remove the hyperedges that don't satisfy AND-logic condition}
            \If {$\Hy_{\rm pruned}$=$\mathcal{H}'$}
            \State $\Hy^{\rm AND}_{\rm s}$ $\leftarrow$ $\Hy_{\rm pruned}$
            \State done  \Comment{Terminate when finding the largest AND-SCC}
            \EndIf
            \If {not done}
            \State \{$\Hy^{\rm OR}_{\rm s}$\} $\leftarrow$ \textsc{Tarjan}($\Hy_{\rm pruned}$) \Comment{Determine OR-SCCs}
  		\State $Q$.add(\{$\Hy^{\rm OR}_{\rm s}$\}) \Comment{These OR-SCCs will be examined}
            \EndIf
                \EndWhile
		\State \textbf{return} $\Hy^{\rm AND}_{\rm s}$
	\end{algorithmic} \label{Al:AND-SCC}
\end{algorithm} 

\begin{algorithm}
	\caption{\textsc{RemoveHyperedges}(Hypergraph $\Hy$, OR-SCC $\mathcal{H}'$, Sub-hypergraph $\Hy_{\rm pruned}$)}
	\begin{algorithmic}[1]
		\State $\Hy_{\rm pruned}$ $\leftarrow$ $\mathcal{H}'$
		\State $\W_{\rm pruned}=\{\alpha|\alpha\in {\W}(\Hy_{\rm pruned})\}$  \Comment{All hyperedges}
		\For {$\alpha\in\W_{\rm pruned}$}  \Comment{Examine all hyperedges}
		\State $\V^{\rm in}_{\alpha}=\{i|i\in\partial^{\rm in}_{\alpha}(\Hy)\}$  \Comment{All its in-neighbours in original hypergraph}
		\For {$i\in\V^{\rm in}_{\alpha}$}  
            \If {$i\notin{\V}(\Hy_{\rm pruned})$}  \Comment{Doesn't satisfy the AND-logic gate}
        		\State $\Hy_{\rm pruned}$ $\leftarrow$ $\alpha$.remove()  \Comment{Remove the hyperedge}
                \EndIf
		\EndFor
		\EndFor
		\State \textbf{return} $\Hy_{\rm pruned}$
	\end{algorithmic} \label{Al:pruning}
\end{algorithm}

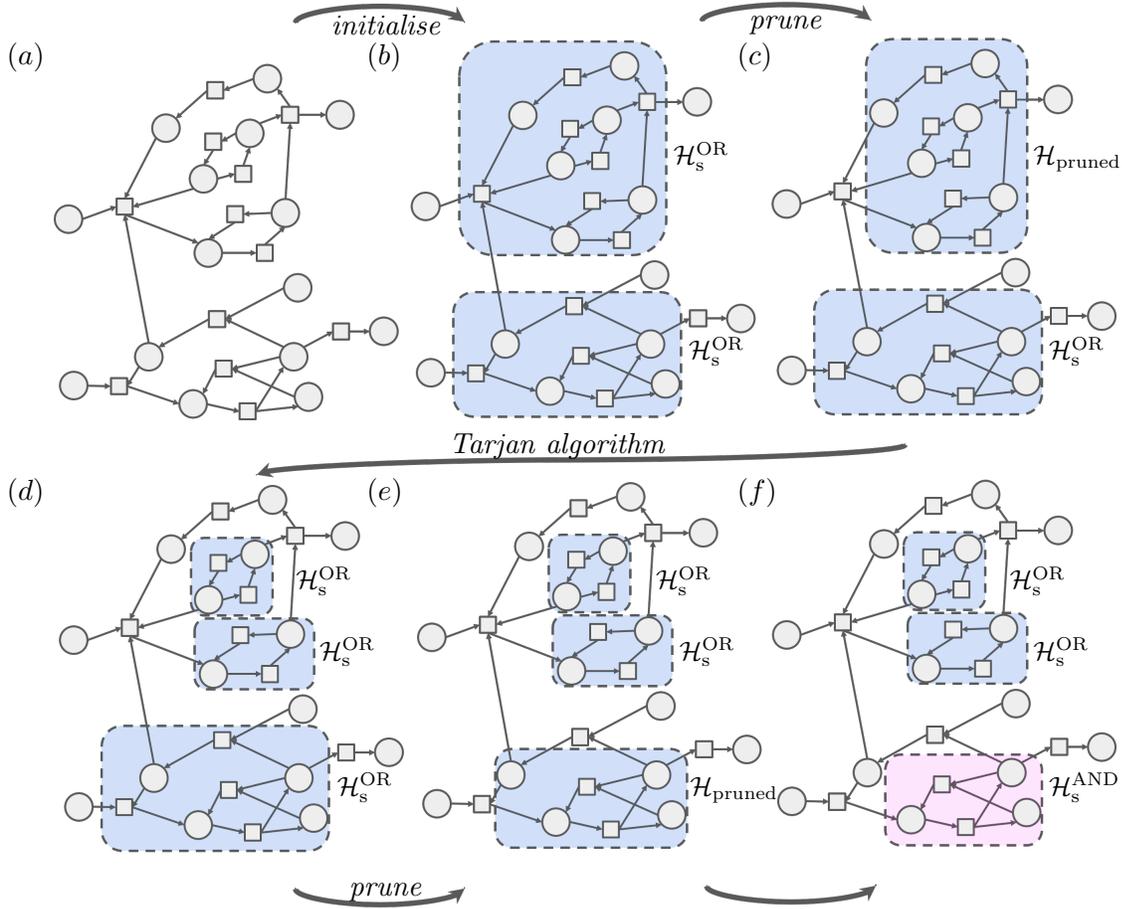
\begin{figure}
 \centering
 \setlength{\unitlength}{0.1\textwidth}
 \includegraphics[width=0.9\textwidth]{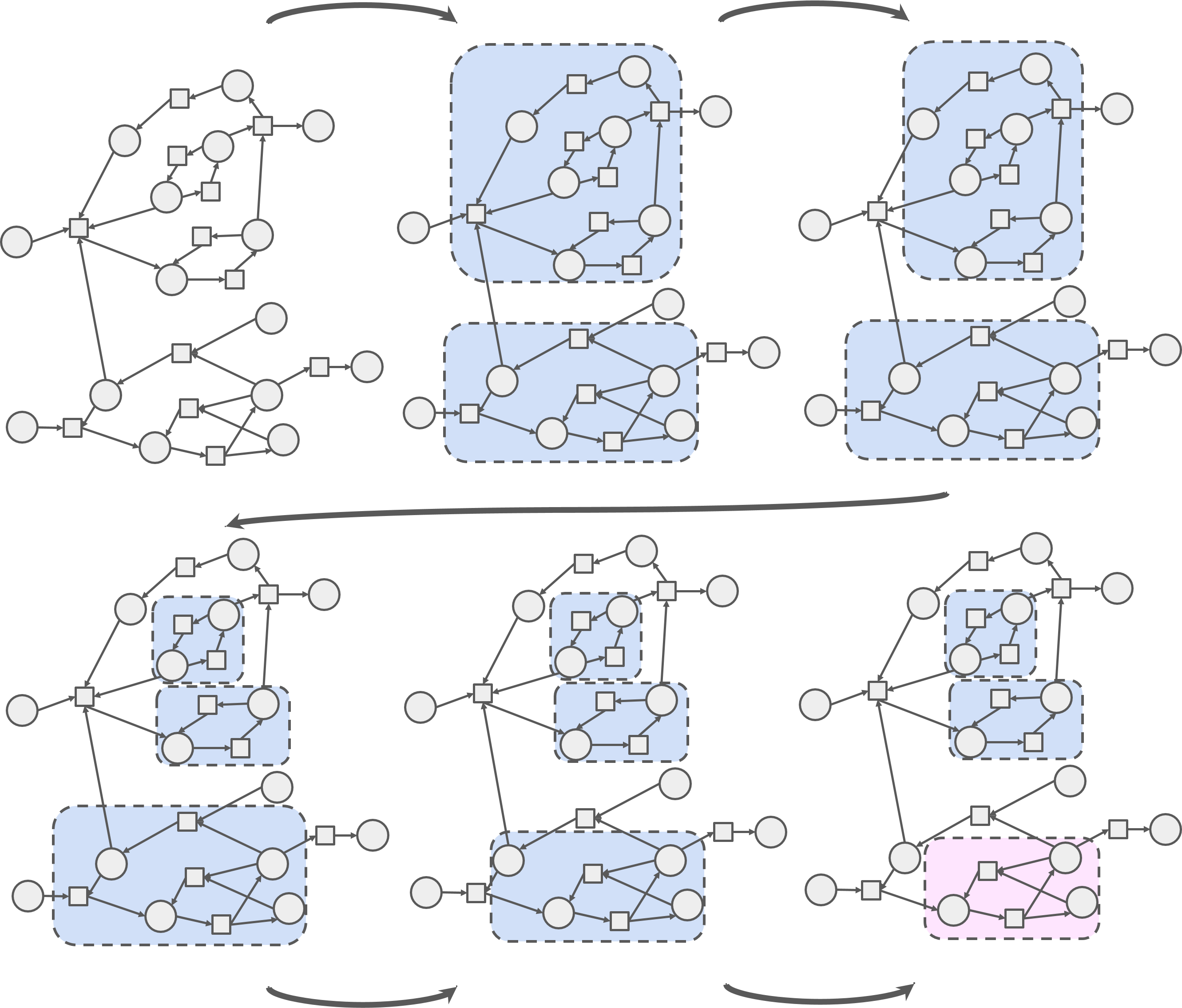}
      \put(-9.4,7.15){\small $(a)$}
      \put(-6.65,7.40){\it\small initialise}
      \put(-6.35,7.15){\small $(b)$}
      \put(-3.11,7.45){\it\small prune}
      \put(-3.21,7.15){\small $(c)$}
      \put(-9.4,3.45){\small $(d)$}
      \put(-5.65,3.85){\it\small Tarjan algorithm}
      \put(-6.35,3.45){\small $(e)$}
      \put(-6.5,0.10){\it\small prune}
      \put(-3.21,3.45){\small $(f)$}
 \put(-3.75,6.3){\footnotesize $\Hy^{\rm OR}_{\rm s}$}
 \put(-3.62,4.65){\footnotesize $\Hy^{\rm OR}_{\rm s}$}
 \put(-.7,6.3){\footnotesize $\Hy_{\rm pruned}$}
 \put(-.57,4.65){\footnotesize $\Hy^{\rm OR}_{\rm s}$}
 \put(-6.95,2.7){\footnotesize $\Hy^{\rm OR}_{\rm s}$}
 \put(-6.75,2.1){\footnotesize $\Hy^{\rm OR}_{\rm s}$}
 \put(-6.6,0.95){\footnotesize $\Hy^{\rm OR}_{\rm s}$}
 \put(-3.9,2.7){\footnotesize $\Hy^{\rm OR}_{\rm s}$}
 \put(-3.7,2.1){\footnotesize $\Hy^{\rm OR}_{\rm s}$}
 \put(-3.61,0.95){\footnotesize $\Hy_{\rm pruned}$}
 \put(-.9,2.7){\footnotesize $\Hy^{\rm OR}_{\rm s}$}
 \put(-.7,2.1){\footnotesize $\Hy^{\rm OR}_{\rm s}$}
 \put(-.57,0.95){\footnotesize $\Hy^{\rm AND}_{\rm s}$}
 \caption{{\it An example of the processing step of the algorithm to determine the largest AND-logic strongly connected component.} $(a)$ Given hypergraph. $(b)$ Tarjan's algorithm determine the OR-logic strongly connected components $\Hy^{\rm OR}_{\rm s}$, from which the two largest ones are highlighted in the figure. $(c)$ The largest $\Hy^{\rm OR}_{\rm s}$ is pruned as $\Hy_{\rm pruned}$. $(d)$  Re-application of the Tarjan algorithm to the pruned sub-hypergraph, resulting in updated OR-logic strongly connected components. $(e)$ Iterative refinement of strongly connected components through additional pruning and connectivity checks. $(f)$ The final sub-hypergraph $\Hy^{\rm AND}_{\rm s}$ representing the largest AND-logic strongly connected component after convergence, where all hyperedges satisfy the AND-logic condition.}
 \label{fig:eg_determine_AND}
\end{figure}

\begin{enumerate}
    \item {\it Initialisation} (pseudo-code line 1-2): Using  Tarjan's algorithm for bipartite graphs~\cite{tarjan1972depth}, all OR-logic strongly connected components $\mathcal{H}^{\rm OR}_{\rm s}$ are identified in the hypergraph $\mathcal{H}$, as illustrated in Figure~\ref{fig:eg_determine_AND}$(b)$. These strongly connected components are sorted by size and stored in the  list $Q$ for iterative processing.   
    \item {\it Hyperedge pruning} (pseudo-code line 4-5):  We extract the  hypergraph  $\mathcal{H}'$ that has the largest number of nodes from the  list $Q$.    For each hyperedge $\alpha\in \W(\mathcal{H}')$, we verify whether it satisfies the condition for AND-logic strongly connectedness, namely, we verify whether for all $i\in \partial^{\rm in}_\alpha(\mathcal{H})$ it holds that $i\in \V(\mathcal{H}')$.     If a hyperedge does not satisfy this condition, it is removed from the hypergraph $\mathcal{H}'$ yielding the sub-hypergraph $\mathcal{H}_{\rm pruned}$ (see Figure~\ref{fig:eg_determine_AND}$(c)$).    Note in this procedure   nodes are not removed, and thus $\V(\Hy') = \V(\Hy^{\rm pruned})$.    If none of the hyperedges have been pruned, then $\mathcal{H}'$   is the largest AND-logic strongly connected component,  we set $\mathcal{H}^{\rm AND}_{\rm s} = \mathcal{H}'$, and the algorithm is terminated here.  
    \item {\it Restoration of OR-logic strongly connectedness } (pseudo-code line 6-13): 
    If one or more  hyperedges have been pruned at the previous (ii) stage, then $\Hy_{\rm pruned}$ is not guaranteed to be an OR-logic strongly connected component.   Therefore, the  algorithm  applies  Tarjan's algorithm to $\Hy_{\rm pruned}$ and finds a new list of  OR-logic strongly connected components, as depicted in Figure~\ref{fig:eg_determine_AND}$(d)$.  These strongly connected components are added to the  list $Q$, and steps (ii) and (iii) of the algorithm are repeated. 
\end{enumerate}  
The pseudo-code of this algorithm is  detailed in the tables entitled Algorithms~\ref{Al:AND-SCC} and~\ref{Al:pruning}, and  Fig.~\ref{fig:eg_determine_AND} illustrates the processing steps.  Figure~\ref{fig:eg_determine_AND}$(f)$ illustrates the final state of the algorithm for an example. 

A modified version of the 
 algorithm determines all the AND-logic strongly connected components of the hypergraph.   In this modified algorithm, instead of pruning only the largest sub-hypergraph and  
 terminating when no hyperedges can be pruned in the largest sub-hypergraph, the algorithm  stores the AND-logic strongly connected sub-hypergraph found in an array and continues processing all the remaining sub-hypergraphs stored in the list $Q$ until no hyperedge can be pruned.

In ~\ref{app:det_AND_OC} we 
provide the pseudocode for
the algorithm that determines the  AND-logic out-component associated with a given AND-logic strongly connected component. 

\subsubsection{Comparing  the AND-logic strongly connected component  with the intersection between its in- and out-components}  \label{sec:ICOC_SCC_AND_logic}$\\$
We discuss a key difference between OR-logic and AND-logic strongly connected components.  Within OR-logic, the strongly connected component is the intersection of its in- and out-components,
\begin{equation}
\mathcal{H}^{\rm OR}_{\rm s}  = (\mathcal{H}^{\rm OR}_{\rm in} \cap  \mathcal{H}^{\rm OR}_{\rm out}), \label{eq:IntersectionOR}
\end{equation}
where as we introduced before in Sec.~\ref{sec:component_dir_AND} the intersection of two hypergraphs is the hypergraph of the intersections of its three sets (vertices, hyperedges, and links).   This    property is important as it is used  to theoretically determine the number of nodes that are part of the strongly connected component in large, random, hypergraphs~\cite{dorogovtsev2001giant,newman2007component, bianconi2024theory}.   

However, with AND-logic 
\begin{equation}
\mathcal{H}^{\rm AND}_{\rm s}  \subseteq (\mathcal{H}^{\rm AND}_{\rm in} \cap  \mathcal{H}^{\rm AND}_{\rm out}), \label{eq:Intersection}
\end{equation}
and in general the equality is not attained in Eq.~(\ref{eq:Intersection}) (see  Fig.~\ref{fig:example_conn_dir} for an example).  Therefore, the size of the AND-logic strongly connected component cannot be determined from the corresponding in- and out-components.

It may still be that for infinitely large random hypergraphs the difference between $\mathcal{H}^{\rm AND}_{\rm s}$ and $(\mathcal{H}^{\rm AND}_{\rm in} \cap  \mathcal{H}^{\rm AND}_{\rm out})$ is negligible.   To resolve this question, we compute the number of nodes that remain in the intersection after all the nodes from the strongly connected component have been removed from it, i.e., 
\begin{equation}
f^{\rm r}_{\rm AND}(\mathbf{I}^{\leftrightarrow}) := \frac{|(\mathcal{V}^{\rm AND}_{\rm in}\cap \mathcal{V}^{\rm AND}_{\rm out})\setminus \mathcal{V}^{\rm AND}_{\rm s}|}{N}. \label{eq:fr}
\end{equation}
If $f^{\rm r}_{\rm AND}$ converges to a nonzero value for large random hypergraphs, then the  difference between  the intersection $\mathcal{V}^{\rm AND}_{\rm in}\cap \mathcal{V}^{\rm AND}_{\rm out}$ and the strongly connected component $\mathcal{V}^{\rm AND}_{\rm s}$ is not a finite size effect, and thus cannot be neglected.

\begin{figure}
     \centering
     \setlength{\unitlength}{0.1\textwidth}
     \hspace*{0.8cm}
     \includegraphics[width=0.4\textwidth]{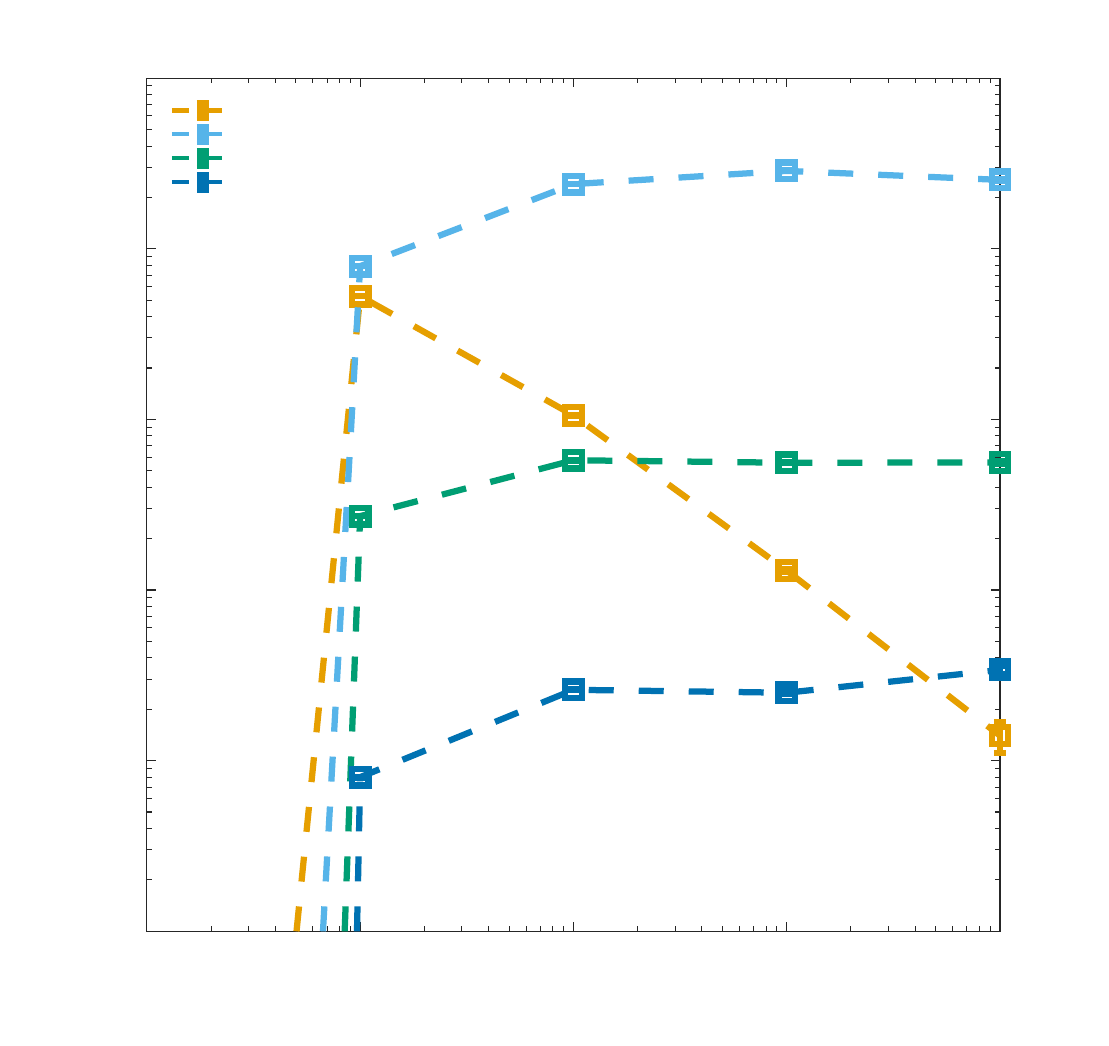}
      \put(-3.2,3.34){\tiny\color{Orange} $\overline{k}^{\rm in}=\overline{k}^{\rm out}=1$}
      \put(-3.2,3.22){\tiny\color{CornflowerBlue} $\overline{k}^{\rm in}=\overline{k}^{\rm out}=3$}
      \put(-3.2,3.10){\tiny\color{PineGreen} $\overline{k}^{\rm in}=\overline{k}^{\rm out}=5$}
      \put(-3.2,2.98){\tiny\color{MidnightBlue} $\overline{k}^{\rm in}=\overline{k}^{\rm out}=7$}
      \put(-3.6,0.15){\footnotesize$10^0$}
      \put(-2.825,0.15){\footnotesize$10^1$}
      \put(-2.05,0.15){\footnotesize$10^2$}
      \put(-2.05,-0.15){\footnotesize$N$}
      \put(-1.275,0.15){\footnotesize$10^3$}
      \put(-0.5,0.15){\footnotesize$10^4$}
      \put(-4,0.35){\footnotesize$10^{-5}$}
      \put(-4,0.95){\footnotesize$10^{-4}$}
      \put(-4,1.57){\footnotesize$10^{-3}$}
      \put(-5.1,1.9){\footnotesize$\langle f^{\rm r}_{\rm AND}(\mathbf{I}^{\leftrightarrow}) \rangle$}
      \put(-4,2.20){\footnotesize$10^{-2}$}
      \put(-4,2.82){\footnotesize$10^{-1}$}
      \put(-4,3.45){\footnotesize$10^0$}
     \caption{{\it The intersection $\V^{\rm AND}_{\rm in}\cap \V^{\rm AND}_{\rm out}$ of in- and out-components is significantly larger than  the strongly connected component $\V^{\rm AND}_{\rm s}$ in Erd\H{o}s-R\'{e}nyi hypergraphs.} The  ensemble average $\langle f^{\rm r}_{\rm AND}(\mathbf{I}^{\leftrightarrow})\rangle$ of $f^{\rm r}_{\rm AND}(\mathbf{I}^{\leftrightarrow})$, as defined in Eq.~(\ref{eq:fr}),    in directed Erd\H{o}s-R\'{e}nyi hypergraphs as a function of the number of nodes $N$, with $M=2N$ and $\overline{k}^{\rm in}=\overline{k}^{\rm out} = \overline{k}$ as indicated in the legend.      Markers are sample  averages over a sufficiently large number of  graph realisations so that  the error bar is smaller than the marker size (except for the last marker of $ \overline{k}=1$).}
     \label{fig:intersec_ICOC}
\end{figure}
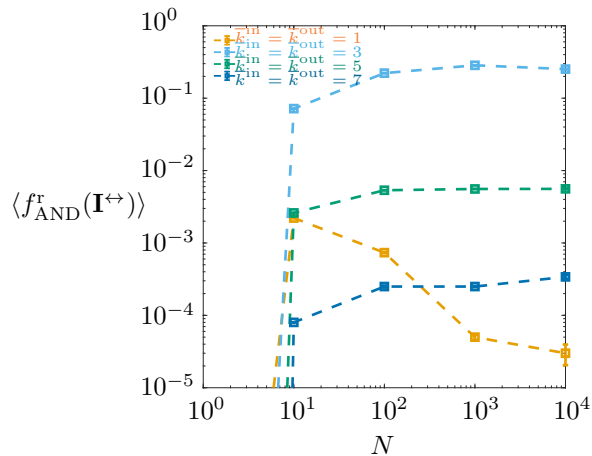

In Fig.~\ref{fig:intersec_ICOC} we plot the average value $\langle f^{\rm r}_{\rm AND}(\mathbf{I}^{\leftrightarrow})\rangle$  as a function of $N$ for directed Erd\H{o}s-R\'{e}nyi hypergraphs  of equal mean in-degree  and out-degree, $\overline{k}^{\rm out} = \overline{k}^{\rm in}=\overline{k}$.    In the  Erd\H{o}s-R\'{e}nyi  ensemble every element of $\mathbf{I}^{\rightarrow}$ (and equivalently in $\mathbf{I}^{\leftarrow}$) is  set independently and with  probability $\overline{k}/M$  to one, and otherwise the element is set to zero.   For the sake of example, we set $M=2N$.   Interestingly,  the results show that for $\overline{k}>1$ the mean value $\langle f^{\rm r}_{\rm AND}(\mathbf{I}^{\leftrightarrow})\rangle$ converges to a  nonzero value  as a function of $N$, and therefore also  for infinitely large random hypergraphs the size of AND-logic strongly connected components cannot be estimated from the intersection between the in- and out-components.    Notice for a mean degree $\overline{k}=1$ the average $\langle f^{\rm r}_{\rm AND}(\mathbf{I}^{\leftrightarrow})\rangle$ converges to zero, as $\overline{k}=1$ corresponds with the percolation transition.

\section{Giant components in nondirected hypergraphs} \label{ch:nondir_GC}
In this section, we develop an exact theory for the giant component of  large, random, nondirected hypergraphs  that have correlations between degrees and cardinalities.   In an infinitely large hypergraph, the giant component is  an infinitely large  connected component, and the probability that a node belongs to the giant component can be computed exactly with the cavity method, see Refs.~\cite{hannam2019percolation,torrisi2020percolation}.  As the largest connected component of large random hypergraphs  approximates well the giant component of an infinite hypergraph, we can use  the cavity method to predict properties of large, finite random hypergraphs, and potentially also real-world networks.   In Sec.~\ref{ch:cavity_nondir} we develop the cavity theory for large, locally tree-like hypergraphs, in Sec.~\ref{ch:cavity_deg_car} we apply the theory to random hypergraphs with prescribed degree-cardinality correlations, and in Sec.~\ref{ch:real_nondir} we compare predictions from the cavity method with real-world hypegraphs.

\subsection{Cavity method for large, locally tree-like  hypergraphs}\label{ch:cavity_nondir}
 For hypergraphs with an OR-logic associated to their hyperedges, a node $i$ does not belong to the giant component if none of the  hyperedges $\alpha\in\partial_{i}$ belong to the giant component.  Analogously, a hyperedge $\alpha$ does not belong to the giant component if none
 of its neighbouring nodes $i
\in\partial_{\alpha}$ 
belong to the giant component.   
To mathematically express the above logic,  we introduce  the indicator variables $\mu_{i}$ and $\sigma_{\alpha}$  for nodes and hyperedges, respectively, with $\mu_i=1$ ($\sigma_\alpha=1$) if node $i$ (hyperedge $\alpha$) does not belong to the giant component, and $\mu_i=0$ ($\sigma_\alpha=0$) if node $i$ (hyperedge $\alpha$) belongs to the giant component.      Using these variables, we can express the OR-logic as  
\begin{equation}
    \mu_{i}(\I)=\prod_{\alpha\in\partial_{i}(\I)}\sigma_{\alpha}(\I), \quad{\rm and}\quad
    \sigma_{\alpha}(\I) = \prod_{i\in\partial_{\alpha}(\I)}\mu_{i}(\I) .
    \label{eq:GC_nondir} 
\end{equation}

For locally tree-like hypergraphs~\cite{hannam2019percolation,torrisi2020percolation}, we can express a set of equations similar to (\ref{eq:GC_nondir}), albeit where the right-hand side contains indicator variables  $\mu^{(\alpha)}_i$ and $\sigma^{(i)}_{\alpha}$ defined on the {\it cavity hypergraphs} $\Hy^{(\alpha)}$ and $\Hy^{(i)}$.   The hypergraph $\Hy^{(\alpha)}$ is constructed from the hypergraph $\Hy$ by removing the hyperedge $\alpha$ from the set $\W$ and by removing all its corresponding links from the set $\E$; analogously, the hypergraph $\Hy^{(i)}$ is obtained from $\Hy$ by removing the node $i$ from the set $\V$ and by removing all its corresponding links from the set $\E$.   Since infinitely large random hypergraphs from the configuration model are locally tree-like, we can write~\cite{bianconi2024theory} 
\begin{equation}
    \mu_{i}(\I)=\prod_{\alpha\in\partial_{i}(\I)}\sigma_{\alpha}^{(i)}(\I), \quad{\rm and}\quad
    \sigma_{\alpha}(\I) = \prod_{i\in\partial_{\alpha}(\I)}\mu_{i}^{(\alpha)}(\I) . 
    \label{eq:GC_nondir_v2}
\end{equation}
In a similar fashion, we get
\begin{equation}
    \mu_{i}^{(\alpha)}(\I)=\prod_{\beta\in\partial_{i}(\I);\atop \beta \neq\alpha}\sigma_{\beta}^{(i)}(\I),\quad{\rm and}\quad
    \sigma_{\alpha}^{(i)}(\I) = \prod_{j\in\partial_{\alpha}(\I);\atop j\neq i}\mu_{j}^{(\alpha)}(\I). 
    \label{eq:cavity_eq}
\end{equation}

Note that the Eqs.~(\ref{eq:GC_nondir_v2}) and (\ref{eq:cavity_eq}) apply to arbitrary locally tree-like hypergraphs, and thus include all possible correlations between degrees and cardinalities of the hypergraph.    However, they need to be solved numerically.  For this notice that the indicator variables $\mu_{i}^{(\alpha)}$  and $ \sigma_{\alpha}^{(i)}$ can be interpreted as messages propagating along the links of the hypergraph; $\mu_{i}^{(\alpha)}$  is  a message  directed from $i$ to $\alpha$ and  $ \sigma_{\alpha}^{(i)}$ is a message directed from $\alpha$ to $i$, and therefore   Eqs.~(\ref{eq:cavity_eq}) are also referred to as   message passing equations ~\cite{newman2023message}.

\subsection{Random hypergraphs with degree-cardinality correlations}\label{ch:cavity_deg_car}
We present a theory for the giant component of  large, random hypergraphs  drawn from the 
configuration model with degree-cardinality correlations~\cite{bollobas1980probabilistic,bassler2015exact}.   In this model, we are provided with a prescribed distribution  $P_{\E}(k,\chi)$, such that  
\begin{equation}
 P_{\E}(k,\chi) =      P_{\E}(k,\chi|\I)  ,
\end{equation}
where 
\begin{equation}
    P_{\E}(k,\chi|\I)=\frac{\sum_{i,\alpha}I_{i\alpha}\delta_{k,k_{i}(\I)}\delta_{\chi,\chi_{\alpha}(\I)}}{\sum_{j,\beta}I_{j\beta}}
    \label{def:deg_car_corr}
\end{equation}  
is the joint distribution of    degree-cardinality pairs $(k,\chi)$ of nodes and hyperedges connected by a link in the hypergraph $\mathbf{I}$.  

The marginal distributions  of $P_{\E}(k,\chi)$ are given by 
\begin{equation}
 \sum_{k\geq 0} P_{\E}(k,\chi) = \frac{\chi P_{\W}(\chi)}{\overline{\chi}} \ \ {\rm and} \  \ \sum^{N}_{\chi\geq 0} P_{\E}(k,\chi) = \frac{P_{\V}(k)k}{\overline{k}},
\end{equation}
where $P_{\V}(k)$ and $P_{\W}(\chi)$ are the degree distribution and the cardinality distribution of nodes and hyperedges, respectively, and where $
\overline{k} = \sum^M_{k=0} P_{\V}(k)k $ and   $
\overline{\chi} = \sum^N_{\chi=0} P_{\W}(\chi)\chi $.

As large, random hypergraphs from the configuration model are locally tree-like, the  cavity  Eqs.~(\ref{eq:cavity_eq}) apply, and we can take their ensemble average.     To this purpose, we define the ensemble averaged quantities  
 \begin{equation}
 y:=\frac{1}{N}\sum^N_{i=1}\langle\mu_{i}(\I)\rangle \ \ {\rm and} \ \  x:=\frac{1}{M}\sum^M_{\alpha=1}\langle\sigma_{\alpha}(\I)\rangle 
\end{equation} 
where $\langle \cdot \rangle$ denotes an average over all infinitely large hypergraphs in the configuration model with prescribed joint distribution $P_\E(k,\chi)$.   As the random variables on the right-hand side of the Eqs.~(\ref{eq:GC_nondir_v2})  are defined on  the cavity hypergraphs $\mathcal{H}^{(i)}$ and $\mathcal{H}^{(\alpha
)}$, these random variables are independent,  i.e., 
\begin{equation}
\hspace{-2cm}\Big\langle \prod_{\alpha\in \partial_i(\I)} \sigma^{(i)}_{\alpha}(\I)\Big\rangle  = \prod_{\alpha\in \partial_i(\I)}\big\langle \sigma^{(i)}_{\alpha}(\I)\big\rangle \quad{\rm and}\quad\Big\langle \prod_{\alpha\in \partial_i(\I)} \mu^{(\alpha)}_{i}(\I)\Big\rangle  =  \prod_{\alpha\in \partial_i(\I)}\big\langle \mu^{(\alpha)}_{i}(\I)\big\rangle.
\end{equation}
Using this independence property we obtain the recursion relations 
\begin{equation}
    y=\sum_{k\geq 0}P_{\V}(k)\tilde{x}_{k}^{k}, \quad {\rm and} \quad 
    x  =\sum_{\chi\geq 0}P_{\W}(\chi)\tilde{y}_{\chi}^{\chi}, 
    \label{eq:GC_nondir_avg}
\end{equation}
where  
\begin{equation}
\tilde{x}_k := \Bigg\langle  \frac{\sum^N_{i=1}   \sum_{\alpha\in \partial_i(\mathbf{I})}\delta_{k,k_i(\mathbf{I})} \sigma^{(i)}_{\alpha}(\I) }{k\sum^N_{i=1}\delta_{k,k_i(\mathbf{I})}  } \Bigg\rangle 
\end{equation}
and 
\begin{equation}
 \tilde{y}_\chi := \Bigg \langle \frac{\sum^M_{\alpha=1} \sum_{i\in \partial_{\alpha}(\mathbf{I})}\delta_{\chi,\chi_{\alpha}(\I)} \mu^{(\alpha)}_{i}(\I)   }{\chi\sum_{\alpha=1}^{M}\delta_{\chi,\chi_{\alpha}(\I)}}\Bigg\rangle  
\end{equation}
are  ensemble averages  of $\sigma^{(i)}_{\alpha}$ and $\mu^{(\alpha)}_{i}$  conditioned on $k_i=k$ and $\chi_\alpha=\chi$, respectively.   Analogously,  using that the random variables on the right-hand side of Eqs.~(\ref{eq:cavity_eq}) are independent, we find  that 
\begin{equation}
    \tilde{x}_{k}=\sum_{\chi\geq1}P_{\E}(\chi|k)\tilde{y}_{\chi}^{\chi-1},\quad {\rm and} \quad 
    \tilde{y}_{\chi}=\sum_{k\geq1}P_{\E}(k|\chi)\tilde{x}_{k}^{k-1}
    \label{eq:tree_structure_corr}
\end{equation}
where $P_{\E}(\chi|k)$ and $P_{\E}(k|\chi)$ are the conditional distributions defined by 
\begin{equation}
    P_{\E}(k|\chi):=\frac{\overline{\chi}}{\chi}\frac{P_{\E}(k,\chi)}{P_{\W}(\chi)}, \quad {\rm and} \quad 
    P_{\E}(\chi|k):=\frac{\overline{k}}{k}\frac{P_{\E}(k,\chi)}{P_{\V}(k)}.
    \label{eq:tree_structure_probdist}
\end{equation}
Note that  on the right-hand side of (\ref{eq:tree_structure_corr}), $P_{\E}(\chi|k)$ is the probability that the cardinality $\chi_\alpha(\mathbf{I})$ of a randomly selected link $(i,\alpha)\in \E$  conditioned on $k_i(\mathbf{I})=k$ equals $\chi$, and in Eq.~(\ref{eq:tree_structure_corr}) there is a   $\tilde{y}_{\chi}^{\chi-1}$  instead of a $\tilde{y}_{\chi}^{\chi}$, as on the right-hand side of the second equality in  (\ref{eq:cavity_eq}) we take the product over all $j\in \partial_\alpha(\mathbf{I})\setminus \left\{i\right\}$.  An analogous argument applies for   $ P_{\E}(\chi|k)$ in the second equation of (\ref{eq:tree_structure_corr}). 

The quantities   
\begin{equation}
f := \langle f(\mathbf{I})\rangle  {\quad } {\rm and} \quad g :=  \langle g(\mathbf{I})\rangle
\end{equation}
denoting the probability that, respectively, a node and a hyperedge belongs to the giant component, are given by  
\begin{equation}
f= 1-y \quad {\rm and} \quad g=1-x
\end{equation}
where $y$ and $x$ are obtained from solving the Eqs.~(\ref{eq:GC_nondir_avg}) and (\ref{eq:tree_structure_corr}).  The cavity Eqs.~(\ref{eq:GC_nondir_avg}) and (\ref{eq:tree_structure_corr}) also provide us with the probabilities  $f(k)$ and $g(\chi)$ that a node or hyperedge with given degree or cardinality, respectively, belongs to the largest connected component, viz., 
\begin{equation}
    f(k)=1-\tilde{x}_{k}^{k}, \quad {\rm and} \quad 
    g(\chi)  =1-\tilde{y}_{\chi}^{\chi}.
    \label{eq:GC_nondir_avg_v2}
\end{equation}

The Eqs.~(\ref{eq:GC_nondir_avg}) and (\ref{eq:tree_structure_corr}) simplify considerably when there are no correlations between degrees and cardinalities.  Indeed, in this case the joint distribution  
\begin{equation}
P_{\E}(k,\chi) = \frac{P_\V(k)k}{\overline{k}}\frac{P_\W(\chi)\chi}{\overline{\chi}}  .\label{eq:factorised}
\end{equation}
Consequently, the probabilities $\tilde{x}_k$ and $\tilde{y}_\chi$ are independent of $k$ and $\chi$, and therefore we can drop the subindex, i.e., 
$\tilde{x}_k = \tilde{x}$ and $\tilde{y}_\chi = \tilde{y}$.  This yields the simpler set 
\begin{equation}
\tilde{y}=\sum_{k\geq 1}\frac{k}{\overline{k}}P_{\V}(k)\tilde{x}^{k-1} \ \ {\rm and} \ \ \tilde{x}=\sum_{\chi\geq 1}\frac{\chi}{\overline{\chi}}P_{\W}(\chi)\tilde{y}^{\chi-1},
\label{eq:tree_structure_wo_corr}
\end{equation}
of self-consistent equations, 
which yield
\begin{equation}
y = \sum_{k\geq 0} P_\V(k)\tilde{x}^k \ \ {\rm and} \ \ x = \sum_{\chi \geq 0}P_{\W}(\chi)\tilde{y}^\chi. \label{eq:tree_structure_wo_corr2}
\end{equation}
 Note that  the Eqs.~(\ref{eq:tree_structure_wo_corr}) can be expressed using the excess degree distribution $q_\V(k)$ and the excess cardinality distribution $q_\W(\chi)$~\cite{kryven2019bond}, which yields
\begin{equation}
\tilde{y}=\sum_{k\geq 0}q_\V(k)\tilde{x}^{k} \ \ {\rm and} \ \ \tilde{x}=\sum_{\chi\geq 0}q_\W(\chi)\tilde{y}^{\chi},
\label{eq:tree_structure_wo_corr_excess}
\end{equation}
where $q_\V(k)$ and $q_\W(\chi)$ are defined  by
\begin{equation}
    q_\V(k):=\frac{k+1}{\overline{k}}P_{\V}(k+1), \quad {\rm and} \quad 
    q_\W(\chi):=\frac{\chi+1}{\overline{\chi}}P_{\W}(\chi+1).
    \label{eq:excess_dist_def}
\end{equation}

\subsection{Application to real-world hypergraphs}\label{ch:real_nondir}
We compare the sizes of the largest connected components of real-world hypergraphs with those predicted by theoretical models.   
We consider six hypergraphs that are built from  real-world  datasets.   These hypergraphs are related to food recipes, sales of items in Walmart,   Youtube channel subscriptions, involvement of criminals in criminal cases, collaborations in Github, and ingredients of the drugs registered in FDA (see  \ref{app:data_randH} for details).

For each of the six hypergraphs we determine the fraction $f(\mathbf{I}_{\rm real})$ of nodes that belong to the giant component, as defined in Eq.~(\ref{eq:largest_component}), and where   $\mathbf{I}_{\rm real}$ denotes the incidence matrix of a real-world hypergraph.  In Table~\ref{tb:f_nondir} we compare the empirical values  $f(\mathbf{I}_{\rm real})$ with theoretical estimates of random hypergraphs with  degree-cardinality correlations [$\langle f(\mathbf{I})\rangle_{\rm corr}$ and $f^{\rm corr}_{\rm th}$ for finite and infinitely large hypergraphs, respectively], and without degree-cardinality correlations [$\langle f(\mathbf{I})\rangle_{\rm un}$ and $f_{\rm th}$ for finite and infinitely large hypergraphs, respectively]: 
\begin{itemize}
\item $\langle f(\mathbf{I})\rangle_{\rm un}$: this is the average of the fraction  $f(\mathbf{I})$ for  random hypergraphs   that have the same degree sequence $\vec{k}(\mathbf{I}) =\vec{k}(\mathbf{I}_{\rm real})$ and cardinality sequence $\vec{\chi}(\mathbf{I}) =\vec{\chi}(\mathbf{I}_{\rm real})$ as the real-world hypergraph of interest (see~\ref{app:gen_hyper} for details).   This hypergraph model  has a prescribed 
joint 
distribution of degrees and cardinalities 
the form 
\begin{equation}
P_{\E}(k,\chi) = \frac{P_\V(k|\mathbf{I}_{\rm real})k}{\overline{k}(\mathbf{I}_{\rm real})}\frac{P_\W(\chi|\mathbf{I}_{\rm real})\chi}{\overline{\chi}(\mathbf{I}_{\rm real})}  .\label{eq:factorised2}
\end{equation}  
Hence, in this model we ignore the correlations between degrees and cardinalities.  
The numbers  in  the second column of Table~\ref{tb:f_nondir}  are estimates of $\langle f(\mathbf{I})\rangle_{\rm un}$ obtained from an  empirical average over $100$ graph realisations.  
\item  $\langle f(\mathbf{I})\rangle_{\rm corr}$: this is the fraction  $f(\mathbf{I})$ averaged over   random hypergraphs that have the same degree and cardinality sequences as the real-world hypergraph of interest, and moreover the number of links connecting nodes of a certain degree and hyperedges of a certain cardinality is identical as in the  real-world hypergraph (see ~\ref{app:gen_hyper} for details).   Hence, in this case the distribution  
\begin{equation}
P_{\E}(k,\chi) = P_\E(k,\chi|\mathbf{I}_{\rm real})  \label{eq:factorised3}
\end{equation} 
does not factorise, and the random graph has degree-cardinality correlations. 
The estimates of $\langle f(\mathbf{I})\rangle_{\rm corr}$ in the table are  empirical averages over $100$ graph realisations using the generating method described in ~\ref{app:gen_hyper}.

\item $f_{\rm th}$: this is the theoretical value   $f=1-y$  for infinitely large, random hypergraphs that do not have degree-cardinality correlations.   Hence,  $y$ is obtained from numerically  solving the equations (\ref{eq:tree_structure_wo_corr}) and (\ref{eq:tree_structure_wo_corr2})  with 
$P_{\V}(k)=P_{\V}(k|\mathbf{I}_{\rm real})$ and $P_{\W}(\chi)=P_{\W}(\chi|\mathbf{I}_{\rm real})$. 
\item $f^{\rm corr}_{\rm th}$:  this is the fraction $f=1-y$  for infinitely large, random hypergraphs with degree-cardinality correlations.   The predicted value of $y$ is obtained from numerically solving the Eqs.~(\ref{eq:GC_nondir_avg}) and (\ref{eq:tree_structure_corr}) with $P_{\E}(k,\chi)=P_{\E}(k,\chi|\mathbf{I}_{\rm real})$.  
\end{itemize}

\begin{table}[b]
\caption{{\it Connected components in nondirected hypergraphs: comparison between  theoretical predictions and real-world data.}   See Sec.~\ref{ch:real_nondir}  for a description of the computed quantities in the table.}\label{tb:f_nondir}
\centering
\begin{tabular}{ccccccccc}
\hline\hline
Dataset & $f(\mathbf{I}_{\rm real})$ & $\langle f(\mathbf{I}))\rangle_{\rm un}$  & $f_{\rm th}$ & $\langle f(\mathbf{I})\rangle_{\rm corr}$ & $f^{\rm corr}_{\rm th}$ \\
\hline
Food recipe & 1.000 & 1.0000 & 0.9999 &  1.0000 & 0.9998 \\
Wallmart & 0.9833 & 0.9973 & 0.9973 & 0.9840 & 0.9925 \\
Youtube & 0.9390 & 0.9731 & 0.9731 & 0.9438  & 0.9341 \\
Crime involvement & 0.9095 & 0.7823 & 0.7810 & 0.9083 &  0.9135 \\
Github & 0.7050 & 0.9121 & 0.9124 & 0.7294 &  0.7199 \\
NDC-substances & 0.6145 & 0.8984 & 0.8979 & 0.8428 &  0.8567 \\
NDC-substances (removed edges) & 0.6145 & 0.9737 & 0.9733 & 0.6401 &  0.6067 \\
\hline\hline
\end{tabular}
\end{table}

From the results in Table~\ref{tb:f_nondir} we can classify the empirical  hypergraphs  under study into three categories.   First, there are  the hypergraphs for which  the theoretical predictions for $f$  are in good correspondence with the empirical value, both for random hypergraphs with and without degree-cardinality correlations.    These are the hypergraphs built from the Food recipe and Wallmart data sets and have  $f\approx1$.   Hence, in these hypergraph models all nodes belong to the  largest connected component.   Second, are the hypergraphs for which theoretical predictions based on random hypergraphs with degree-cardinality correlations provide a significant improvement upon estimates  without degree-cardinality correlations. The three examples here are the hypergraphs built from the Crime involvement, Youtube and the Github data sets. Thirdly, we have the NDC-substances hypergraph for which the theoretical predictions for $f$ are not in  good correspondence with empirical data, even when these include degree-cardinality correlations.    For this hypergraph, the discrepancy between the empirical and theoretical value are caused by a large number of duplicated hyperedges that connect the same nodes.   Removing those duplicated hyperedges we find a  good agreement  between theory and real-world data (see last line of Table~\ref{tb:f_nondir}).

With the cavity method we can also determine the probability  $f(k)$  that a node with degree $k$ belongs to the giant component, which is defined by
  \begin{equation}
    f(k;\I)\define\frac{\sum^N_{i=1}(1-\mu_{i}(\I))\delta_{k,k_{i}(\I)}}{\sum^N_{i=1}\delta_{k,k_{i}(\I)}},
  \end{equation}
where $\mu_{i}(\I)$ are the indicator variables with $\mu_i=1$ if node $i$ does not belong to the largest connected component of $\mathbf{I}$, and $\mu_i=0$ otherwise.

\begin{figure}
     \centering
     \setlength{\unitlength}{0.1\textwidth}
     \hspace*{0.8cm}
     \includegraphics[width=\textwidth]{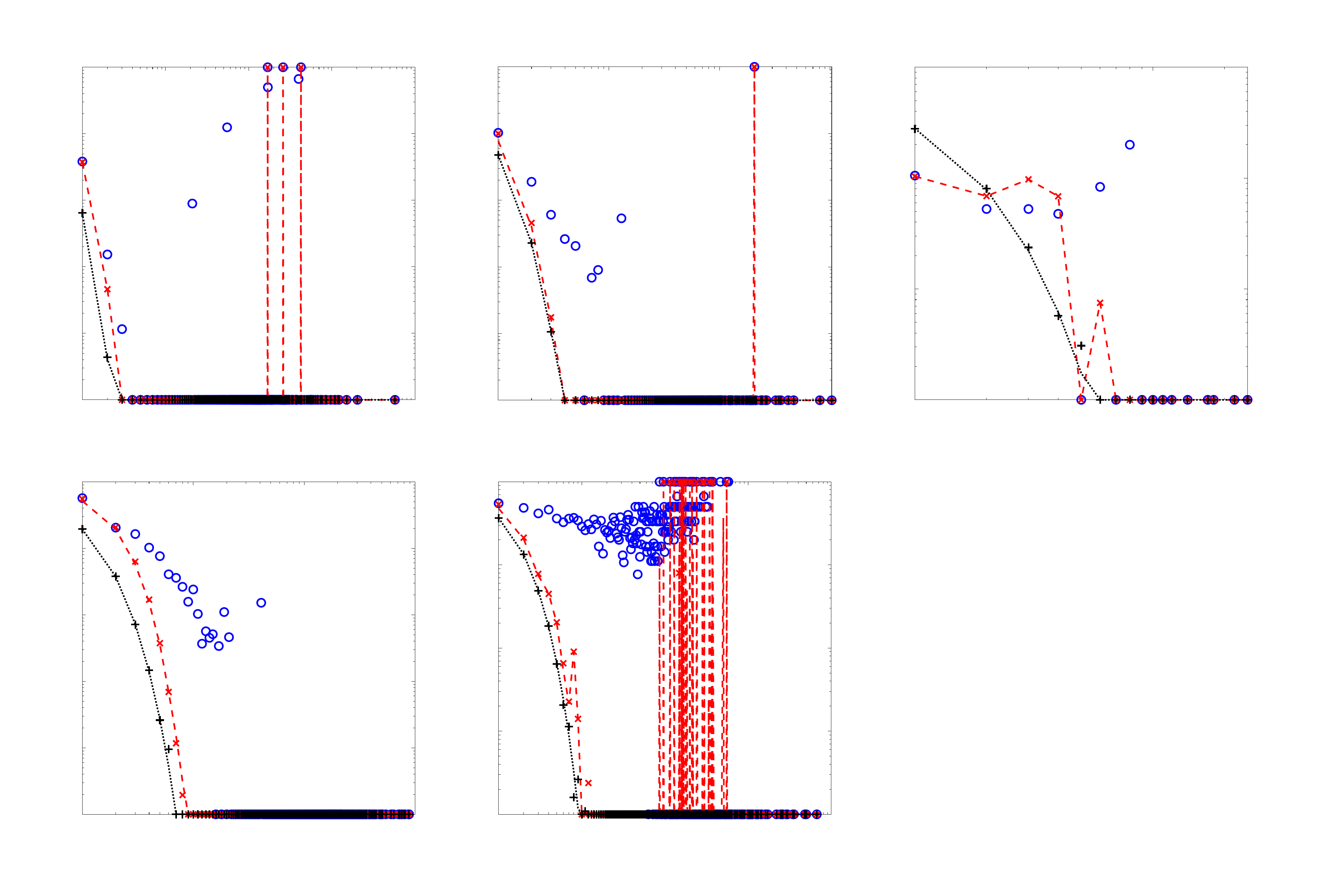}
      \put(-10.56,6.15){\small $(a)$}
      \put(-6.85,6.15){\small $(b)$}
      \put(-3.71,6.15){\small $(c)$}
      \put(-10.56,3.05){\small $(d)$}
      \put(-6.85,3.05){\small $(e)$}
      \put(-10.55,4.91){\small $1-f(k)$}
      \put(-10.55,1.85){\small $1-f(k)$}
      \put(-8.25,3.2){\small $k$}
      \put(-8.25,0.){\small $k$}
      \put(-5.2,0.){\small $k$}
      \put(-5.2,3.2){\small $k$}
      \put(-2.1,3.2){\small $k$}
      \put(-9.4,0.3){\footnotesize$1$}
      \put(-8.65,0.3){\footnotesize$10$}
      \put(-7.85,0.3){\footnotesize$10^{2}$}
      \put(-7.05,0.3){\footnotesize$10^3$}
      \put(-6.25,0.3){\footnotesize$1$}
      \put(-5.75,0.3){\footnotesize$10$}
      \put(-5.1,0.3){\footnotesize$10^{2}$}
      \put(-4.5,0.3){\footnotesize$10^{3}$}
      \put(-3.9,0.3){\footnotesize$10^{4}$}
      \put(-9.4,3.4){\footnotesize$1$}
      \put(-8.8,3.4){\footnotesize$10$}
      \put(-8.2,3.4){\footnotesize$10^2$}
      \put(-7.65,3.4){\footnotesize$10^3$}
      \put(-7.05,3.4){\footnotesize$10^4$}
      \put(-6.3,3.4){\footnotesize$1$}
      \put(-5.5,3.4){\footnotesize$10$}
      \put(-4.75,3.4){\footnotesize$10^2$}
      \put(-3.9,3.4){\footnotesize$10^3$}
      \put(-3.15,3.4){\footnotesize$1$}
      \put(-1.45,3.4){\footnotesize$10$}
      \put(-10,3.7){\scriptsize$<10^{-5}$}
      \put(-9.85,4.15){\footnotesize$10^{-4}$}
      \put(-9.85,4.67){\footnotesize$10^{-3}$}
      \put(-9.85,5.15){\footnotesize$10^{-2}$}
      \put(-9.85,5.65){\footnotesize$10^{-1}$}
      \put(-9.71,6.15){\footnotesize$10^0$}
      \put(-6.87,3.7){\scriptsize$<10^{-5}$}
      \put(-6.72,4.15){\footnotesize$10^{-4}$}
      \put(-6.72,4.67){\footnotesize$10^{-3}$}
      \put(-6.72,5.15){\footnotesize$10^{-2}$}
      \put(-6.72,5.65){\footnotesize$10^{-1}$}
      \put(-6.58,6.15){\footnotesize$10^0$}
      \put(-3.74,3.69){\scriptsize$<10^{-3}$}
      \put(-3.59,4.49){\footnotesize$10^{-2}$}
      \put(-3.59,5.33){\footnotesize$10^{-1}$}
      \put(-3.45,6.18){\footnotesize$10^0$}
      \put(-9.99,0.6){\scriptsize$<10^{-5}$}
      \put(-9.85,1.07){\footnotesize$10^{-4}$}
      \put(-9.85,1.57){\footnotesize$10^{-3}$}
      \put(-9.85,2.07){\footnotesize$10^{-2}$}
      \put(-9.85,2.58){\footnotesize$10^{-1}$}
      \put(-9.6,3.05){\footnotesize$1$}
      \put(-6.86,0.6){\scriptsize$<10^{-4}$}
      \put(-6.72,1.2){\footnotesize$10^{-3}$}
      \put(-6.72,1.8){\footnotesize$10^{-2}$}
      \put(-6.72,2.45){\footnotesize$10^{-1}$}
      \put(-6.4,3.05){\footnotesize$1$}
     \caption{Comparison between $f(k;\mathbf{I}_{\rm real})$  for five   real-world hypergraphs (blue circles) and various of its  theoretical estimates: $f_{\rm th}(k)$ and $f^{\rm corr}_{\rm th}(k)$ for infinitely large random hypergraphs without degree-cardinality correlations  (black, dotted line) and  with degree-cardinality correlation (red, dashed line), respectively; $\langle f(k;\I)\rangle_{\rm un}$  and  $\langle f(k;\I)\rangle_{\rm corr}$  for  synthetic random hypergraphs  without degree-cardinality  correlations (black plus signs)  and with degree-cardinality correlations (red crosses), respectively.  Estimates of $\langle f(k;\I)\rangle$ are based on 100 hypergraph realisations.  The real-world hypergraphs considered are: $(a)$ {\it Wallmart},   $(b)$ {\it Youtube},   $(c)$ {\it Crime involvement},   $(d)$ {\it Github}, and   $(e)$ {\it NDC-substances} (original).}
     \label{fig:gc_belonging_k}
\end{figure}
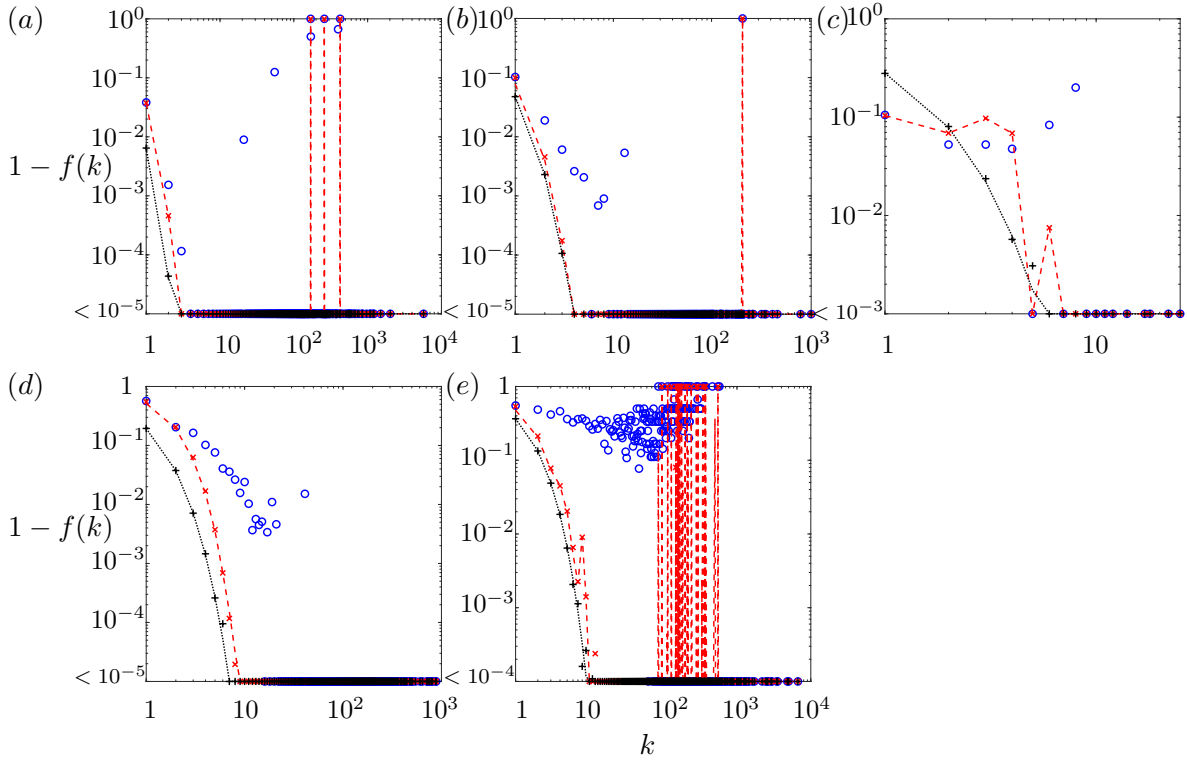

Figure~\ref{fig:gc_belonging_k} compares the fraction  $f(k;\I_{\rm real})$ in the real-world hypergraphs under study (blue circles) with  theoretical predictions 
with and without degree-cardinality correlations: $\langle f(k;\I)\rangle_{\rm corr}$ (red cross) is the average of  $f(k;\I)$ for finite, random hypergraphs that have the same joint distribution of degrees and cardinalities as the real-world hypergraph  and $\langle f(k;\I)\rangle_{\rm un}$ (black plus sign) is the corresponding quantity when neglecting degree-cardinality correlations.   We also compare the empirical values with theoretical estimations for infinitely large hypergraphs, given by $f^{\rm corr}_{\rm th}(k)$  and $f_{\rm th}(k)$ for hypergraphs with and without degree cardinality correlations.   For infinitely large hypergraphs with degree-cardinality correlations, we solve the Eqs.~(\ref{eq:tree_structure_corr}) and  (\ref{eq:GC_nondir_avg_v2}) for a distribution $P_\E(k,\chi)$ that is equal to the one in the real-world hypergraphs of interest yielding $f^{\rm corr}_{\rm th}(k)$ (red dashed line); analogously,    $f_{\rm th}(k)$ (black dotted line)  is  obtained from solving the Eqs.~(\ref{eq:tree_structure_wo_corr}) and (\ref{eq:GC_nondir_avg_v2}).

We highlight a few noteworthy features of these plots.  First, we find that including degree-cardinality correlations in the hypergraph model  improves the theoretical predictions for $f$ .
Second, the nodes that belong to the giant component are high degree nodes (see the predominance of blue circles along the $k$-axis), except for a few exceptions that we discuss below.   Both models with and without degree-cardinality correlations accurately predict when $f(k;\I_{\rm real})=1$.     Third, we observe that there exist nodes of high degree with  $f(k;\I_{\rm real})=0$ (see for example 
the real-world hypergraphs (a), (b), and (e)).  These peaks  are due to nodes in the hypergraph that have large degree but are exclusively connected to hyperedges with cardinality $1$, and therefore the model with   degree-cardinality correlations accurately predicts that they do not belong to the largest connected component.

\section{Giant components in  directed hypergraphs}  \label{ch:dir_GC}
We extend the cavity approach of the previous section to the case of directed hypergraphs.  
In Sec.~\ref{ch:cavity_dir} we develop a cavity theory for  the  OR-logic connected components   on large,  locally-tree like directed hypergraphs, and in  
Sec.~\ref{ch:dir_cavity_deg_car} 
we apply the theory to random directed hypergraphs with prescribed correlations between degrees and cardinalities of linked nodes and hyperedges.   In 
\ref{app:cavity_dir_AND} we  present the theory for AND-logic connected components.    In Sec.~\ref{ch:real_dir} we compare theoretical results with real-world
hypergraphs.

\subsection{Cavity method for locally tree-like directed hypergraphs with OR-logic}\label{ch:cavity_dir}

Within  OR-logic, a node $i$ does not belong to the in-component (out-component) if 
none of 
its neighbouring  hyperedges $\alpha\in\partial^{\rm out}_{i}$ ($\alpha\in\partial^{\rm in}_{i}$) 
belong to the in-component (out-component). Analogously, a hyperedge $\alpha$ does not belong to the in-component (out-component) if 
none
of its neighbouring nodes $i
\in\partial^{\rm out}_{\alpha}$ ($i
\in\partial^{\rm in}_{\alpha}$)  
belong to the in-component (out-component).  
To express the above relations, we introduce  indicator variables $\mu^{\rm ic}_{i}$ ($\mu^{\rm oc}_{i}$) and $\sigma^{\rm ic}_{\alpha}$ ($\sigma^{\rm oc}_{\alpha}$) for nodes and hyperedges.  We set $\mu^{\rm ic}_{i}=1$ $(\mu^{\rm oc}_{i}=1)$ and $\sigma^{\rm ic}_{\alpha}=1$ ($\sigma^{\rm oc}_{\alpha}=1$) if node $i$ and hyperedge $\alpha$, respectively, do not belong to the in-component (out-component).  Conversely, we set $\mu^{\rm ic}_{i}=0$ $(\mu^{\rm oc}_{i}=0)$ and $\sigma^{\rm ic}_{\alpha}=0$ ($\sigma^{\rm oc}_{\alpha}=0$) if node $i$ and hyperedge $\alpha$, respectively,  belong to the in-component (out-component). 
Using these variables, we can express the OR-logic relations between neighbouring nodes and hyperedges as  
\begin{eqnarray}
    \mu^{\rm oc}_{i}(\I^\leftrightarrow)=\prod_{\alpha\in\partial_{i}^{\rm in}(\I^{\leftrightarrow})}\sigma^{\rm oc}_{\alpha}(\I^\leftrightarrow), \quad
    \sigma^{\rm ic}_{\alpha}(\I^\leftrightarrow) = \prod_{i\in\partial_{\alpha}^{\rm out}(\I^{\leftrightarrow})}\mu^{\rm ic}_{i}(\I^\leftrightarrow),\nonumber \\
    \mu^{\rm ic}_{i}(\I^\leftrightarrow)=\prod_{\alpha\in\partial_{i}^{\rm out}(\I^{\leftrightarrow})}\sigma_{\alpha}^{\rm ic}(\I^\leftrightarrow), \quad
    \sigma^{\rm oc}_{\alpha}(\I^\leftrightarrow) = \prod_{i\in\partial_{\alpha}^{\rm in}(\I^{\leftrightarrow})}\mu^{\rm oc}_{i}(\I^\leftrightarrow).
    \label{eq:ICOC_dir_basic} 
\end{eqnarray}
Analogously as we did in the nondirected case, we can use the locally tree-like topology to express   the indicator variables on the hypergraph $\mathcal{H}$ in terms of corresponding variables on the  cavity hypergraphs $\mathcal{H}^{(\alpha)}$ and $\mathcal{H}^{(i)}$ obtained from $\mathcal{H}$ by removing the corresponding node and hyperedge.     This yields the sets of equations   
\begin{eqnarray}
    \mu^{\rm oc}_{i}(\I^\leftrightarrow)=\prod_{\alpha\in\partial_{i}^{\rm in}(\I^\leftrightarrow)}\sigma^{{\rm oc},(i)}_{\alpha}(\I^\leftrightarrow), \quad
    \sigma^{\rm ic}_{\alpha}(\I^\leftrightarrow) = \prod_{i\in\partial_{\alpha}^{\rm out}(\I^\leftrightarrow)}\mu^{{\rm ic},(\alpha)}_{i}(\I^\leftrightarrow), \nonumber\\
    \mu^{\rm ic}_{i}(\I^\leftrightarrow)=\prod_{\alpha\in\partial_{i}^{\rm out}(\I^\leftrightarrow)}\sigma_{\alpha}^{{\rm ic},(i)}(\I^\leftrightarrow), \quad
    \sigma^{\rm oc}_{\alpha}(\I^\leftrightarrow) = \prod_{i\in\partial_{\alpha}^{\rm in}(\I^\leftrightarrow)}\mu^{{\rm oc},(\alpha)}_{i}(\I^\leftrightarrow).
    \label{eq:ICOC_dir} 
\end{eqnarray} 
Repeating this procedure, and using the locally-tree like topology, we find the  message passing equations 
\begin{eqnarray}
    \mu^{{\rm oc},(\alpha)}_{i}(\I^\leftrightarrow)=\prod_{\beta\in\partial^{\rm in}_{i}(\I^\leftrightarrow);\atop\beta\neq\alpha}\sigma^{{\rm oc},(i)}_{\beta}(\I^\leftrightarrow),\quad
    \sigma_{\alpha}^{{\rm ic},(i)}(\I^\leftrightarrow) = \prod_{j\in\partial^{\rm out}_{\alpha}(\I^\leftrightarrow);\atop i\neq j}\mu^{{\rm ic},(\alpha)}_{j}(\I^\leftrightarrow), \nonumber\\
    \mu^{{\rm ic},(\alpha)}_{i}(\I^\leftrightarrow)=\prod_{\beta\in\partial^{\rm out}_{i}(\I^\leftrightarrow);\atop\beta\neq\alpha}\sigma^{{\rm ic},(i)}_{\beta}(\I^\leftrightarrow),\quad
    \sigma_{\alpha}^{{\rm oc},(i)}(\I^\leftrightarrow) = \prod_{j\in\partial^{\rm in}_{\alpha}(\I^\leftrightarrow);\atop i\neq j}\mu^{{\rm oc},(\alpha)}_{j}(\I^\leftrightarrow), 
    \label{eq:cavity_ICOC}
\end{eqnarray}
where in the first line $\alpha\in \partial^{\rm out}_i$ and $i\in \partial^{\rm in}_\alpha$, and in the second line $\alpha\in \partial^{\rm in}_i$ and $i\in \partial^{\rm out}_\alpha$.

As the strongly connected component is the intersection of the in-component and the out-component, a node  $i$ belongs to the strongly connected component if $\mu^{\rm ic}_i\mu^{\rm oc}_i=0$ (and analogously for hyperedges).

Note that the weakly connected component can be obtained from  the  cavity Eqs.~(\ref{eq:GC_nondir_v2}) and (\ref{eq:cavity_eq}) for the nondirected case with now $\partial_i = \partial^{\rm in}_i\cup \partial^{\rm out}_i$,  $\partial_\alpha = \partial^{\rm in}_\alpha\cup \partial^{\rm out}_\alpha$, and $[\mathbf{I}]_{ij} = \Theta([\IRight]_{ij}+[\ILeft]_{ij})$ (where $\Theta$ is the Heaviside function)~\cite{kryven2016emergence}.

\subsection{Random directed hypergraphs with degree-cardinality correlations}\label{ch:dir_cavity_deg_car} 

We consider large random directed hypergraphs extracted from the configuration model with two prescribed, joint distributions $P_{\E}^{\rightarrow}(k^{\rm in},k^{\rm out},\chi^{\rm in},\chi^{\rm out})$ and $P_{\E}^{\leftarrow}(k^{\rm in},k^{\rm out},\chi^{\rm in},\chi^{\rm out})$ 
for the directed hypergraph observables 
\begin{equation}
 \hspace{-2cm}   P_{\E}^{\rightarrow}(k^{\rm in},k^{\rm out},\chi^{\rm in},\chi^{\rm out}|\I^{\leftrightarrow})=\frac{\sum_{i,\alpha}I^{\rightarrow}_{i\alpha}\delta_{k^{\rm in},k^{\rm in}_{i}(\ILeft)}\delta_{k^{\rm out},k^{\rm out}_{i}(\IRight)}\delta_{\chi^{\rm in},\chi^{\rm in}_{\alpha}(\IRight)}\delta_{\chi^{\rm out},\chi^{\rm out}_{\alpha}(\ILeft)}}{\sum_{j,\beta}I^{\rightarrow}_{j\beta}},    \label{def:deg_car_corr1}
    \end{equation}
    and 
        \begin{equation}
 \hspace{-2cm}   P_{\E}^{\leftarrow}(k^{\rm in},k^{\rm out},\chi^{\rm in},\chi^{\rm out}|\I^{\leftrightarrow})=\frac{\sum_{i,\alpha}I^{\leftarrow}_{i\alpha}\delta_{k^{\rm in},k^{\rm in}_{i}(\ILeft)}\delta_{k^{\rm out},k^{\rm out}_{i}(\IRight)}\delta_{\chi^{\rm in},\chi^{\rm in}_{\alpha}(\IRight)}\delta_{\chi^{\rm out},\chi^{\rm out}_{\alpha}(\ILeft)}}{\sum_{j,\beta}I^{\leftarrow}_{j\beta}},
    \label{def:deg_car_corr2}
\end{equation}
respectively.

Note that marginalising  $P_{\E}^{\rightarrow}(k^{\rm in},k^{\rm out},\chi^{\rm in},\chi^{\rm out})$ and $P_{\E}^{\leftarrow}(k^{\rm in},k^{\rm out},\chi^{\rm in},\chi^{\rm out})$ we obtain  

\begin{eqnarray}
 \sum_{k^{\rm in},k^{\rm out}\geq0} P_{\E}^{\rightarrow}(k^{\rm in},k^{\rm out},\chi^{\rm in},\chi^{\rm out}) = \frac{\chi^{\rm in}}{\overline{\chi^{\rm in}}}P_{\W}(\chi^{\rm in},\chi^{\rm out}),\nonumber\\
 \sum_{\chi^{\rm in},\chi^{\rm out}\geq0} P_{\E}^{\rightarrow}(k^{\rm in},k^{\rm out},\chi^{\rm in},\chi^{\rm out}) = \frac{k^{\rm out}}{\overline{k^{\rm out}}}P_{\V}(k^{\rm in},k^{\rm out}),\nonumber\\
 \sum_{k^{\rm in},k^{\rm out}\geq0} P_{\E}^{\leftarrow}(k^{\rm in},k^{\rm out},\chi^{\rm in},\chi^{\rm out}) = \frac{\chi^{\rm out}}{\overline{\chi^{\rm out}}}P_{\W}(\chi^{\rm in},\chi^{\rm out}),\nonumber\\
 \sum_{\chi^{\rm in},\chi^{\rm out}\geq0} P_{\E}^{\leftarrow}(k^{\rm in},k^{\rm out},\chi^{\rm in},\chi^{\rm out}) = \frac{k^{\rm in}}{\overline{k^{\rm in}}}P_{\V}(k^{\rm in},k^{\rm out}),
\end{eqnarray}

where $P_{\V}(k^{\rm in},k^{\rm out})$ ($P_{\W}(\chi^{\rm in},\chi^{\rm out})$) are the joint distributions of degrees (cardinalities) of randomly selected nodes (hyperedges) in the hypergraph.    The quantities

\begin{equation}
 \hspace{-2cm} \overline{k^{\rm in}}:= \sum_{k^{\rm in},k^{\rm out}\geq0}P_{\V}(k^{\rm in},k^{\rm out})k^{\rm in}  \quad {\rm and} \quad  \overline{k^{\rm out}}:= \sum_{k^{\rm in},k^{\rm out}\geq0}P_{\V}(k^{\rm in},k^{\rm out})k^{\rm out}
\end{equation}
are, respectively, the  mean  in-degree and out-degree.    Analogously,  

\begin{equation}
 \hspace{-2cm} \overline{\chi^{\rm in}}:= \sum_{\chi^{\rm in},\chi^{\rm out}\geq0}  P_{\W}(\chi^{\rm in},\chi^{\rm out})\chi^{\rm in} \quad  {\rm and}  \quad \overline{\chi^{\rm out}}:= \sum_{\chi^{\rm in},\chi^{\rm out}\geq0}  P_{\W}(\chi^{\rm in},\chi^{\rm out})\chi^{\rm out}
\end{equation}
are, respectively, the mean in-cardinality and out-cardinality.

Next we take an ensemble average over large hypergraphs from the configuration model with prescribed distributions $P_{\E}^{\leftarrow}$ and $P_{\E}^{\rightarrow}$.    Using the notations
\begin{equation}
y^{\rm ic}:=\frac{1}{N}\sum^{N}_{i=1}\langle\mu^{\rm ic}_{i}(\I^\leftrightarrow)\rangle  \quad {\rm and} \quad x^{\rm ic}:=\frac{1}{M}\sum^{M}_{\alpha=1}\langle\sigma^{\rm ic}_{\alpha}(\I^\leftrightarrow)\rangle,
\end{equation}
and analogously defining
\begin{equation}
y^{\rm oc}:=\frac{1}{N}\sum^{N}_{i=1}\langle\mu^{\rm oc}_{i}(\I^\leftrightarrow)\rangle  \quad {\rm and} \quad x^{\rm oc}:=\frac{1}{M}\sum^{M}_{\alpha=1}\langle\sigma^{\rm oc}_{\alpha}(\I^\leftrightarrow)\rangle,
\end{equation}
 we 
 obtain  from Eqs.~(\ref{eq:ICOC_dir}) the recursions
\begin{eqnarray}
  y^{\rm ic}=\sum_{k^{\rm in},k^{\rm out}}P_{\V}(k^{\rm in},k^{\rm out})\left(\tilde{x}^{\rm ic}_{(k^{\rm in},k^{\rm out})}\right)^{k^{\rm out}},\nonumber \\
    x^{\rm ic}=\sum_{\chi^{\rm in},\chi^{\rm out}}P_{\W}(\chi^{\rm in},\chi^{\rm out})\left(\tilde{y}^{\rm ic}_{(\chi^{\rm in},\chi^{\rm out})}\right)^{\chi^{\rm out}},\nonumber\\
  y^{\rm oc}=\sum_{k^{\rm in},k^{\rm out}}P_{\V}(k^{\rm in},k^{\rm out})\left(\tilde{x}^{\rm oc}_{(k^{\rm in},k^{\rm out})}\right)^{k^{\rm in}},\nonumber\\
    x^{\rm oc}=\sum_{\chi^{\rm in},\chi^{\rm out}}P_{\W}(\chi^{\rm in},\chi^{\rm out})\left(\tilde{y}^{\rm oc}_{(\chi^{\rm in},\chi^{\rm out})}\right)^{\chi^{\rm in}},
    \label{eq:ICOC_fraction_dir}
\end{eqnarray}
where 
\begin{equation}
\tilde{x}^{\rm ic}_{(k^{\rm in},k^{\rm out})}:= \Bigg\langle  \frac{\sum^N_{i=1}   \sum_{\alpha\in \partial^{\rm out}_i(\mathbf{I}^{\leftrightarrow})}\delta_{k^{\rm in},k^{\rm in}_i(\I^\leftarrow)} \delta_{k^{\rm out},k^{\rm out}_i(\I^{\rightarrow})}\sigma^{{\rm ic},(i)}_{\alpha}(\I^\leftrightarrow) }{k^{\rm out}\sum^N_{i=1}\delta_{k^{\rm in},k^{\rm in}_i(\I^\leftarrow)} \delta_{k^{\rm out},k^{\rm out}_i(\I^{\rightarrow})}  } \Bigg\rangle 
\end{equation}
and 
\begin{equation}
 \tilde{y}^{\rm ic}_{(\chi^{\rm in},\chi^{\rm out})} := \Bigg \langle \frac{\sum^M_{\alpha=1} \sum_{i\in \partial^{\rm out}_{\alpha}(\mathbf{I}^{\leftrightarrow})}\delta_{\chi^{\rm in},\chi^{\rm in}_{\alpha}(\I^{\rightarrow})}\delta_{\chi^{\rm out},\chi^{\rm out}_{\alpha}(\I^\leftarrow)} \mu^{{\rm ic},(\alpha)}_{i}(\I^\leftrightarrow)   }{\chi^{\rm out}\sum_{\alpha=1}^{M}\delta_{\chi^{\rm in},\chi^{\rm in}_{\alpha}(\I^{\rightarrow})}\delta_{\chi^{\rm out},\chi^{\rm out}_{\alpha}(\I^\leftarrow)}}\Bigg\rangle 
\end{equation}
and a similar definition applies for the out-component probabilities $\tilde{x}^{\rm oc}_{(k^{\rm in},k^{\rm out})}$ and  $\tilde{y}^{\rm oc}_{(\chi^{\rm in},\chi^{\rm out})}$.
Taking the esemble average of Eq.~(\ref{eq:cavity_ICOC}) we find
\begin{eqnarray}
    \tilde{y}^{\rm ic}_{(\chi^{\rm in},\chi^{\rm out})}=\sum_{k^{\rm in},k^{\rm out}}P_{\E}^{\leftarrow}(k^{\rm in},k^{\rm out}|\chi^{\rm in},\chi^{\rm out})\left(\tilde{x}^{\rm ic}_{(k^{\rm in},k^{\rm out})}\right)^{k^{\rm out}}, \nonumber\\
    \tilde{x}^{\rm ic}_{(k^{\rm in},k^{\rm out})}=\sum_{\chi^{\rm in},\chi^{\rm out}}P_{\E}^{\rightarrow}(\chi^{\rm in},\chi^{\rm out}|k^{\rm in},k^{\rm out})\left(\tilde{y}^{\rm ic}_{(\chi^{\rm in},\chi^{\rm out})}\right)^{\chi^{\rm out}},\nonumber\\
    \tilde{y}^{\rm oc}_{(\chi^{\rm in},\chi^{\rm out})}=\sum_{k^{\rm in},k^{\rm out}}P_{\E}^{\rightarrow}(k^{\rm in},k^{\rm out}|\chi^{\rm in},\chi^{\rm out})\left(\tilde{x}^{\rm oc}_{(k^{\rm in},k^{\rm out})}\right)^{k^{\rm in}}, \nonumber\\
    \tilde{x}^{\rm oc}_{(k^{\rm in},k^{\rm out})}=\sum_{\chi^{\rm in},\chi^{\rm out}}P_{\E}^{\leftarrow}(\chi^{\rm in},\chi^{\rm out}|k^{\rm in},k^{\rm out})\left(\tilde{y}^{\rm oc}_{(\chi^{\rm in},\chi^{\rm out})}\right)^{\chi^{\rm in}},
    \label{eq:tree_structure_dir}
\end{eqnarray}
where the conditional probabilities are defined by 

\begin{eqnarray}
    P_{\E}^{\rightarrow}(k^{\rm in},k^{\rm out}|\chi^{\rm in},\chi^{\rm out}):=\frac{\overline{\chi^{\rm in}}P_{\E}^{\rightarrow}(k^{\rm in},k^{\rm out}\chi^{\rm in},\chi^{\rm out})}{\chi^{\rm in} P_{\W}(\chi^{\rm in},\chi^{\rm out})}, \nonumber\\
    P_{\E}^{\rightarrow}(\chi^{\rm in},\chi^{\rm out}|k^{\rm in},k^{\rm out}):=\frac{\overline{k^{\rm out}} P_{\E}^{\rightarrow}(k^{\rm in},k^{\rm out}\chi^{\rm in},\chi^{\rm out})}{k^{\rm out} P_{\V}(k^{\rm in},k^{\rm out})}, \nonumber\\
    P_{\E}^{\leftarrow}(k^{\rm in},k^{\rm out}|\chi^{\rm in},\chi^{\rm out}):=\frac{\overline{\chi^{\rm out}} P_{\E}^{\leftarrow}(k^{\rm in},k^{\rm out}\chi^{\rm in},\chi^{\rm out})}{\chi^{\rm out} P_{\W}(\chi^{\rm in},\chi^{\rm out})},\nonumber\\
    P_{\E}^{\leftarrow}(\chi^{\rm in},\chi^{\rm out}|k^{\rm in},k^{\rm out}):=\frac{\overline{k^{\rm in}}P_{\E}^{\leftarrow}(k^{\rm in},k^{\rm out}\chi^{\rm in},\chi^{\rm out})}{k^{\rm in} P_{\V}(k^{\rm in},k^{\rm out})}.
    \label{def:cond_deg_car_jointcorr}
\end{eqnarray}

Note that, differently from   the nondirected case [see Eqs.~(\ref{eq:tree_structure_corr})], the degrees and cardinalities  in the  exponents   on the right-hand sides of the Eqs.~(\ref{eq:tree_structure_dir}) are not substracted  by one.  This is  because  in the Eqs.~(\ref{eq:cavity_ICOC})   the probability that a vertex belongs to both the in-neighbourhood and the out-neighbourhood of another vertex is negligible  for large, random, directed hypergraphs.   
Solving the 
set of 
Eqs. (\ref{eq:ICOC_fraction_dir}) together with 
(\ref{eq:tree_structure_dir}) for  given distributions  $P_{\E}^{\leftarrow}$ and  $P_{\E}^{\rightarrow}$ we obtain the probabilities $f^{\rm ic}_{\rm OR}=1-y^{\rm ic}$ and $f^{\rm oc}_{\rm OR}=1-y^{\rm oc}$ that  a node belongs to the in- and out-component, respectively.

The strongly connected component  is the intersection of the in-component and the out-component.   Using that the fraction of nodes that belong to the union of in-component and out-component is given by 
\begin{equation}
1-\sum_{k_{\rm in},k_{\rm out}}P_{\V} \left(\tilde{x}^{\rm ic}_{(k^{\rm in},k^{\rm out})}\right)^{k^{\rm out}}\left(\tilde{x}^{\rm oc}_{(k^{\rm in},k^{\rm out})}\right)^{k^{\rm in}}
\end{equation}
and using the inclusion-exclusion principle, we find that 
\begin{equation}
    f^{\rm sc}_{\rm OR}=\sum_{k_{\rm in},k_{\rm out}}P_{\V}(k_{\rm in},k_{\rm out})\left[1-\left(\tilde{x}^{\rm ic}_{(k^{\rm in},k^{\rm out})}\right)^{k^{\rm out}}\right]\left[1-\left(\tilde{x}^{\rm oc}_{(k^{\rm in},k^{\rm out})}\right)^{k^{\rm in}}\right].
    \label{eq:SC_form}
\end{equation}

Analogous with nondirected hypergraph (see Eqs.(\ref{eq:GC_nondir_avg_v2})), for  OR-logic directed hypergraphs, the cavity  Eqs.~(\ref{eq:ICOC_fraction_dir}) and (\ref{eq:tree_structure_dir}) give us  the probabilities $f^{\rm ic}(k^{\rm in},k^{\rm out})$, $f^{\rm oc}(k^{\rm in},k^{\rm out})$ and $f^{\rm sc}(k^{\rm in},k^{\rm out})$ that, respectively, a node with degrees $k^{\rm in}$ and $k^{\rm out}$ belongs to the in-component, out-component and strongly connected component, viz.,
\begin{eqnarray}
    f^{\rm ic}(k^{\rm in},k^{\rm out})=1-\left(\tilde{x}^{\rm ic}_{(k^{\rm in},k^{\rm out})}\right)^{k^{\rm out}},\nonumber\\
    f^{\rm oc}(k^{\rm in},k^{\rm out})=1-\left(\tilde{x}^{\rm sc}_{(k^{\rm in},k^{\rm out})}\right)^{k^{\rm in}},\nonumber\\
    f^{\rm sc}(k^{\rm in},k^{\rm out})=\left[1-\left(\tilde{x}^{\rm ic}_{(k^{\rm in},k^{\rm out})}\right)^{k^{\rm out}}\right]\left[1-\left(\tilde{x}^{\rm oc}_{(k^{\rm in},k^{\rm out})}\right)^{k^{\rm in}}\right].
    \label{eq:ICOC_fraction_dir_v2}
\end{eqnarray}

For random hypergraphs without degree-cardinality correlations it holds that

\begin{eqnarray}
P^{\rightarrow}_{\E}(k^{\rm in},k^{\rm out},\chi^{\rm in},\chi^{\rm out}) &= \frac{P_\V(k^{\rm in},k^{\rm out})k^{\rm out}}{\overline{k^{\rm out}}}\frac{P_\W(\chi^{\rm in},\chi^{\rm out})\chi^{\rm in}}{\overline{\chi^{\rm in}}},\nonumber\\
P^{\leftarrow}_{\E}(k^{\rm in},k^{\rm out},\chi^{\rm in},\chi^{\rm out}) &= \frac{P_\V(k^{\rm in},k^{\rm out})k^{\rm in}}{\overline{k^{\rm in}}}\frac{P_\W(\chi^{\rm in},\chi^{\rm out})\chi^{\rm out}}{\overline{\chi^{\rm out}}},\label{eq:dir_factorised}
\end{eqnarray}

and consequently  $\tilde{x}_{(k^{\rm in},k^{\rm out})} = \tilde{x}$ and $\tilde{y}_{(\chi^{\rm in},\chi^{\rm out})} = \tilde{y}$, independent of $k^{\rm in}$, $k^{\rm out}$, $\chi^{\rm in}$ and $\chi^{\rm out}$.    This yields the simpler set of self-consistent equations

\begin{eqnarray}
\hspace{-2cm}    \tilde{y}^{\rm ic}=\sum_{k^{\rm in},k^{\rm out}}\frac{P_\V(k^{\rm in},k^{\rm out})k^{\rm in}}{\overline{k^{\rm in}}}\left(\tilde{x}^{\rm ic}\right)^{k^{\rm out}}, \quad
    \tilde{x}^{\rm ic}=\sum_{\chi^{\rm in},\chi^{\rm out}}\frac{P_\W(\chi^{\rm in},\chi^{\rm out})\chi^{\rm in}}{\overline{\chi^{\rm in}}}\left(\tilde{y}^{\rm ic}\right)^{\chi^{\rm out}},\nonumber\\
\hspace{-2cm}     \tilde{y}^{\rm oc}=\sum_{k^{\rm in},k^{\rm out}}\frac{P_\V(k^{\rm in},k^{\rm out})k^{\rm out}}{\overline{k^{\rm out}}}\left(\tilde{x}^{\rm oc}\right)^{k^{\rm in}}, \quad
    \tilde{x}^{\rm oc}=\sum_{\chi^{\rm in},\chi^{\rm out}}\frac{P_\W(\chi^{\rm in},\chi^{\rm out})\chi^{\rm out}}{\overline{\chi^{\rm out}}}\left(\tilde{y}^{\rm oc}\right)^{\chi^{\rm in}},
\label{eq:tree_structure_dir_uncorr}
\end{eqnarray}
and
\begin{eqnarray}
\hspace{-1.5cm}    y^{\rm ic}=\sum_{k^{\rm out}}P_{\V}(k^{\rm out})\left(\tilde{x}^{\rm ic}\right)^{k^{\rm out}}, \quad
    x^{\rm ic}=\sum_{\chi^{\rm out}}P_{\W}(\chi^{\rm out})\left(\tilde{y}^{\rm ic}\right)^{\chi^{\rm out}},\nonumber\\
\hspace{-1.5cm}    y^{\rm oc}=\sum_{k^{\rm in}}P_{\V}(k^{\rm in})\left(\tilde{x}^{\rm oc}\right)^{k^{\rm in}}, \quad
    x^{\rm oc}=\sum_{\chi^{\rm in}}P_{\W}(\chi^{\rm in})\left(\tilde{y}^{\rm oc}\right)^{\chi^{\rm in}},
    \label{eq:ICOC_fraction_dir_uncorr}
\end{eqnarray}
where we have used the single-variable marginal probabilities $P_{\V}(k^{\rm in}) := \sum_{k^{\rm out}}P_{\V}(k^{\rm in},k^{\rm out})$,  $P_{\V}(k^{\rm out}) := \sum_{k^{\rm in}}P_{\V}(k^{\rm in},k^{\rm out})$, $P_{\W}(\chi^{\rm in}) := \sum_{\chi^{\rm out}}P_{\W}(\chi^{\rm in},\chi^{\rm out})$ and $P_{\W}(\chi^{\rm out}) := \sum_{\chi^{\rm in}}P_{\W}(\chi^{\rm in},\chi^{\rm out})$.

\subsection{Application to real-world hypergraphs} \label{ch:real_dir}
We compare theoretical predictions for the size of the largest strongly-connected component (and the corresponding in-components, out-components, etc.) with  data from  real-world directed hypergraphs. We consider 
three  real-world datasets corresponding with distinct domains: human metabolic pathways (biological network), email-sending patterns (social network), and synonyms in the English language (information network); see \ref{app:data_randH} for further details.

\subsubsection{OR-logic}\label{Sec:realWorldOR}$\\$
First, we consider OR-logic connected  components. 
For each of the three hypergraphs we determine the fractions $f^{\mathfrak{a}}_{\rm OR}(\mathbf{I}^{\leftrightarrow}_{\rm real})$ of nodes that belong to the largest connected components with $\mathfrak{a}\in \left\{{\rm sc}, {\rm ic}, {\rm oc}, {\rm wc}, {\rm t}\right\}$, see Eq.~(\ref{eq:fSc}).  We use  Tarjan's algorithm  for bipartite networks to determine the OR-logic strongly connected components in directed hypergraphs, and we use  breadth first search algorithm to determine the remaining components (weakly connected, in- and out-components~\cite{hartmann2006phase}).

 Table~\ref{tb:f_dir} compares these empirical values   with theoretical estimates of random hypergraphs with  degree-cardinality correlations, and without degree-cardinality correlations: 
\begin{itemize}
\item $\langle f^{\mathfrak{a}}_{\rm OR}(\mathbf{I}^{\leftrightarrow})\rangle_{\rm un}$: this is the average of   $f^{\mathfrak{a}}_{\rm OR}(\mathbf{I}^{\leftrightarrow})$, the fraction of nodes that belong to the largest $\mathfrak{a}$-component, for  random hypergraphs   that have the same  in-degree and out-degree sequences as the real-world hypergraph of interest, i.e., $\vec{k}^{\rm in}(\mathbf{I}^{\leftarrow}) =\vec{k}^{\rm in}(\mathbf{I}^{\leftarrow}_{\rm real})$ and $\vec{k}^{\rm out}(\mathbf{I}^{\rightarrow}) =\vec{k}^{\rm out}(\mathbf{I}^{\rightarrow}_{\rm real})$, and that have the same  in-cardinality and out-cardinality sequences of the real-world hypergraph of interest, i.e., $\vec{\chi}^{\rm in}(\mathbf{I}^{\rightarrow}) =\vec{\chi}^{\rm in}(\mathbf{I}^{\rightarrow}_{\rm real})$, $\vec{\chi}^{\rm out}(\mathbf{I}^{\leftarrow}) =\vec{\chi}^{\rm out}(\mathbf{I}^{\leftarrow}_{\rm real})$ (see~\ref{app:gen_hyper} for details). This hypergraph model  has a prescribed distribution of the form 

\begin{equation}
\hspace{-1cm}    P_{\E}^{\rightarrow}(k^{\rm in},k^{\rm out},\chi^{\rm in},\chi^{\rm out})=      \frac{P_{\V}(k^{\rm in},k^{\rm out}|\mathbf{I}^{\leftrightarrow}_{\rm real}) k^{\rm out}}{\overline{k^{\rm out}}} \frac{P_{\W}(\chi^{\rm in},\chi^{\rm out}|\mathbf{I}^{\leftrightarrow}_{\rm real}) \chi^{\rm in}}{\overline{\chi^{\rm in}}}   \label{eq:factorisedDir1}
\end{equation} 
and 
\begin{equation}
\hspace{-1cm}    P_{\E}^{\leftarrow}(k^{\rm in},k^{\rm out},\chi^{\rm in},\chi^{\rm out})=      \frac{P_{\V}(k^{\rm in},k^{\rm out}|\mathbf{I}^{\leftrightarrow}_{\rm real}) k^{\rm in}}{\overline{k^{\rm in}}} \frac{P_{\W}(\chi^{\rm in},\chi^{\rm out}|\mathbf{I}^{\leftrightarrow}_{\rm real}) \chi^{\rm out}}{\overline{\chi^{\rm out}}}.   \label{eq:factorisedDir2}
\end{equation} 
Thus, in this model we ignore the correlations between degrees and cardinalities.   The estimates in Table~\ref{tb:f_dir}  are obtained from  empirical  averages over $100$ graph realisations:
\item  $\langle f^{\mathfrak{a}}_{\rm OR}(\mathbf{I}^{\leftrightarrow})\rangle_{\rm corr}$: this is the fraction   $f^{\mathfrak{a}}_{\rm OR}(\mathbf{I}^{\leftrightarrow})$ averaged over   finite and random hypergraphs that
have the same degree sequences and cardinality sequences as the real-world hypergraph of interest, and in addition the number of
links that point from nodes to hyperedges (and from hyperedges to nodes) for given degrees and cardinalities at their end points is the same as in the real-world hypergraph under study (see ~\ref{app:gen_hyper} for details).
Hence, in this case we set  $P^{\rightarrow}_{\E}(k^{\rm in},k^{\rm out},\chi^{\rm in},\chi^{\rm out})$ and $P^{\leftarrow}_{\E}(k^{\rm in},k^{\rm out},\chi^{\rm in},\chi^{\rm out})$ equals to the corresponding empirical distributions  as defined in (\ref{def:deg_car_corr1}) and (\ref{def:deg_car_corr2}) for $\mathbf{I}^{\leftrightarrow}_{\rm real}$.
The estimates of $\langle f^{\mathfrak{a}}(\mathbf{I}^{\leftrightarrow})\rangle_{\rm corr}$ in the table are again empirical averages over $100$ graph realisations.

\item $f^{\mathfrak{a}}_{\rm th}$: these are the theoretical values   $f^{\mathfrak{a}}_{\rm OR}$  for  infinitely large, random hypergraphs that do not have degree-cardinality correlations (for notation simplicity, we omitted OR in $f^{\mathfrak{a}}_{\rm th}$).   Notably, $f^{\rm in}_{\rm th}=1-y^{\rm ic}$ and $f^{\rm out}_{\rm th} = 1-y^{\rm oc}$, where $y^{\rm ic}$ and $y^{\rm oc}$ are obtained from solving the Eqs.~(\ref{eq:tree_structure_dir_uncorr}) and (\ref{eq:ICOC_fraction_dir_uncorr}) for $P_\V(k^{\rm in},k^{\rm out}) = P_{\V}(k^{\rm in},k^{\rm out}|\mathbf{I}^{\leftrightarrow}_{\rm real})$ and $P_\W(\chi^{\rm in},\chi^{\rm out}) = P_{\W}(\chi^{\rm in},\chi^{\rm out}|\mathbf{I}^{\leftrightarrow}_{\rm real})$.    The value of $f^{\rm sc}_{\rm th}$ follows from Eq.~(\ref{eq:SC_form}) and setting  $\tilde{x}^{\rm ic}_{(k^{\rm in},k^{\rm out})} = \tilde{x}^{\rm ic}$ and $\tilde{x}^{\rm oc}_{(k^{\rm in},k^{\rm out})} = \tilde{x}^{\rm oc}$, with $\tilde{x}^{\rm ic}$ and $\tilde{x}^{\rm oc}$ the solutions to (\ref{eq:tree_structure_dir_uncorr}).     To obtain the  value of $f^{\rm wc}_{\rm th}$, we use the same approach  as for $f_{\rm th}$ in Sec.~\ref{ch:real_nondir}.  Laslty, $f^{\rm t}_{\rm th} = f^{\rm wc}_{\rm th}-f^{\rm in}_{\rm th} - f^{\rm out}_{\rm th} + f^{\rm sc}_{\rm th}$. 

\item $f^{\mathfrak{a}, \rm corr}_{\rm th}$:   these are the theoretical values   $f^{\mathfrak{a}}_{\rm OR}$  for  infinitely large, random hypergraphs that   have degree-cardinality correlations.    We obtain $f^{\rm in,\rm corr}_{\rm th}$ and $f^{\rm out,\rm corr}_{\rm th}$ from solving the Eqs.~(\ref{eq:ICOC_fraction_dir}) together with (\ref{eq:tree_structure_dir}) for distributions $P^{\rightarrow}_{\E}$ and $P^{\leftarrow}_{\E}$  that are equal to those of the real-world hypergraphs of interest.  The fraction of nodes that occupy the strongly connected component, $f^{\rm sc,\rm corr}_{\rm th}$ are determined by Eq.~(\ref{eq:SC_form}).  For  $f^{\rm wc, \rm corr}_{\rm th}$ we use the same procedure as for $f^{\rm corr}_{\rm th}$ with nondirected hypergraphs, see Sec.~\ref{ch:real_nondir}, and again  $f^{\rm t}_{\rm th, \rm corr} = f^{\rm wc, \rm corr}_{\rm th}-f^{\rm in, \rm corr}_{\rm th} - f^{\rm out, \rm corr}_{\rm th} + f^{\rm sc, \rm corr}_{\rm th}$. 
\end{itemize}

Note that unlike nondirected hypergraphs the theoretical predictions without degree-cardinality correlations, $f_{\rm th}$, correspond  well with the empirical values obtained from real-world data.  Hence, we obtain the unexpected result that degree-cardinality correlations are not necessary to  describe connected components in directed hypergraphs.

\begin{table}
\caption{{\it OR-logic connected components in directed hypergraphs:  comparison between  theoretical predictions and real-world data}.  See Sec.~\ref{Sec:realWorldOR} for a description of the computed quantities in the table.  }
\centering
\begin{tabular}{c|c|cccccc}
\hline\hline
Dataset& $\mathfrak{a}$ & $f^{\mathfrak{a}}_{\rm OR}(\mathbf{I}^{\leftrightarrow}_{\rm real})$ & $\langle f^{\mathfrak{a}}_{\rm OR}(\mathbf{I}^{\leftrightarrow})\rangle_{\rm un}$ & $f^{\mathfrak{a}}_{\rm th}$ & $\langle f^{\mathfrak{a}}(\mathbf{I}^{\leftrightarrow})_{\rm OR}\rangle_{\rm corr}$ & $f^{\mathfrak{a}, \rm corr}_{\rm th}$ \\
\hline
Metabolic pathways&\begin{tabular}{@{}c@{}c@{}c@{}c@{}}wc \\ ic \\ oc\\ sc\\t\end{tabular}& \begin{tabular}{@{}c@{}c@{}c@{}c@{}} 0.9721 \\ 0.6439 \\ 0.6969 \\ 0.4072 \\ 0.0385 \end{tabular}& \begin{tabular}{@{}c@{}c@{}c@{}c@{}}0.9976 \\ 0.6754 \\ 0.7154\\0.4102\\0.0169\end{tabular}& \begin{tabular}{@{}c@{}c@{}c@{}c@{}}0.9975 \\ 0.6756 \\ 0.7156 \\ 0.4104 \\0.0167\end{tabular}& \begin{tabular}{@{}c@{}c@{}c@{}c@{}}0.9950 \\ 0.6573 \\ 0.7033\\0.3928\\0.0272\end{tabular}& \begin{tabular}{@{}c@{}c@{}c@{}c@{}}0.9967 \\ 0.6576 \\ 0.7031\\ 0.3927\\0.0287\end{tabular}\\\hline
DNC-email&\begin{tabular}{@{}c@{}c@{}c@{}c@{}}wc \\ ic \\ oc\\ sc\\t\end{tabular}& \begin{tabular}{@{}c@{}c@{}c@{}c@{}} 0.9693 \\ 0.5003 \\ 0.6774 \\ 0.2750 \\ 0.0666 \end{tabular}& \begin{tabular}{@{}c@{}c@{}c@{}c@{}}0.9969 \\ 0.5303 \\ 0.6758\\0.2779\\0.0682\end{tabular}& \begin{tabular}{@{}c@{}c@{}c@{}c@{}}0.9965 \\ 0.5298 \\ 0.6755 \\ 0.2782 \\0.0694\end{tabular}& \begin{tabular}{@{}c@{}c@{}c@{}c@{}}0.9964 \\ 0.5122 \\ 0.6860\\0.2729\\0.0709\end{tabular}& \begin{tabular}{@{}c@{}c@{}c@{}c@{}}0.9899 \\ 0.5121 \\ 0.6861 \\ 0.2728 \\ 0.0645\end{tabular}\\\hline
English Synonyms&\begin{tabular}{@{}c@{}c@{}c@{}c@{}}wc \\ ic \\ oc\\ sc\\t\end{tabular}& \begin{tabular}{@{}c@{}c@{}c@{}c@{}} 0.8148 \\ 0.3582 \\ 0.6883 \\ 0.3060 \\ 0.0744 \end{tabular}& \begin{tabular}{@{}c@{}c@{}c@{}c@{}}0.9966 \\ 0.4816 \\ 0.8520\\0.3887\\0.0517\end{tabular}& \begin{tabular}{@{}c@{}c@{}c@{}c@{}}0.9960 \\ 0.4816 \\ 0.8518\\0.3887\\0.0513\end{tabular}& \begin{tabular}{@{}c@{}c@{}c@{}c@{}}0.9681 \\ 0.4433 \\ 0.8467\\0.3666\\0.0446\end{tabular}& \begin{tabular}{@{}c@{}c@{}c@{}c@{}}0.9595 \\ 0.4432 \\ 0.8468 \\ 0.3664 \\0.0359\end{tabular}\\
\hline\hline
\end{tabular}\label{tb:f_dir}
\end{table}

\begin{figure}[t]
     \centering
     \setlength{\unitlength}{0.1\textwidth}
     \hspace*{1.7cm}
     \includegraphics[width=0.93\textwidth]{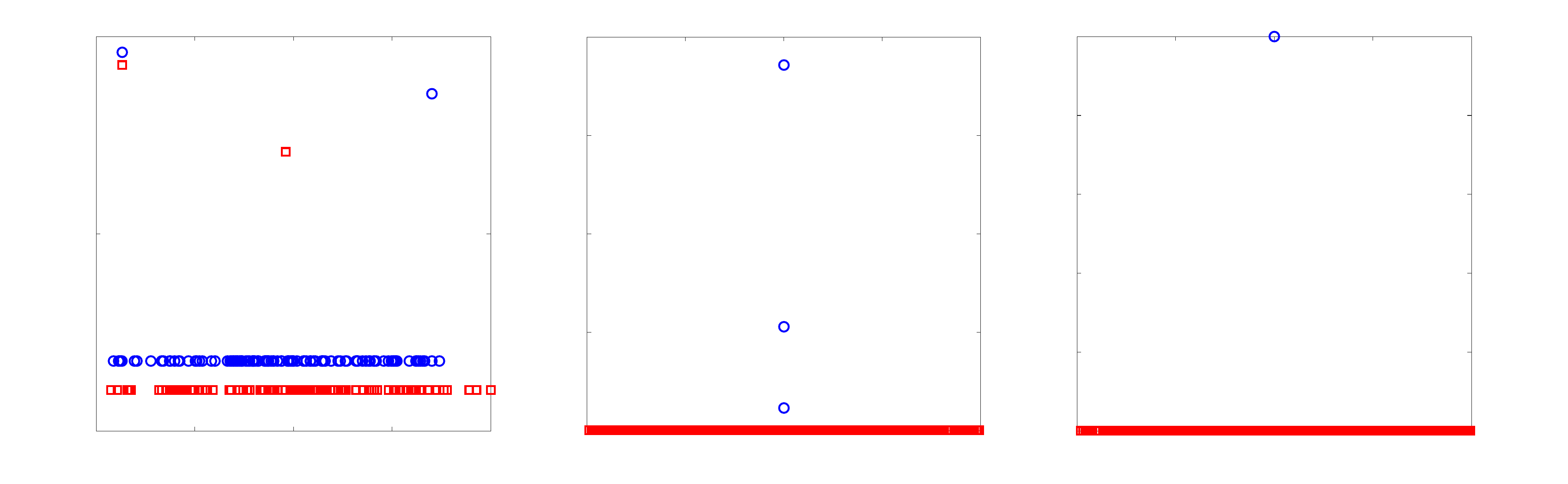}
      \put(-10.3,2.9){\small $(a)$}
      \put(-6.3,2.9){\small $(b)$}
      \put(-3.3,2.9){\small $(c)$}
      \put(-8.52,2.53){\tiny\color{blue} $\mathfrak{a}=\rightarrow$ }
      \put(-8.52,2.43){\tiny\color{red} $\mathfrak{a}=\leftarrow$ }
      \put(-10.1,1.7){\footnotesize $\rho^{\mathfrak{a}}\left(\lambda|\mathbf{I}^{\leftrightarrow}_{\rm real}\right)$}
      \put(-4.71,-0.3){\normalsize $\lambda$}
      \put(-7.62,0.12){\footnotesize$1$}
      \put(-6.45,0.12){\footnotesize$2$}
      \put(-9.3,0.25){\footnotesize$0.006$}
      \put(-8.75,0.12){\footnotesize$0$}
      \put(-9.3,1.43){\footnotesize$0.012$}
      \put(-9.3,2.60){\footnotesize$0.018$}
      \put(-6.,0.12){\footnotesize$0$}
      \put(-6.17,0.84){\footnotesize$0.2$}
      \put(-6.17,1.43){\footnotesize$0.4$}
      \put(-6.17,2.01){\footnotesize$0.6$}
      \put(-6.17,2.60){\footnotesize$0.8$}
      \put(-4.71,0.12){\footnotesize$1$}
      \put(-3.6,0.12){\footnotesize$2$}
      \put(-3.07,0.12){\footnotesize$0$}
      \put(-3.24,0.71){\footnotesize$0.2$}
      \put(-3.24,1.19){\footnotesize$0.4$}
      \put(-3.24,1.66){\footnotesize$0.6$}
      \put(-3.24,2.13){\footnotesize$0.8$}
      \put(-3.07,2.6){\footnotesize$1$}
      \put(-1.8,0.12){\footnotesize$1$}
      \put(-0.70,0.12){\footnotesize$2$}
    \caption{Plot of $\rho^{\mathfrak{a}}\left(\lambda|\mathbf{I}^{\leftrightarrow}_{\rm real}\right)$  as defined in Eq.~(\ref{eq:rhoalambda}) with $\mathfrak{a}\in \left\{\leftarrow,\rightarrow\right\}$  for the three real-world datasets considered:  {\it Human metabolic pathways} (Panel (a)), {\it DNC-email} (Panel (b)), and {\it English thesaurus} (Panel (c)).}
    \label{fig:dir_W_ratio}
\end{figure}
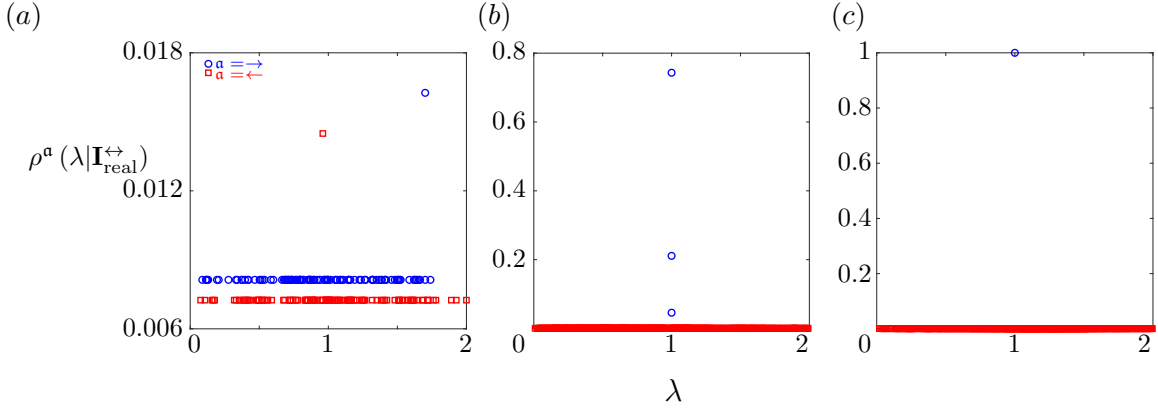

The good corresponence between random graphs models without degree-cardinality correlations and real-world directed hypergraphs relies on the fact that the real-world hypergraphs considered do not have significant correlations between degrees and cardinalities.    We confirm that this is indeed the case by calculating the quantity 
\begin{equation}
    \rho^{\mathfrak{a}}\left(\lambda|\mathbf{I}^{\leftrightarrow}_{\rm real}\right)=\frac{\sum_{k\in \kappa^{\mathfrak{a}}(\mathbf{I}^{\leftrightarrow}_{\rm real})} \sum_{\chi\in \xi^{\mathfrak{a}}(\mathbf{I}^{\leftrightarrow}_{\rm real})} \delta\left(\frac{P^{\mathfrak{a}}_{\E}(k,\chi|\mathbf{I}^{\leftrightarrow}_{\rm real})}{P^{\mathfrak{a}}_{\E}(k|\mathbf{I}^{\leftrightarrow}_{\rm real})P^{\mathfrak{a}}_{\E}(\chi|\mathbf{I}^{\leftrightarrow}_{\rm real})},\lambda\right)}{\sum_{k\in \kappa^{\mathfrak{a}}(\mathbf{I}^{\leftrightarrow}_{\rm real})}\sum_{\chi\in \xi^{\mathfrak{a}}(\mathbf{I}^{\leftrightarrow}_{\rm real})}1}, \label{eq:rhoalambda}
\end{equation}
where $\mathfrak{a}\in \left\{\rightarrow,\leftarrow\right\}$, $P^{\leftarrow}_{\E}(k,\chi|\mathbf{I}^{\leftrightarrow}_{\rm real})=\sum_{k^{\rm out},\chi^{\rm in}} P_{\E}^{\leftarrow}(k,k^{\rm out},\chi^{\rm in},\chi|\I^{\leftrightarrow})$ and $P^{\rightarrow}_{\E}(k,\chi|\mathbf{I}^{\leftrightarrow}_{\rm real})=\sum_{k^{\rm in},\chi^{\rm out}} P_{\E}^{\rightarrow}(k^{\rm in},k,\chi,\chi^{\rm out}|\I^{\leftrightarrow}_{\rm real})$, where $\kappa^{\leftarrow}(\mathbf{I}^{\leftrightarrow}_{\rm real}) = \left\{k^{\rm in}_i(\mathbf{I}^{\leftrightarrow}_{\rm real}):i\in \V\right\}$, $\kappa^{\rightarrow}(\mathbf{I}^{\leftrightarrow}_{\rm real}) = \left\{k^{\rm out}_i(\mathbf{I}^{\leftrightarrow}_{\rm real}):i\in \V\right\}$, $\xi^{\leftarrow}(\mathbf{I}^{\leftrightarrow}_{\rm real}) = \left\{\chi^{\rm out}_a(\mathbf{I}^{\leftrightarrow}_{\rm real}):a\in \W\right\}$,  $\xi^{\rightarrow}(\mathbf{I}^{\leftrightarrow}_{\rm real}) = \left\{\chi^{\rm in}_a(\mathbf{I}^{\leftrightarrow}_{\rm real}):a\in \W\right\}$ and where $\delta(\cdot,\cdot)$ is the Kronecker delta function.   The results presented in Fig.~\ref{fig:dir_W_ratio} suggest that indeed degree-cardinality correlations are relatively weak across all directed hypergraphs considered in this work, which clarifies why in Table~\ref{tb:f_dir} the real-world data is well characterised by random hypergraphs without degree-cardinality correlations.

\begin{figure}[t]
     \centering
     \setlength{\unitlength}{0.1\textwidth}
     \hspace*{1.5cm}
     \includegraphics[width=0.95\textwidth]{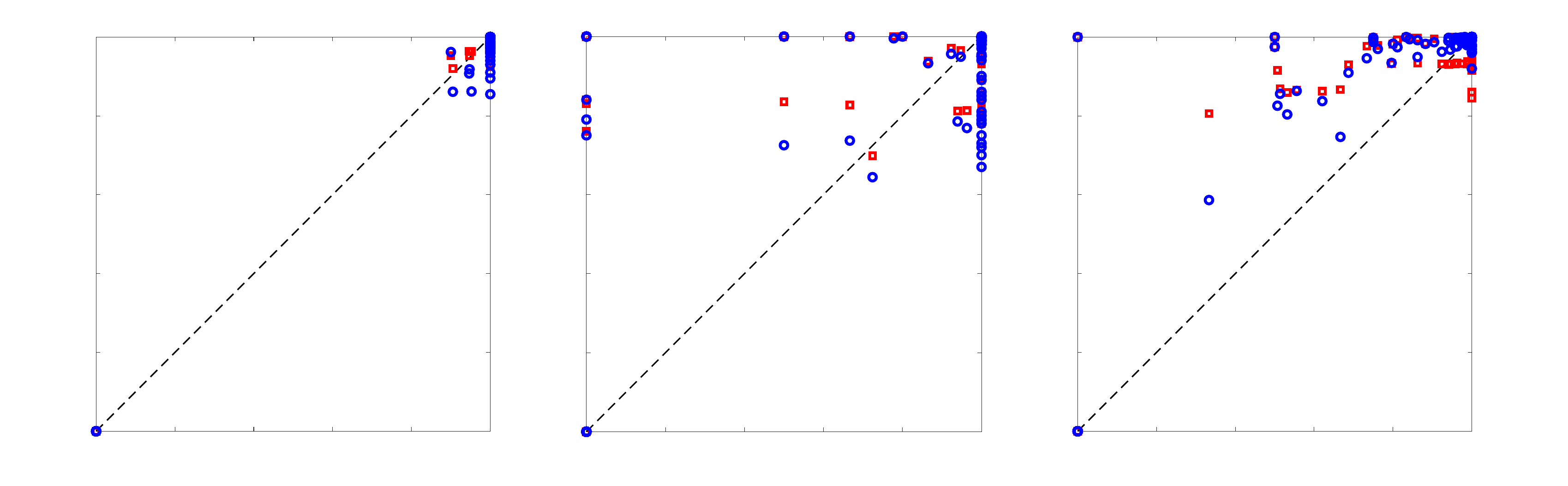}
      \put(-10.4,2.8){\small $(a)$}
      \put(-6.3,2.8){\small $(b)$}
      \put(-3.3,2.8){\small $(c)$}
      \put(-10.35,1.8){\small $\langle f^{\rm sc}(k^{\rm in},$}
      \put(-10.1,1.55){\small $k^{\rm out};\I^{\leftrightarrow})\rangle_{ \mathfrak{a}}$}
      \put(-5.7,-0.2){\small $f^{\rm sc}(k^{\rm in},k^{\rm out};\I^{\leftrightarrow}_{\rm real})$}
      \put(-8.59,0.12){\footnotesize$0.2$}
      \put(-8.11,0.12){\footnotesize$0.4$}
      \put(-7.63,0.12){\footnotesize$0.6$}
      \put(-7.15,0.12){\footnotesize$0.8$}
      \put(-6.59,0.12){\footnotesize$1$}
      \put(-6.11,0.12){\footnotesize$0$}
      \put(-5.6,0.12){\footnotesize$0.2$}
      \put(-5.12,0.12){\footnotesize$0.4$}
      \put(-4.64,0.12){\footnotesize$0.6$}
      \put(-4.16,0.12){\footnotesize$0.8$}
      \put(-3.6,0.12){\footnotesize$1$}
      \put(-3.17,0.12){\footnotesize$0$}
      \put(-2.66,0.12){\footnotesize$0.1$}
      \put(-2.16,0.12){\footnotesize$0.4$}
      \put(-1.68,0.12){\footnotesize$0.6$}
      \put(-1.20,0.12){\footnotesize$0.8$}
      \put(-0.64,0.12){\footnotesize$1$}
      \put(-9.1,0.12){\footnotesize$0$}
      \put(-9.25,0.8){\footnotesize$0.2$}
      \put(-9.25,1.3){\footnotesize$0.4$}
      \put(-9.25,1.8){\footnotesize$0.6$}
      \put(-9.25,2.3){\footnotesize$0.8$}
      \put(-9.1,2.65){\footnotesize$1$}
    \caption{Comparison between the fraction  $f^{\rm sc}(k^{\rm in},k^{\rm out};\I^{\leftrightarrow}_{\rm real})$ in the real-world hypergraph and the empirical probability $\langle f^{\rm sc}(k^{\rm in},k^{\rm out};\I^{\leftrightarrow})\rangle_{\mathfrak{a}}$ in synthetic hypergraphs ensemble with $\mathfrak{a}\in\{\rm corr, un\}$. The blue circles compare with random hypergraphs with degrees-cardinalities correlation $\langle f^{\rm sc}(k^{\rm in},k^{\rm out};\I^{\leftrightarrow})\rangle_{\rm corr}$, and the red squares compare with random hypergraphs without correlation $\langle f^{\rm sc}(k^{\rm in},k^{\rm out};\I^{\leftrightarrow})\rangle_{\rm un}$. The black dashed line denotes $y=x$. Each plots are extracted from $(a)$ {\it Human metabolic pathways}, $(b)$ {\it DNC-email}, and $(c)$ {\it English thesaurus}.}
    \label{fig:dir_gc_belonging_k}
\end{figure}
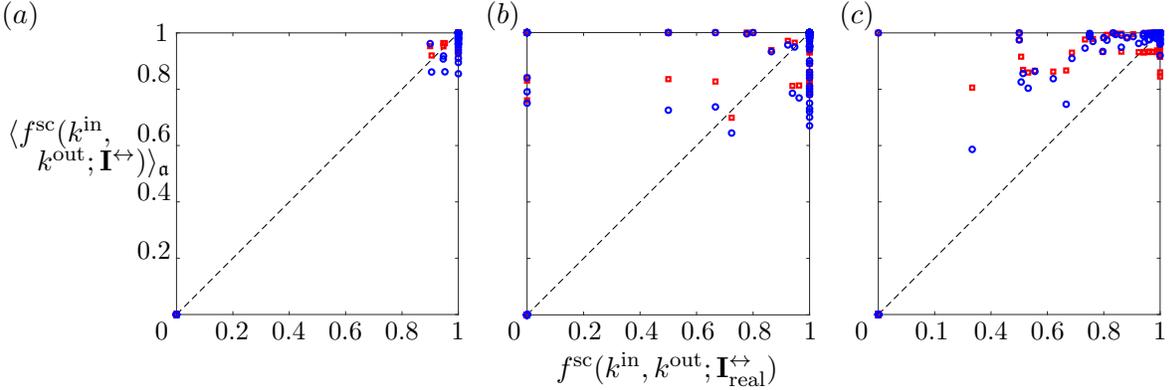

To further validate these findings  we consider the probability 
\begin{equation}
    f^{\rm sc}(k^{\rm in},k^{\rm out};\I^{\leftrightarrow}_{\rm real})\define\frac{\sum^N_{i=1}(1-\mu^{\rm sc}_{i}(\I^{\leftrightarrow}_{\rm real}))\delta_{k^{\rm in},k^{\rm in}_{i}(\I^{\leftarrow}_{\rm real})}\delta_{k^{\rm out},k^{\rm out}_{i}(\I^{\rightarrow}_{\rm real})}}{\sum^N_{i=1}\delta_{k^{\rm in},k^{\rm in}_{i}(\I^{\leftarrow}_{\rm real})}\delta_{k^{\rm out},k^{\rm out}_{i}(\I^{\rightarrow}_{\rm real})}}, \label{eq:deffsc}
\end{equation}
that a node $i\in \V$ with degrees  $(k^{\rm in}_i(\I^{\leftrightarrow}_{\rm real}), k^{\rm out}_i(\I^{\leftrightarrow}_{\rm real})) = (k^{\rm in},k^{\rm out})$ belongs to the largest strongly connected component.   In Eq.~(\ref{eq:deffsc}) the indicator variable $\mu^{\rm sc}_{i}(\I^{\leftrightarrow})=0$ if $i$ is part of the largest strongly connected component, and it is one otherwise.   In  
Fig.~\ref{fig:dir_gc_belonging_k} we compare the empirical values of $f^{\rm sc}(k^{\rm in},k^{\rm out};\I^{\leftrightarrow}_{\rm real})$ for the three real-world networks studied with the expected values $\langle f^{\rm sc}(k^{\rm in},k^{\rm out};\I^{\leftrightarrow})\rangle_{\rm corr}$ and $\langle f^{\rm sc}(k^{\rm in},k^{\rm out};\I^{\leftrightarrow})\rangle_{\rm un}$ in the configuration model without and with degree-cardinality correlations.   The findings in Fig.~\ref{fig:dir_gc_belonging_k} show, consitent with those in Fig.~\ref{fig:dir_W_ratio}, that   degree-cardinality correlations  are small in the real-world networks considered in this study.

\subsubsection{AND-logic}\label{sec:ANDlogicf}$\\$
Next, we investigate the properties of the largest AND-logic connected components in the metabolic pathways hypergraph.  We do not consider the  DNC-email hypergraph or the English synonyms hypergraph, as for these two hypergraphs all hyperedges have  in-cardinality equal to one, and therefore the OR-logic and AND-logic connected components are identical.  

We determine the fractions $f^{\mathfrak{a}}_{\rm AND}(\mathbf{I}^{\leftrightarrow}_{\rm real})$ of nodes that belong to the largest connected components with $\mathfrak{a}\in \left\{{\rm sc}, {\rm ic}, {\rm oc}, {\rm inter}, {\rm wc}, {\rm t}\right\}$, as defined in Eq.~(\ref{eq:fSc}). Note that for AND-logic we also calculate  the intersection $f^{\rm inter}_{\rm AND}(\mathbf{I}^{\leftrightarrow}_{\rm real})$ of the in- and out-components,  
since in AND-logic the strongly connected component differs from the intersection of in- and out-components.

To determine the largest AND-logic connected component, we use the algorithm developed in Sec.~\ref{sec:component_algo}, and for the corresponding out-components we use   the algorithm described in~\ref{app:det_AND_OC}.  Since the in-component  of the largest AND-logic strongly connected component equals the in-component of the largest OR-logic strongly connected component, we use for the in-component the algorithm for this latter.     Analogously,  the AND-logic weakly connected  component equals the OR-logic weakly connected component, and thus we use the  algorithm  for the latter to obtain the largest weakly connected component.

Table~\ref{tb:f_dir_AND} compares the empirical values   $f^{\mathfrak{a}}_{\rm AND}(\mathbf{I}^{\leftrightarrow}_{\rm real})$  with the corresponding theoretical estimates for random hypergraphs with and without degree-cardinality correlations:
\begin{itemize}
\item $\langle f^{\mathfrak{a}}_{\rm AND}(\mathbf{I}^{\leftrightarrow})\rangle_{\rm un}$: this quantity is computed with AND-logic for  the same ensemble of random hypergraphs as we computed $\langle f^{\mathfrak{a}}_{\rm OR}(\mathbf{I}^{\leftrightarrow})\rangle_{\rm un}$  (see previous section). As before, the estimates in Table~\ref{tb:f_dir_AND} are obtained from  empirical  averages over $100$ graph realisations.  
\item  $\langle f^{\mathfrak{a}}_{\rm AND}(\mathbf{I}^{\leftrightarrow})\rangle_{\rm corr}$: we compute this quantity for the same ensemble of hypergraphs as we computed     $\langle f^{\mathfrak{a}}_{\rm OR}(\mathbf{I}^{\leftrightarrow})\rangle_{\rm corr}$.
The estimates of $\langle f^{\mathfrak{a}}_{\rm AND}(\mathbf{I}^{\leftrightarrow})\rangle_{\rm corr}$ in the table are as before empirical averages over $100$ graph realisations.
\item $f^{\mathfrak{a}}_{\rm th}$: these are the theoretical values   $f^{\mathfrak{a}}_{\rm AND}$ with $\mathfrak{a}\in \left\{{\rm ic}, {\rm oc}, {\rm inter}, {\rm wc}, {\rm t}\right\}$ for  infinitely large, random hypergraphs that do not have degree-cardinality correlations; notice that again for notational simplicity  we omitted the AND in  $f^{\mathfrak{a}}_{\rm th}$.    As the AND-logic in-component equals the OR-logic in-component, we obtain the fractions $f^{\rm in}_{\rm th}=1-y^{\rm ic}$  from  solving the Eqs.~(\ref{eq:tree_structure_dir_uncorr}) and (\ref{eq:ICOC_fraction_dir_uncorr}) for $P_\V(k^{\rm in},k^{\rm out}) = P_{\V}(k^{\rm in},k^{\rm out}|\mathbf{I}^{\leftrightarrow}_{\rm real})$ and $P_\W(\chi^{\rm in},\chi^{\rm out}) = P_{\W}(\chi^{\rm in},\chi^{\rm out}|\mathbf{I}^{\leftrightarrow}_{\rm real})$.  On  the other hand, for $y^{\rm oc}$ we solve the  Eqs.~(\ref{eq:tree_structure_dir_uncorr_nonlinear}) and
(\ref{eq:ICOC_fraction_dir_uncorr_nonlinear}) together with the first three equations in (\ref{eq:tree_structure_dir_uncorr}) and (\ref{eq:ICOC_fraction_dir_uncorr}).    
The size of the intersection between the in-component and the out-component, $f^{{\rm inter}}_{\rm th}$ equals the right-hand side of 
Eq.~(\ref{eq:SC_form}) if   $\tilde{x}^{\rm ic}_{(k^{\rm in},k^{\rm out})} = \tilde{x}^{\rm ic}$ and $\tilde{x}^{\rm oc}_{(k^{\rm in},k^{\rm out})} = \tilde{x}^{\rm oc}$, with $\tilde{x}^{\rm ic}$ and $\tilde{x}^{\rm oc}$ the solutions to the first three equations (\ref{eq:tree_structure_dir_uncorr}) and (\ref{eq:tree_structure_dir_uncorr_nonlinear}).        For $f^{\rm wc}_{\rm th}$ we use the same approach  as for $f_{\rm th}$ in Sec.~\ref{ch:real_nondir}.  Lastly, $f^{\rm t}_{\rm th} = f^{\rm wc}_{\rm th}-f^{\rm in}_{\rm th} - f^{\rm out}_{\rm th} + f^{\rm inter}_{\rm th}$.   Note that in AND-logic we do not have a theoretical expression for $f^{\rm sc}_{\rm th}$, as the right-hand side of Eq.~(\ref{eq:SC_form})  provides us with the intersection between in- and out-components, which is different from the strongly connected component.

\item $f^{\mathfrak{a}, \rm corr}_{\rm th}$:   these are the theoretical values   $f^{\mathfrak{a}}$  with $\mathfrak{a}\in \left\{{\rm ic}, {\rm oc}, {\rm inter}, {\rm wc}, {\rm t}\right\}$ for  infinitely large, random hypergraphs that do  have degree-cardinality correlations.  Just as for the uncorrelated case, we do not have a theoretical estimate for $f^{\rm sc, corr}_{\rm th}$, as   Eq.~(\ref{eq:SC_form})  provides us  with the intersection instead of the largest strongly connected component.   The value of $f^{\rm in,\rm corr}_{\rm th}= 1-y^{\rm ic}$ where $y^{\rm ic}$  is found as  the solution to the  Eqs.~(\ref{eq:ICOC_fraction_dir}) and
(\ref{eq:tree_structure_dir}) for  distributions $P^{\rightarrow}_{\E}$ and $P^{\leftarrow}_{\E}$  that are equal to the ones of the metabolic pathway hypergraph; notice that these are the same equations as for the OR-logic in-component.   On the other hand, the size of the out-component, $f^{\rm out,\rm corr}_{\rm th}= 1-y^{\rm oc}$, is different from the one within OR-logic.   In AND-logic we obtain $y^{\rm oc}$ from the solution to  the set of equations consisting of (\ref{eq:ICOC_fraction_dir_nonlinear}), 
(\ref{eq:tree_structure_dir_nonlinear}), and the first three  equations of (\ref{eq:ICOC_fraction_dir}) and
(\ref{eq:tree_structure_dir}).
The fraction of nodes that occupy the intersection of the in- and out-components, $f^{\rm inter,\rm corr}_{\rm th}$ is given by the right-hand side of Eq.~(\ref{eq:SC_form}).  For  $f^{\rm wc, \rm corr}_{\rm th}$ we use the same procedure as for $f^{\rm corr}_{\rm th}$ with nondirected hypergraphs, see Sec.~\ref{ch:real_nondir}, and as before   $f^{\rm t}_{\rm th, \rm corr} = f^{\rm wc, \rm corr}_{\rm th}-f^{\rm in, \rm corr}_{\rm th} - f^{\rm out, \rm corr}_{\rm th} + f^{\rm inter, \rm corr}_{\rm th}$.

\end{itemize}

\begin{table}
\caption{{\it AND-logic connected components in directed hypergraphs:  comparison between  theoretical predictions and real-world data}.  See Sec.~\ref{sec:ANDlogicf} for a description of the computed quantities in the table.    }
\centering
\begin{tabular}{c|c|ccccccc}
\hline\hline
Dataset& $\mathfrak{a}$ & $f^{\mathfrak{a}}_{\rm AND}(\mathbf{I}^{\leftrightarrow}_{\rm real})$ & $\langle f^{\mathfrak{a}}_{\rm AND}(\mathbf{I}^{\leftrightarrow}))\rangle_{\rm un}$ & $f^{\mathfrak{a}}_{\rm th}$ &$\langle f^{\mathfrak{a}}_{\rm AND}(\mathbf{I}^{\leftrightarrow})\rangle_{\rm corr}$ & $f^{\mathfrak{a}, \rm corr}_{\rm th}$ \\
\hline
\multirow{1}{5em}{\centering Metabolic pathways}&\begin{tabular}{@{}c@{}c@{}c@{}c@{}}wc \\ ic \\ oc\\inter\\sc\\t\end{tabular}& \begin{tabular}{@{}c@{}c@{}c@{}c@{}} 0.9721 \\ 0.6439 \\ 0.2812 \\ 0.2222 \\0.2155\\ 0.2692 \end{tabular}& \begin{tabular}{@{}c@{}c@{}c@{}c@{}}0.9976 \\ 0.6754 \\ 0.3178\\0.1979\\0.1333\\0.2023\end{tabular}& \begin{tabular}{@{}c@{}c@{}c@{}c@{}}0.9808 \\ 0.6756 \\ 0.3010\\0.1884\\ \notableentry \\ 0.1926 \end{tabular}& \begin{tabular}{@{}c@{}c@{}c@{}c@{}} 0.9950 \\ 0.6573 \\ 0.3387 \\0.2066\\0.2054\\0.2056\end{tabular}& \begin{tabular}{@{}c@{}c@{}c@{}c@{}}0.9680 \\ 0.6576 \\ 0.3243\\0.1982\\  \notableentry \\  0.1843 \end{tabular}\\
\hline\hline
\end{tabular}\label{tb:f_dir_AND}
\end{table}

Interestingly, from the results in Table~\ref{tb:f_dir_AND} we conclude that $\langle f^{\rm sc}_{\rm AND}(\mathbf{I}^{\leftrightarrow})\rangle_{\rm corr}$ predicts well the real-world value $f^{\rm sc}_{\rm AND}(\mathbf{I}^{\leftrightarrow}_{\rm real})$, while $\langle f^{\rm sc}_{\rm AND}(\mathbf{I}^{\leftrightarrow})\rangle_{\rm un}$ provides a poor prediction of the same quantity.    This is unexpected as all other topological properties of the metabolic pathway hypergraph are well predicted by the configuration model without degree-cardinality correlations, including the value of $f^{\rm sc}_{\rm OR}(\mathbf{I}^{\leftrightarrow}_{\rm real})$ for OR-logic strongly connected components.     This example suggests  that degree-cardinality correlations have a stronger impact on percolation properties when these involve cooperative interactions.

\section{Discussion} \label{ch:discussion}
In the theory of random graphs, much attention goes to the study of connected components.  These are subgraphs consisting of nodes that are interconnected by paths.    The challenge in generalising connected components to hypergraphs is in accounting for the higher-order nature of the hyperedges representing interactions between system variables.  Indeed, the  most straightforward approach is to represent the hypergraph as a bipartite graph of nodes and hyperedges, and then use the usual definition of  connected components on this bipartite graph.    This yields what we have called OR-logic connected components.   However,  for OR-logic connected components the hyperedge represents a noncooperative interaction, which is not what we  in general want  when modelling systems with higher-order interactions~\cite{bianconi2024theory}.   Therefore, we have considered a second model of connected components in hypergraphs that we call the AND-logic connected components and that consider hyperedges as ``proper" higher interactions.   

We have shown that for nondirected hypergraphs  both definitions of connected components are equivalent, while for directed hypergraphs the AND-logic strongly connected component is  a subset of the OR-logic strongly connected component.  For directed hypergraphs, we have characterised the topological properties of AND-logic strongly connected components and have found that they are different from those of OR-logic strongly connected components, as illustrated in Figs.~\ref{fig:AND_component} and \ref{fig:example_conn_dir}.  Notably, in contrast with OR-logic connected components, for AND-logic the intersection between in- and out-components is in general not equal to the strongly connected component, which complicates the analytical analysis of AND-logic strongly connected components.   We also  developed  a numerical algorithm to determine the AND-logic strongly connected components of a hypergraph. 

Next, we have developed a theory for the size of connected components in infinitely large random hypergraphs, and we have used this theory to predict the size of connected components in real-world hypergraphs.    For nondirected hypergraphs, we have found that degree-cardinality correlations significantly improve  the predictions from the theory, as shown in Table~\ref{tb:f_nondir}.   For directed hypergraphs, we have found that connected components within OR-logic are well described by random hypergraphs without degree-cardinality correlations, see Table~\ref{tb:f_dir}.   However, for  AND-logic strongly connected components, we have found that degree-cardinality correlation  are essential  to describe the  size of the strongly connected component, see Table~\ref{tb:f_dir_AND}.   Note that the good agreement between cavity theory and real-world networks is unexpected, as  the former assumes the graph is locally tree-like, while the latter contains numerous loops,  community structure, and correlations beyond nearest neighbours.

We end the paper with a perspective and a few open problems that follow from this work.  We have used the cavity method to determine the nodes that belong to the connected components of large hypergraphs.    This approach works for OR-logic (strongly) connected components, in-components, and out-components.  However, determining the AND-logic strongly connected component remains an open problem.   This is because   the AND-logic strongly connected component is not the intersection between the in-component and the out-component, and this property is used by the cavity method to determine the strongly connected component of large, random, directed graphs.  

Finding the largest, AND-logic, strongly connected component of a directed hypergraph numerically is by itself an interesting discrete optimisation problem for which, to the best of our knowledge, little is known.    In particular, it remains to be understood to which complexity class the AND-logic strongly connected component problem belongs.    Preliminary  numerical results indicate that for random graphs the complexity scales, on average, as the square of the number of nodes (results  not shown).

In the hypergraph literature, there exist other definitions of strongly connected components in directed hypergraphs with AND-logic, which we have only recently become aware of.     Notably, Refs.~\cite{gallo1993directed,mohring2004scheduling,allamigeon2011strongly} define  strongly connected components using the notion of B-connectedness.    It remains to be understood how these other notions of strongly connected components in hypergraphs   are related to the AND-logic strongly connected components  defined in the present paper, and which definition is the most relevant one for the study of   dynamical systems on hypergraphs. 

In this Paper we have used OR-logic and AND-logic to define connected components  in  hypergraphs.   In both cases, the connected components  are the equivalence classes associated with   an equivalence relation defined on the set $\V\cup \W$.  
Although both OR-logic and AND-logic, requiring, respectively,  at least one or all in-neighbours of an hyperedge to be present, are  natural choices,  one can consider other logics associated with the hyperedges.   In this regard, the case studied in this paper with AND-logic should be seen as a first example that can inspire definitions of more general  models of connected components in hypergraphs.  It remains also to be understood whether giant connected components in hypergraphs play an important role for the dynamics of processes governed through them, which can be investigated with the dynamical version of the cavity method~
\cite{neri2009cavity,torrisi2022uncovering,hurry2022dynamics}.

\ack
G.-G. Ha thanks D.-S. Lee and M. Ha. This work was supported by the Engineering and Physical Sciences Research Council, part of the EPSRC DTP, Grant Ref No.: EP/V520019/1.

\section*{Data Availability Statement}

We used the databases {\it NDC-substances}\cite{benson2018simplicial}, {\it Youtube}\cite{kunegis2013konect,mislove2009online}, {\it Food recipe}\cite{whats-cooking}, {\it Github}\cite{kunegis2013konect,Scott2009GitHub}, {\it Crime involvement}\cite{kunegis2013konect} and  {\it Wallmart}\cite{amburg2020clustering} as the real-world undirected hypergraph. And as a directed hypergraph, we used {\it DNC-email}\cite{kunegis2013konect}, {\it English thesaurus}\cite{ward2002moby} and {\it Human metabolic pathways}\cite{karp2019biocyc} database. See~\ref{app:data_randH} for more detailed information about these datasets. And we have made the  numerical code  for  the AND-logic  connected components publicly available on  \url{https://github.com/Gyeong-GyunHa/andgc}.

\appendix

\section{Algorithm for  the AND-logic out-component}\label{app:det_AND_OC}

\begin{figure}
 \centering
 \setlength{\unitlength}{0.1\textwidth}
 \includegraphics[width=0.9\textwidth]{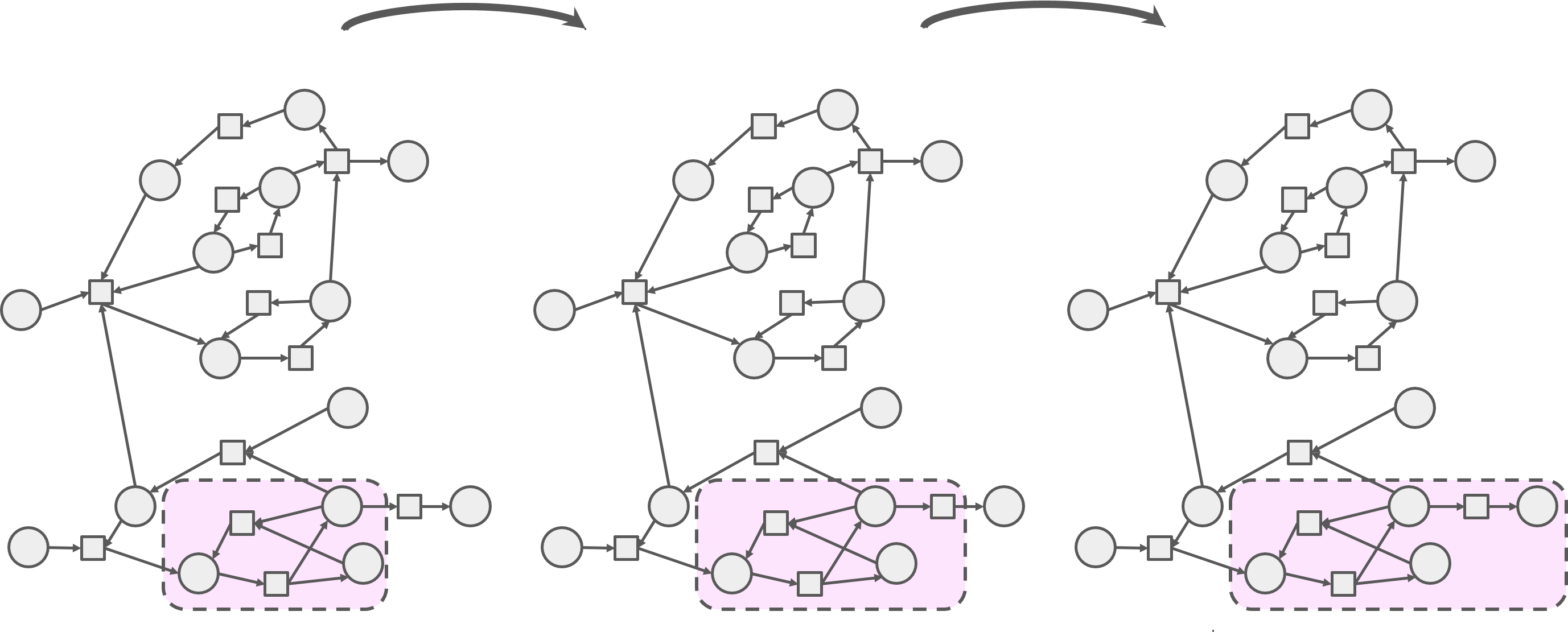}
      \put(-9.4,3.35){\small $(a)$}
      \put(-7.35,3.65){\it\small expand in hyperedge}
      \put(-6.15,3.35){\small $(b)$}
      \put(-3.8,3.65){\it\small expand in node}
      \put(-3.01,3.35){\small $(c)$}
 \put(-6.9,0.95){\footnotesize $\Hy^{\rm AND}_{\rm s}$}
 \put(-.6,1.0){\footnotesize $\Hy^{\rm AND}_{\rm out}$}
 \caption{{\it Illustration of the algorithm for determining the AND-logic out-component of a given AND-logic strongly connected component.} $(a)$ Initialisation: we include all nodes and hyperedges of  the given strongly connected component  $\Hy^{\rm AND}_{{\rm s}}$  into the out-component. $(b)$ Hyperedge expansion: hyperedges that are direct out-neighbours of nodes in  $\Hy^{\rm AND}_{{\rm out}}$ are added to   $\Hy^{\rm AND}_{{\rm out}}$ if   all of their in-neighbours are part of  $\Hy^{\rm AND}_{{\rm out}}$. $(c)$ Node expansion: all nodes that are out-neighbours of hyperedges in $\Hy^{\rm AND}_{{\rm out}}$ are added to  $\Hy^{\rm AND}_{{\rm out}}$.}
 \label{fig:eg_determine_AND_OC}
\end{figure}

We present an algorithm for determining the AND-logic out-component associated with a given AND-logic strongly connected component in a hypergraph. 
The pseudo-code of this algorithm is detailed in the tables entitled Algorithms~\ref{Al:AND-OC}, \ref{Al:OCnode} and~\ref{Al:OChyperedge}, and Fig.~\ref{fig:eg_determine_AND_OC} illustrates the processing steps.   The algorithm constructs iteratively the out-component by adding nodes and hyperedges to the sub-hypergraph $\mathcal{H}^{\rm AND}_{\rm out}$, until    $\mathcal{H}^{\rm AND}_{\rm out}$ equals the out-component of the hypergraph.  
The algorithm starts with including all  the nodes that belong to the  AND-logic strongly connected component of graph,  which is given as input the algorithm, to the AND-logic out-component, i.e., $\mathcal{H}^{\rm AND}_{\rm out} = \mathcal{H}^{\rm AND}_{\rm s}$.   Subsequently, the algorithm  iterates through two main phases, viz.,  the node expansion phase (described in Algorithm~\ref{Al:OCnode}) and the hyperedge expansion phase (described in Algorithm~\ref{Al:OChyperedge}):

\begin{algorithm} 
	\caption{\textsc{FindAND-OC}(Hypergraph $\Hy$, AND-SCC $\Hy^{\rm AND}_{\rm s}$, AND-OC $\Hy^{\rm AND}_{\rm out}$)}
	\begin{algorithmic}[1]
    
        \State $\Hy^{\rm AND}_{\rm out}$ $\leftarrow$ $\Hy^{\rm AND}_{\rm s}$ \Comment{Initialisation}
		\While {not done} 
        \State $\Hy^{\rm AND \ast}_{\rm out}$ $\leftarrow$\textsc{MoveNodes}($\Hy$,$\Hy^{\rm AND}_{\rm out}$) \Comment{add nodes}
        \State $\Hy^{\rm AND}_{\rm out}$ $\leftarrow$\textsc{CheckHyperedges}($\Hy$,$\Hy^{\rm AND  \ast}_{\rm out}$) \Comment{add hyperedges}
            \If {$\Hy^{\rm AND \ast}_{\rm out}$=$\Hy^{\rm AND}_{\rm out}$}
            \State done  \Comment{Termination}
            \EndIf
                \EndWhile
		\State \textbf{return} $\Hy^{\rm AND}_{\rm out}$
	\end{algorithmic} \label{Al:AND-OC}
\end{algorithm} 

\begin{algorithm}
	\caption{\textsc{MoveNodes}(Hypergraph $\Hy$, Current AND-OC $\Hy^{\rm AND}_{\rm out}$, Updated AND-OC $\Hy^{\rm AND \ast}_{\rm out}$)}
	\begin{algorithmic}[1]
		\State $\Hy^{\rm AND \ast}_{\rm out}$ $\leftarrow$ $\Hy^{\rm AND}_{\rm out}$
		\State $\W^{\rm AND}_{\rm out}=\{\alpha|\alpha\in {\W}(\Hy^{\rm AND \ast}_{\rm out})\}$  \Comment{all hyperedges}
		\For {$\alpha\in\W^{\rm AND}_{\rm out}$}  \Comment{Examine all hyperedges}
		\State $\V^{\rm out}_{\alpha}=\{i|i\in\partial^{\rm out}_{\alpha}(\Hy)\}$  \Comment{all its out-neighbours in original hypergraph}
		\For {$i\in\V^{\rm out}_{\alpha}$}  
            \If {$i\notin{\V}(\Hy^{\rm AND \ast}_{\rm out})$}  \Comment{$i$ is reachable node}
        		\State $\Hy^{\rm AND \ast}_{\rm out}$ $\leftarrow$ $i$.add()  \Comment{add the node}
                \EndIf
		\EndFor
		\EndFor
		\State \textbf{return} $\Hy^{\rm AND \ast}_{\rm out}$
	\end{algorithmic} \label{Al:OCnode}
\end{algorithm} 

\begin{algorithm}
	\caption{\textsc{CheckHyperedges}(Hypergraph $\Hy$, Current AND-OC $\Hy^{\rm AND \ast}_{\rm out}$, Updated AND-OC $\Hy^{\rm AND}_{\rm out}$)}
	\begin{algorithmic}[1]
		\State $\Hy^{\rm AND}_{\rm out}$ $\leftarrow$ $\Hy^{\rm AND \ast}_{\rm out}$
		\State $\W=\{\alpha|\alpha\in {\W}(\Hy) ~{\rm and}~ \alpha\notin {\W}(\Hy^{\rm AND}_{\rm out}) \}$  \Comment{every hyperedges not belong to $\Hy^{\rm AND}_{\rm out}$}
		\For {$\alpha\in\W$}  
		\State $\V=\{i|i\in\partial^{\rm in}_{\alpha}(\Hy)\}$  \Comment{all its in-neighbours in original hypergraph}
            \If {$\V\subset{\V}(\Hy^{\rm AND}_{\rm out})$}    \Comment{check whether the hyperedge satisfies AND-logic}
        		\State $\Hy^{\rm AND}_{\rm out}$ $\leftarrow$ $\alpha$.add()  \Comment{add the hyperedge}
                \EndIf
		\EndFor
		\State \textbf{return} $\Hy^{\rm AND}_{\rm out}$
	\end{algorithmic} \label{Al:OChyperedge}
\end{algorithm}

\begin{enumerate}
    \item {\it Node expansion} (Algorithm~\ref{Al:OCnode}): we add  to $\mathcal{H}^{\rm AND}_{\rm out}$ all nodes in $\mathcal{H}$ that belong to the out-neighbourhood sets $\partial^{\rm out}_{\alpha}$ of a hyperedge $\alpha$ that is part of  the sub-hypergraph $\mathcal{H}^{\rm AND}_{\rm out}$.  This step ensures that the out-component contains all reachable nodes.
    \item {\it Hyperedge expansion} (Algorithm~\ref{Al:OChyperedge}): we examine all hyperedges $\alpha$ in the original hypergraph that are out-neighbours of nodes in $\mathcal{H}^{\rm AND}_{\rm out}$ (and do not yet belong to  $\mathcal{H}^{\rm AND}_{\rm out}$). A hyperedge $\alpha$ is added to $\mathcal{H}^{\rm AND}_{\rm out}$ if all  of the  in-neighbours $i\in \partial^{\rm in}_\alpha$ of  the original hypergraph $\mathcal{H}$ are part of $\mathcal{H}^{\rm AND}_{\rm out}$.  This process is depicted in Figure~\ref{fig:eg_determine_AND_OC}$(b)$.
\end{enumerate}

The algorithm iterates  through these two phases until $\mathcal{H}^{\rm AND}_{\rm out}$ has converged, at which point we identify it as the  AND-logic out-component (see Figure~\ref{fig:eg_determine_AND_OC}$(c)$).

\section{Datasets for real-world hypergraphs}\label{app:data_randH}

In Sec.~\ref{ch:real_nondir} of this Paper,   
we have considered the  six nondirected hypergraphs based on the following data sets: 
   \begin{enumerate}
     \item {\it NDC-substances}~\cite{benson2018simplicial}: The nodes are substances, and the hyperedges are commercial drugs registered in by the U.S. Food and Drug Administration in the National Drug Code (NDC).   A node is linked to a hyperedge whenever the corresponding substance is used to synthesise the drug.  
     \item {\it Youtube}~\cite{kunegis2013konect,mislove2009online}: Nodes represent YouTube users and  hyperedges represent Youtube channels with paid subscription.   A user is linked to a hyperedge when the user pays for the  membership service.
     \item {\it Food recipe}~\cite{whats-cooking}: Nodes are ingredients and  hyperedges are recipes for food dishes. 
     \item {\it Github}~\cite{kunegis2013konect,Scott2009GitHub}: Nodes are  GitHub users and  hyperedges are GitHub projects.  A node is linked to a hyperedge whenever the corresponding user contributes to the GitHub project.   
     \item {\it Crime involvement}~\cite{kunegis2013konect}: The nodes are  suspects, and the hyperedges are  crime cases.   Nodes are linked to hyperedges whenever the corresponding suspects are involved with the crime investigation.  
     \item  {\it Wallmart}~\cite{amburg2020clustering}: Nodes are products sold by Walmart, and the  hyperedges represent  purchase orders.  Nodes are linked to hyperedges whenever the corresponding products are  part of the purchased order.  
   \end{enumerate}

\begin{table}
\caption{Characteristics of the  real-world hypergraphs considered in this Paper: number of nodes, $N$; number of   hyperedges, $M$; mean degree, $\overline{k}$; and mean cardinality, $\overline{\chi}$.  The last line of the table is a NDC-substance network for which all multiple  hyperedges have been removed, yielding a hypergraph.  }\label{tb:data}
\centering
\begin{tabular}{cccccccccccc}
\hline\hline
Dataset & $N$ & $M$ & $\overline{k}$ & $\overline{\chi}$\\
\hline
Food recipe & 6,714 & 39,774 & 63.8 & 10.8 \\
Wallmart & 88,860 & 69,906 & 5.2 & 6.6 \\
Youtube & 94,238 & 30,087 & 3.1 & 9.8 \\
Crime involvement & 829 & 551 & 1.8 & 2.7 \\
Github & 56,519 & 120,867 & 7.8 & 3.6 \\
NDC-substances & 5,556 & 112,919 & 12.2 & 2.0 \\
NDC-substances (removed edges) & 5,556 & 10,273 & - & - \\
\hline\hline
\end{tabular}
\end{table}

In Sec.~\ref{ch:real_dir}, we have considered three directed hypergraphs: 
   \begin{enumerate}
     \item {\it DNC-email}~\cite{kunegis2013konect}:  Nodes are  users sending and receiving emails and  hyperedges are  emails that are part of the 2016 Democratic National Committee (DNC) email leak.  Hyperedges are directed from the sender to its recipients.    Since an email always has  a single sender, the in-cardinality  of each  hyperedge equals  one.   
     \item {\it Human metabolic pathways}~\cite{karp2019biocyc}: Nodes represent metabolic compounds in the human metabolism, and  hyperedges are   metabolic reactions.   A hyperedge is directed from the reactants towards the products of the metabolic reaction.  Since many reactions are irreversible, this hypergraph is directed.
     \item {\it English thesaurus}~\cite{ward2002moby}: Nodes are English words and  hyperedges represent synonym relations between words.   Hyperedges are directed from a root word to target words.  Since not all words occur as root words, the hypergraph is directed.    The in-cardinality of each hyperedge equals to one.   
   \end{enumerate}

\begin{table}
\caption{
Network characteristics of the real-world directed hypergraphs: number of nodes, $N$; and hyperedges, $M$.}\label{tb:dirdata}
\centering
\begin{tabular}{cccccccc}
\hline\hline
Dataset & $N$ & $M$\\
\hline
Metabolic pathways & 1,508 & 1,451 \\
DNC-email & 2,029 & 5,598 \\
English thesaurus & 40,963 & 35,104 \\
\hline\hline
\end{tabular}
\end{table}

\section{Generating random hypergraphs with prescribed degree-cardinality correlations}\label{app:gen_hyper}

This Appendix presents the algorithms we use in Secs.~\ref{ch:real_nondir} and \ref{ch:real_dir} for generating  synthetic,  random hypergraphs that have the same degree-cardinality correlations as those of a given real-world hypergraph.  The algorithm is based on the stub-matching   method~\cite{britton2006generating,bassler2015exact}.   We consider in detail the case of nondirected hypergraphs, and at the end of the appendix we briefly discuss how to generate directed hypergraphs with degree-cardinality correlations.

\begin{figure}[t]
 \centering
 \setlength{\unitlength}{0.1\textwidth}
 \begin{subfigure}{0.23\textwidth}
 \includegraphics[width=\textwidth]{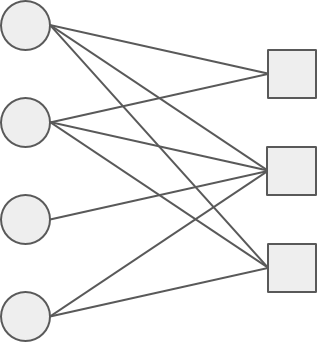}
\end{subfigure}
 \begin{subfigure}{0.30\textwidth}
\centering
        \begin{tabular}{@{} ccc@{}}
\hline\hline
            $k$ & $\chi$ & $\mathcal{T}_{k,\chi}$ \\
\hline
            1&   4&   1\\
            2&   3&   1\\
            2&   4&   1\\
            3&   2&   2\\
            3&   3&   2\\
            3&   4&   2\\
\hline\hline
        \end{tabular}
\end{subfigure}
      \put(-5.8,1.3){\normalsize $(a)$}
      \put(-2.8,1.3){\normalsize $(b)$}
 \caption{{\it Example joint degree-cardinality matrix $\mathcal{T}$ for a given hypergraph of interest.} $(a)$ Illustration of the  given hypergraph. $(b)$ The entries $\mathcal{T}_{k,\chi}$ of the joint degree-cardinality matrix of the hypergraph  $\mathbf{I}$ shown in $(a)$ equal $\mathcal{T}_{k,\chi}$, with $\mathcal{T}_{k,\chi} = \left\{(a,\alpha)\in \mathcal{E}: k_a(\mathbf{I}) =k \quad {\rm and}  \quad \chi_\alpha(\mathbf{I})=\chi\right\}$.}
 \label{fig:random hypergraph}
\end{figure}

First we extract the 
degree sequence $\vec{k}(\mathbf{I}_{\rm real})$, the cardinality sequence  $\vec{\chi}(\mathbf{I}_{\rm real})$, and the joint degree-cardinality matrix $\mathcal{T}(\mathbf{I}_{\rm real})$ of the hypergraph $\mathbf{I}_{\rm real}$, where we used  $\mathbf{I}_{\rm real}$ for the incidence matrix of the  real-world hypergraph of interest.   The entries $\mathcal{T}_{k,\chi}(\mathbf{I}_{\rm real})$ of this matrix denotes  the total number of links in the hypergraph that connect nodes of degree $k$ with hyperedges of cardinality $\chi$.  An example of a joint degree-cardinality matrix is shown in Fig.~\ref{fig:random hypergraph}.   

Next, the algorithm assigns to each node $a$ and each hyperedge  $\alpha$ a number $k_a(\mathbf{I}_{\rm real})$ and $\chi_{\alpha}(\mathbf{I}_{\rm real})$ of stubs, respectively.   A stub is  an ``unconnected" link, in the sense that one of its end points is connected to a vertex but the other endpoint is free.  We call stubs connected to nodes, node-stubs; and stubs connected to hyperedges, edge-stubs.   
The generation of the hypergraph is completed by matching each node-stub with  a unique edge-stub in a manner that  preserves the  degree-cardinality correlations as prescribed by $\mathcal{T}$.      

This procedure implements the following steps for each degree $k\in\{1,2,\dots,M\}$:

\begin{enumerate}
    \item {\it Extracting all the node-stubs of  degree $k$}: we retrieve all node-stubs attached to nodes of a given degree $k$.
    \item {\it Extracting stubs with relevant cardinality}: For each value of $\chi\in\{1,2,\dots,N\}$, we uniformly and randomly select a number $\mathcal{T}_{k,\chi}$ of edge-stubs attached to hyperedges of cardinality $\chi$. 
    \item {\it Matching stubs}:    We uniformly and randomly match the  $\sum_\chi\mathcal{T}_{k,\chi}$ node-stubs extracted in (i) with the  $\sum_\chi\mathcal{T}_{k,\chi}$  edge-stubs extracted in (ii).    The matched node and edge-stubs are removed from the hypergraph, as they have formed links.   
\end{enumerate}

For directed hypergraphs, a similar approach applies, but in this case there are two joint degree matrices, viz., $\mathcal{T}^{\rightarrow}_{(k^{\rm in},k^{\rm out}),(\chi^{\rm in},\chi^{\rm out})}$  and $\mathcal{T}^{\leftarrow}_{(k^{\rm in},k^{\rm out}),(\chi^{\rm in},\chi^{\rm out})}$, corresponding with links that are directed from nodes to hyperedges or from hyperedges to nodes, respectively.   The algorithm assigns directed stubs to the nodes and edges, and these are then matched with each other according to the statistics provided by the two joint degree matrices.

\section{Cavity method for AND-logic giant components}\label{app:cavity_dir_AND}

In this Appendix we develop  the cavity method for  giant components in random hypergraphs under AND-logic constraints. While the general framework follows the approach developed for OR-logic in Sec.~\ref{ch:cavity_dir}, the AND-logic  implies a different update rule for the variables $\sigma^{\rm oc}_\alpha$ in Eqs.~(\ref{eq:ICOC_dir}) and the variables $\sigma^{{\rm oc},(i)}_{\alpha}$ in (\ref{eq:cavity_ICOC}).   Indeed, in the OR-logic case, a node is considered part of a connected component if it can reach or be reached through at least one hyperedge. In contrast, under AND-logic, a hyperedge belongs to a connected component if  all its in-neighbours are also part of the connected component. Therefore, for AND-logic the fourth equation in  Eq.~(\ref{eq:ICOC_dir})  should be replaced by
\begin{equation}
    \sigma_{\alpha}^{\rm oc}(\I^\leftrightarrow) = 1-\prod_{i\in\partial_{\alpha}^{\rm in}}\left(1-\mu^{{\rm oc},(\alpha)}_{i}(\I^\leftrightarrow) \right),
    \label{eq:ICOC_dir_nonlinear} 
\end{equation} 
and the fourth equation of (~\ref{eq:cavity_ICOC}) should be replaced by \begin{equation}
    \sigma_{\alpha}^{{\rm oc},(i)}(\I^\leftrightarrow) = 1-\prod_{j\in\partial^{\rm in}_{\alpha}(\I);\atop j\neq i}\left(1-\mu^{{\rm oc},(\alpha)}_{j}(\I^\leftrightarrow)\right).
    \label{eq:cavity_ICOC_nonlinear}
\end{equation}
 Note that the right-hand side  of   Eqs.~(\ref{eq:ICOC_dir_nonlinear})  
 states that $\sigma_{\alpha}^{\rm oc}(\I^\leftrightarrow)=0$ 
  if   all the in-neighbours of $\alpha$ are  part of the out-component, i.e., $\mu^{{\rm oc},(\alpha)}_{i}=0$ for all $i\in\partial_{\alpha}^{\rm in}$, and similarly for the right-hand side of (\ref{eq:cavity_ICOC_nonlinear}).

To determine the  number of nodes and hyperedges that are part of the largest out-component and in-component in infinitely larger random hypergraphs with  two  prescribed  joint distributions $P^{\rightarrow }_\E$ and $P^{\leftarrow}_\E$, we derive equations for the ensemble averaged quantities $y^{\rm ic}=\langle\mu^{\rm ic}_{i}(\I^\leftrightarrow)\rangle$, $y^{\rm oc}=\langle\mu^{\rm oc}_{i}(\I^\leftrightarrow)\rangle$, $x^{\rm ic}=\langle\sigma^{\rm ic}_{\alpha}(\I^\leftrightarrow)\rangle$ and $x^{\rm oc}=\langle\sigma^{\rm oc}_{\alpha}(\I^\leftrightarrow)\rangle$.    This yields the same equations as in ~(\ref{eq:ICOC_fraction_dir}) and (\ref{eq:tree_structure_dir}), apart from 
\begin{equation}
    x^{\rm oc}=1-\sum_{\chi^{\rm in},\chi^{\rm out}}P_{\W}(\chi^{\rm in},\chi^{\rm out})\left(1-\tilde{y}^{\rm oc}_{(\chi^{\rm in},\chi^{\rm out})}\right)^{\chi^{\rm in}}.
    \label{eq:ICOC_fraction_dir_nonlinear}
\end{equation}
and
\begin{equation}
    \tilde{x}^{\rm oc}_{(k^{\rm in},k^{\rm out})}=1-\sum_{\chi^{\rm in},\chi^{\rm out}}P_{\E}^{\leftarrow}(\chi^{\rm in},\chi^{\rm out}|k^{\rm in},k^{\rm out})\left(1-\tilde{y}^{\rm oc}_{(\chi^{\rm in},\chi^{\rm out})}\right)^{\chi^{\rm in}}.
    \label{eq:tree_structure_dir_nonlinear}
\end{equation}
Solving the Eqs.~(\ref{eq:ICOC_fraction_dir_nonlinear}) and
(\ref{eq:tree_structure_dir_nonlinear})  together with the three first equations in (\ref{eq:ICOC_fraction_dir}) and 
(\ref{eq:tree_structure_dir}), we obtain the fraction of nodes that occupy the largest out-component and in-component of a large hypergraph through $f^{\rm oc}_{\rm AND}=1-y^{\rm oc}$ and $f^{\rm ic}_{\rm AND} = 1-y^{\rm ic}$, respectively.  

In the simpler case when there are no correlations between degrees and cardinalities, the Eqs.~(\ref{eq:tree_structure_dir_nonlinear}) and
(\ref{eq:ICOC_fraction_dir_nonlinear}) simplify into 
\begin{equation}
    \tilde{x}^{\rm oc}=1-\sum_{\chi^{\rm in},\chi^{\rm out}}\frac{P_\W(\chi^{\rm in},\chi^{\rm out})\chi^{\rm out}}{\overline{\chi^{\rm out}}}\left(1-\tilde{y}^{\rm oc}\right)^{\chi^{\rm in}}
    \label{eq:tree_structure_dir_uncorr_nonlinear}
\end{equation}
and
\begin{equation}
    x^{\rm oc}=1-\sum_{\chi^{\rm in}}P_{\W}(\chi^{\rm in})\left(1-\tilde{y}^{\rm oc}\right)^{\chi^{\rm in}}.
    \label{eq:ICOC_fraction_dir_uncorr_nonlinear}
\end{equation}

Differently from the OR-logic case, the strongly connected component within AND-logic is {\it not} the intersection between the largest in- and out-component.   Therefore, Eq.~(\ref{eq:SC_form})  does not apply for the AND-logic strongly connected component.   Nevertheless, the right-hand side of  Eq.~(\ref{eq:SC_form})  provides us  the  relative size of the intersection between in- and out-components.

\section*{References}
\bibliographystyle{ieeetr}
\bibliography{biblio}

\end{document}